%% file: tach_pap.tex
\hoffset -1 true cm
\overfullrule=0 pt
\raggedbottom
\newcount\equationno      \equationno=0
\newtoks\chapterno \xdef\chapterno{}
\newdimen\tabledimen  \tabledimen=\hsize
%
%
%

\def\eqname#1{\global \advance \equationno by 1 \relax
\xdef#1{{\noexpand{\rm}(\chapterno\number\equationno)}}#1}
%
%
%
%
\def\tablet#1#2{
\vbox{\tabskip=1em plus 4em minus 0.9em
\halign to #1{#2}} }
%
\def\tabmidrule{\noalign{\smallskip\hrule\smallskip}}             
\input epsf
\input mn

\def\jcd{Christensen-Dalsgaard}

\begintopmatter
\title{Solar internal rotation rate and the latitudinal variation of 
the tachocline}
\author{H. M. Antia$^1$,  Sarbani Basu$^2$ and S. M. Chitre$^1$}

\affiliation{$^1$ Tata Institute of Fundamental Research, Homi Bhabha Road,
Mumbai 400 005, India}
\vskip 4pt
\affiliation{$^2$ Teoretisk Astrofysik Center,  Danmarks Grundforskningsfond,
Institut for Fysik og
Astronomi, \hfill\break
Aarhus Universitet, 
DK-8000 Aarhus C, Denmark}
\acceptedline{Accepted \ . Received \ }

\abstract{A new set of 
accurately measured frequencies of solar oscillations
are used to infer the rotation rate inside the Sun, as a function of
radial distance as well as latitude. We have adopted a regularized
least squares technique with iterative refinement for
both 1.5D inversion using
the splitting coefficients and 2D inversion using individual $m$
splittings. The inferred rotation rate agrees well
with earlier estimates showing a shear layer
just below the surface and another one around the base of the convection
zone. The tachocline or the transition layer where
the rotation rate changes from differential rotation in the convection zone
to almost latitudinally independent rotation rate in the radiative interior
is studied in detail. 
No compelling evidence for any latitudinal
variation in position and width of tachocline is found though it appears
that the tachocline probably shifts to slightly larger radial
distance at higher latitudes and possibly becomes thicker also.
However, these variations are within the estimated errors
and more accurate data would be needed to make a definitive statement
about latitudinal variations.
}


\maketitle

\section{Introduction}

The measured splittings of solar oscillation frequencies offer
us a valuable tool for studying the rotation rate inside the Sun.
It is possible
to obtain both radial and latitudinal variation in the rotation rate
(Schou, \jcd\ \& Thompson 1994). 
Various techniques have been employed
for inverting the splitting coefficients or even individual splittings
in a multiplet (Brown et al.~1989; Gough \& Thompson 1991;
Pijpers \& Thompson 1992; Sekii 1993; Wilson \& Burtonclay 1995;
Corbard et al.~1997).

The results obtained so far suggest that the observed surface differential
rotation of the Sun persists through the convection
zone (CZ). The rotation rate is nearly constant along different
latitudes in most of the CZ, while in the radiative interior it
is almost like rigid body rotation with a value
intermediate between that of the solar equator and pole at the surface
(cf., Thompson et al.~1996; Kosovichev et al.~1997).
The transition occurs over a fairly thin layer, which
is referred to as the ``tachocline'' (Spiegel \& Zahn 1992).  The thickness
of the transition layer seems to be smaller  than the best resolution
that is currently achieved by inversion methods. This layer contains
a substantial radial gradient of rotation velocity, of opposite signs in low
and high latitudes. It is widely believed that the tachocline with
its angular velocity gradients could be the seat of the dynamo
responsible for the solar magnetic cycle (Weiss 1994; Gilman \& Fox 1997).
The introduction of a toroidal magnetic field in this layer with
latitudinal differential rotation is naturally expected to lead to
interesting consequences for the operation of the dynamo and its
resultant manifestation in the solar surface activity.
The strong gradient in the 
rotation rate is also expected to produce turbulence which is likely 
to mix material just below the convection zone --- a phenomenon needed
to match the structure of solar models with the helioseismically
determined structure of the Sun (Richard et al.~1996; Basu 1997).
The accurate
measurement of solar internal rotation rate also provides strong
constraints on the theory of angular momentum transport in the stellar
interior (R\"udiger \& Kitchatinov 1996), which should contribute
to our understanding of the solar spin down over its lifetime.

In this work we investigate the internal
rotation rate of the Sun, with particular emphasis on the 
tachocline.  We have adopted the 1.5D inversion technique
for inverting the rotation rate from the measured splitting coefficients
(Ritzwoller \& Lavely 1991; Schou, \jcd\ \& Thompson 1994;
Antia \& Chitre 1996). We have also used a two-dimensional
inversion of the individual frequency splittings themselves.

The location and structure of the tachocline is thus crucial in 
many models of the solar dynamo and it has been the subject of several
detailed studies.  Using a simple forward
modelling Kosovichev~(1996) found that the tachocline is centered at a radial
distance of $(0.692\pm0.005)R_\odot$ and with a width of
$(0.09\pm0.04)R_\odot$, while Charbonneau et al.~(1997)
found it to be centered at $(0.704\pm0.003)R_\odot$ and with a width of
$(0.050\pm0.012)R_\odot$. Similar results were
also found by Basu~(1997), who found that the tachocline is centered
at $(0.7050\pm 0.0027)R_\odot$ with a half-width of $(0.0098\pm 0.0026)R_\odot$
which is $(0.0480\pm 0.0127)R_\odot$ when scaled to the width as
define by  Kosovichev~(1996) and Charbonneau et al.~(1997)
(see Section 3.1 for the definition of the width). 
Wilson, Burtonclay \& Li~(1996) find the
tachocline to be somewhat deeper at $r=0.68\pm0.01R_\odot$ and also
slightly thicker ($0.12R_\odot$).
While all these results are roughly in agreement with one another, the
differences in thickness are quite significant from the point of view
of dynamo models as well as the hydrodynamical stability. Similarly,
the exact location of tachocline with respect to the base of the convection
zone is also crucial.
Further, all these studies effectively assumed
that the position and thickness of tachocline are independent of latitude.
In this work, we attempt to find latitudinal variations in properties
of the tachocline and also use improved data from GONG network to obtain
better estimate for the tachocline properties. 
We use forward modelling techniques to detect
possible latitudinal variation in properties of the tachocline.
This includes the calibration method used by Basu (1997) and 
another technique based on simulated annealing. 

For this work we have used a number of different data sets obtained
by the Global Oscillations Network Group (GONG) project
(Hill et al.~1996). These data are
very precise and scan a large range of frequency and degree of
modes. Apart from this we also use the data from BBSO (Woodard
\& Libbrecht 1993) combined with splitting coefficients for low degree
modes as measured by BiSON (Elsworth et al.~1995).

The rest of the paper is organized as follows. In Section~2 we outline
the methods used to invert the data to obtain the solar rotation rate
and describe the inversion results. 
Techniques for determining whether there is any  latitudinal 
variation in the tachocline are summarized in Section~3.
In Section~4 we discuss results of inversion
when the contribution from the tachocline is removed from the data before
inversion and our conclusions are stated in Section~5.

\section {Inversions to determine the rotation rate}

The different modes of solar oscillations can be described by
three integers: the radial order $n$, the angular degree $\ell$ and the
azimuthal order $m$. The integers $\ell$ and $m$ are the degree
and order respectively of the spherical harmonic function used 
to describe the angular behaviour of the mode.
In a spherically symmetric, non-rotating star,
the frequency $\omega_{n,\ell,m}$
of an eigenmode is independent of $m$ and the mode is $(2\ell+1)$-fold
degenerate.
The spherical symmetry of the Sun is  broken by rotation, lifting
the degeneracy of the modes. The differences in frequency of modes
with the same $n$ and $\ell$, but different $m$, can be related to the
rotation rate in the Sun by
$$\eqalign{D_{n,\ell,m}&={\omega_{n,\ell,m}-\omega_{n,\ell,-m}\over 2m}\cr
&=\int_0^{R_\odot}\int_{-1}^1dr\;d\cos\theta\; K_{n,\ell,m}(r,\theta)
\Omega(r,\theta),\cr}\eqno\eqname\splitm$$
where the kernels $K_{n,\ell,m}(r,\theta)$ are defined by Pijpers~(1997).

Most helioseismic data sets do not contain frequencies of individual
modes or the individual splittings $D_{n,\ell,m}$ as defined by
equation~\splitm, but rather
frequencies of modes for a given $(n,\ell)$ are 
expressed as sum of polynomials in $m$, namely,
$$\omega_{n,\ell,m}= {\omega_{n,\ell }}+
\sum_{s=1}^{s_{\rm max}} c_s^{(n,\ell )}{\cal P}^{(\ell )}_s(m),\eqno\eqname\expa$$
where ${\cal P}^{(\ell )}_s(m)$ are suitable polynomials of
degree $s$, and generally, $s_{\rm max}<2\ell$.
For a proper choice of the polynomials, the individual inversion problems
for each splitting coefficient $c_s^{(n,\ell)}$ becomes decoupled from the rest
(Ritzwoller \& Lavely 1991).

The data from the GONG instrument are available as frequencies of all
individual modes, as well as splitting coefficients for the polynomials
as defined by Ritzwoller \& Lavely~(1991). We adopt both these forms
of data, and
use the so-called 1.5D inversion method (described below) to invert
the data in the form of splitting coefficients. 
This method has the advantage of being efficient in terms of
computing resources.
However, in order to exploit the full potential of the data we need to
invert the individual frequency splittings directly using a  two dimensional
inversion method.

\subsection{The 1.5D inversion}

The rotational splitting coefficients are sensitive only to the
component of rotation velocity that is symmetric about the equator and we
therefore assume the rotation velocity to be symmetric.
In order to determine the latitudinal dependence of the rotation rate, we
follow Ritzwoller and Lavely~(1991) and express the rotation velocity as
$$v_{\rm rot}(r,\theta)=\Omega(r,\theta)r\sin\theta=-\sum_{s=0}^\infty 
w_{2s+1}(r){\partial\over\partial\theta}Y_{2s+1}^0(\theta),
\eqno\eqname\rot  $$
where $\theta$ is the colatitude, $Y_k^0(\theta)$ are the spherical
harmonics and $w_s(r)$ are expansion coefficients
which are related to the splitting coefficients $c_s^{(n,\ell)}$
(cf., equation~\expa) by
$$c_s^{(n,\ell)}=\int_0^{R_{\odot}} w_s(r)K_s^{(n,\ell)}(r) r^2\;dr,
\eqno\eqname\split$$
where the kernels $K_s^{(n,\ell)}(r)$ are given by (Ritzwoller \& Lavely 1991)
$$K_s^{(n,\ell)}=-{\rho_0\over r}{(\xi_r^2+\ell(\ell+1)\xi_h^2-2\xi_r\xi_h
-{1\over2}s(s+1)\xi_h^2)\over\int_0^{R_\odot}\rho_0(\xi_r^2+
\ell(\ell+1)\xi_h^2)r^2\;dr}.\eqno\eqname\kernel$$
Here, $\rho_0(r)$ is the density in the equilibrium solar model, while 
$\xi_r$ and $\xi_h$ are respectively, the radial and horizontal components of
displacement eigenfunctions.
Using the splitting coefficients $c_s^{(n,\ell)}$ from the GONG data,
equation \split\ can be inverted to obtain $w_s(r)$.
The advantage of this choice for
expansion is that the resulting inverse problems for determining
the individual components $w_1(r)$, $w_3(r), \ldots$  get decoupled and
each component can be estimated independently.
The components $w_s(r)$ are
calculated by solving separate one-dimensional inversion problems with
the iterative refinement of the regularized least squares solution
(Antia, Chitre \& Thompson 1996). The rotation rate at any given radial
distance and colatitude can then be computed using equation~\rot.

We use cubic B-spline basis functions to represent the rotation
rate and the regularized least squares inversion is performed using the
singular value decomposition. The B-splines are defined over a set of
50 knots which are uniformly spaced in acoustic depth.
We have used only the first 6 terms of the expansion \rot,
as the higher splitting coefficients in the GONG data appear to be
dominated by random noise.
The observed rotational splitting coefficients
from GONG data for $\ell\le150$ and $1\le\nu\le3.5$~mHz
are used for inversion.
Further, in actual practice we directly find the
individual  components of rotation rate as defined by
$$
\Omega_s(r)={\sqrt{2s+1\over4\pi}}{w_s(r)\over r},
\eqno\eqname\compo
$$
instead of $w_s(r)$.

Since the inversion problem defined by equation~\split\ is in general
ill-conditioned, some regularization or smoothing is required to obtain
any meaningful solution in the presence of errors in observed data sets.
In order to study the sensitivity of inversion results on
smoothing prescription,
we have tried three different prescriptions for the regularized least
squares inversion:
(i) First derivative smoothing, where we minimize
$$
\eqalignno{
&\sum_{n,\ell}(\sigma_s^{(n,\ell)})^{-2}\left[c_s^{({n,\ell})}-
\int_0^{R_\odot} K_s^{(n,\ell)}(r) w_s(r)r^2\;dr\right]^2\cr
&\qquad\qquad+\lambda\int_0^{R_\odot}{1\over r}\left(dw_s\over dr\right)^2\;dr,
&\eqname\smooth\cr}
$$
Here $c_s^{(n,\ell)}$ are the observed splitting coefficients and
$\sigma_s^{(n,\ell)}$ the corresponding error
and $\lambda$ is the regularization parameter.
For this case, the rotation rate in the solar core --- where the amount 
of information is rather meager as the splitting data for low degree modes 
have large errors ---  tends to a constant value when sufficient
 smoothing is applied.

\beginfigure{1}
\hbox to 0 pt{\hskip -1.5cm
\vbox to 9.5 true cm{\vskip -0.5 true cm
\epsfysize=10.50 true cm\epsfbox{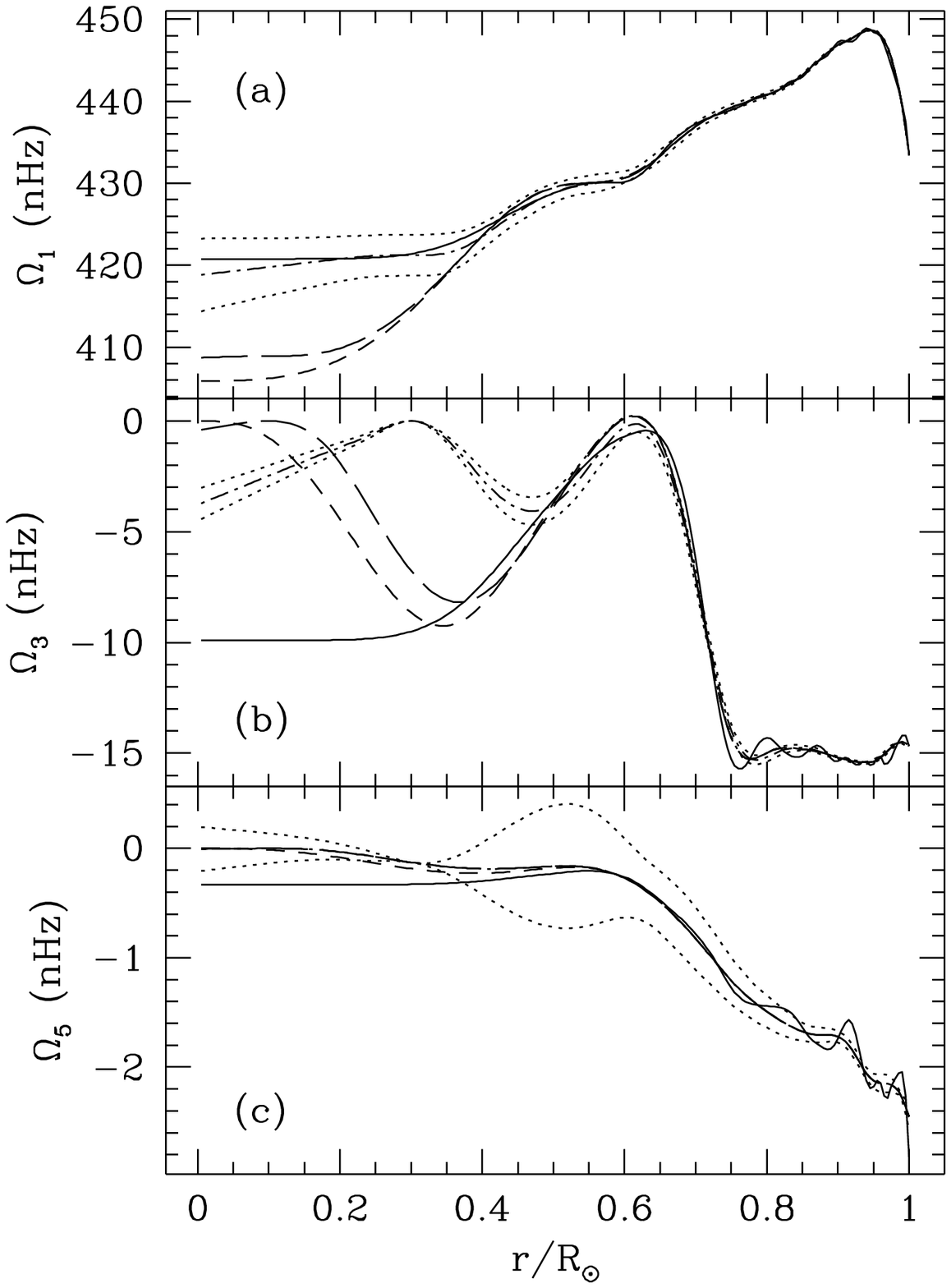}\vskip -0.5 true cm}}
\caption{\bf Figure 1. \rm  The first three components of the rotation
rate obtained with different prescriptions of smoothing. In all three panels
the continuous line is the result obtained with  first derivative 
smoothing. The short-dashed, long-dashed  and dot dashed line are  for 
second-derivative smoothing with the boundary conditions (equation (8)) applied
at $r=0$, $0.1$ and $0.3 R_\odot$
respectively. The dotted lines show 1$\sigma$ errors on solution
obtained with the second derivative smoothing and boundary conditions
applied at $0.3R_\odot$.
}
\endfigure

(ii) Second derivative smoothing, 
where we use the second derivative instead of the first in
the second term of equation~\smooth.
In this case, the rotation rate in the solar core tends to a monotonic
linear profile, which is perhaps
unrealistic as it may keep rising or falling depending on the gradient.
In order to overcome this problem we apply the following boundary conditions
at the center:
$$\eqalign{
&{d\Omega_i\over dr}=0,\cr
&  \Omega_i(0)=0\quad (i>1).\cr}\eqno\eqname\bc$$
The second boundary condition may have some justification, as there is
no information available
to determine the higher order coefficients $\Omega_i(r)$ ($i>1$) in the
solar core from the splitting data. This is because 
only the first splitting coefficient $c_1$ is known from observations
for the low degree modes
which sample the solar core .
This condition is also required to ensure regularity of rotation velocity
at the origin (Corbard et al.~1997).

(iii) This is same as (ii) except that the boundary conditions given by
equation~\bc\ are applied at a radial distance $r=0.3R_\odot$
(or $r=0.1R_\odot$).
These boundary conditions essentially try to constrain the rotation rate to
become constant in the region inside where the boundary conditions are applied.
This may be justified as we do not appear to have enough information
to determine the gradient in rotation rate in the core
(cf., Chaplin et al.~1996).

In all these cases, the method of iterative refinement effectively chooses the
regularization parameter $\lambda$ as explained by Antia, Chitre \&
Thompson~(1996). It is found that the regularization parameter increases
with the order $s$ of the splitting coefficient, which probably reflects
the fact that higher order coefficients are dominated by noise in most
regions.

\subsubsection{Results of 1.5D inversion}

We use the 1.5D inversion technique outlined above to infer the rotation
rate in the solar interior using the GONG data for months 4--14, which
consists of splitting coefficients for modes with $\ell\le150$.
The $\chi^2$ per degree of freedom is found to be close to unity
(between 1.1--1.2) in all cases.
In order to study the influence of smoothing on the inversion results
we perform inversion using different smoothing prescriptions given
in Section~2.1 and the results are displayed in Fig.~1. This
shows the rotation rate corresponding to the first three splitting
coefficients $c_1$, $c_3$ and $c_5$.
From the figure it is clear that the results in the convection zone are
not very sensitive to the choice of smoothing or to the point at which the
boundary conditions given in equation~\bc\ are applied.
Noticeable differences are seen
only for those parts of the Sun where the data have large errors.
The maximum difference of about 10 nHz occurs in the core in the 
latitudinally independent component $\Omega_1$. Similar differences
are seen for the component $\Omega_3$ also. These differences are
comparable to the error estimate in the inversion results arising from
errors in observed splitting coefficients. It may be noted that the error
estimates shown in the figure for the second derivative smoothing with
the boundary conditions applied at $r=0.3R_\odot$ show a decrease in the
core. This is artificial, and is entirely the result of boundary conditions.
In reality the errors should increase rapidly as $r$ decreases in the core.
Some of the differences in the core between the results using first and
second derivative smoothing are due to absence of boundary conditions in
the first derivative case. However, a part of the difference may also be due to
possible systematic errors in splitting coefficients of the low degree modes.
From this figure it is clear that for $r>0.5R_\odot$ the choice of smoothing
or the point where the boundary conditions are applied does not make
significant difference and in most of the work we have confined ourselves
to this region.
All the following results using 1.5D inversion have been obtained
using   second derivative smoothing
with the boundary conditions applied at $0.3R_\odot$.

\beginfigure{2}
\hbox to 0 pt{\hskip -1.5cm
\vbox to 9.5 true cm{\vskip -0.5 true cm
\epsfysize=10.50 true cm\epsfbox{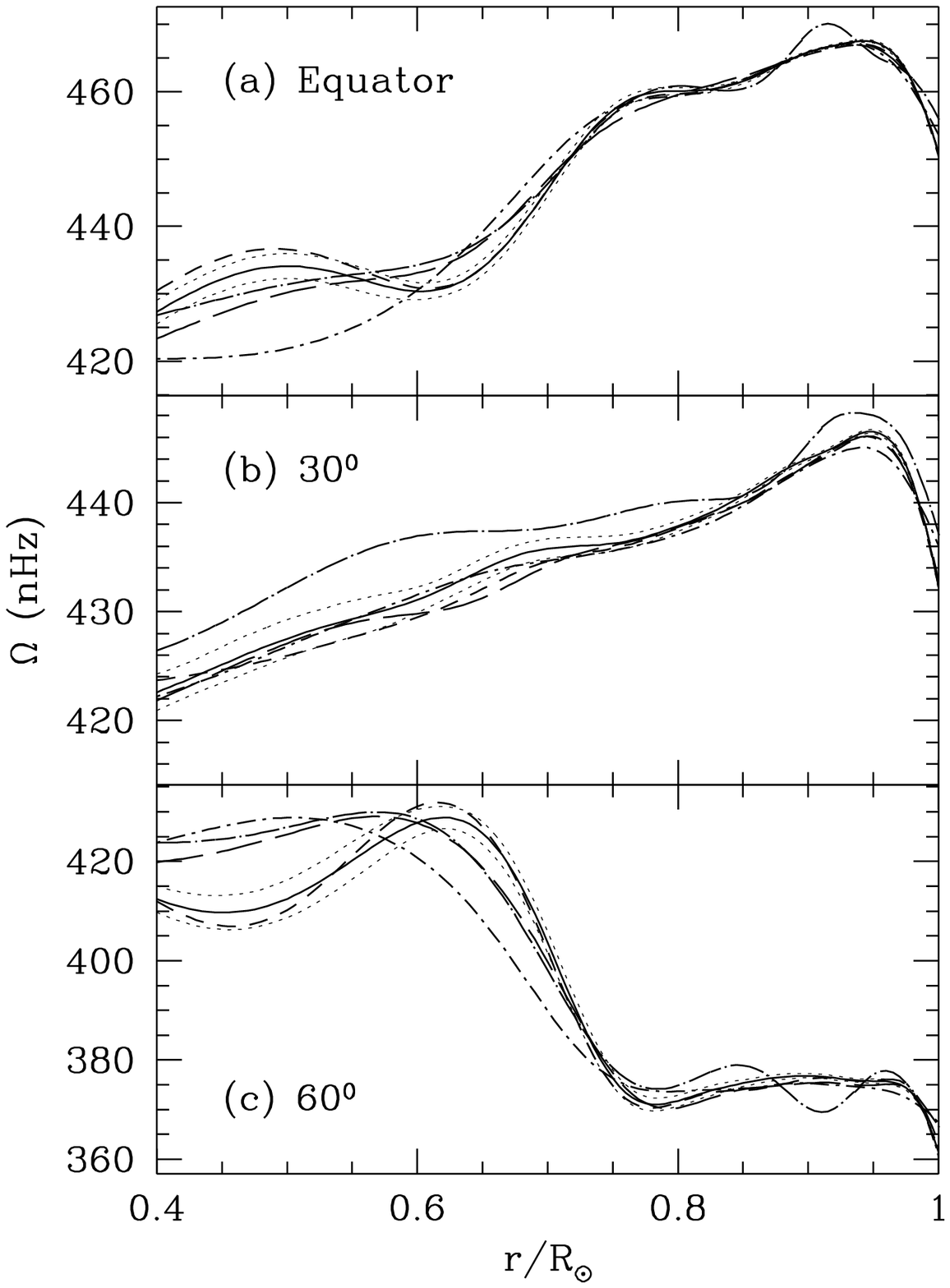}\vskip -0.5 true cm}}
\caption{\bf Figure 2. \rm  Solar rotation rate at the equator, $30^\circ$
and $60^\circ$  latitudes obtained using different data sets.
In all three panels, the continuous lines are the results obtained using the
GONG months 4--14 data with the dotted lines showing the $1\sigma$ error
limits. The short-dashed, long-dashed and dot-short dashed  lines are for
GONG months 4--10, months 4--7 and  month 10 data respectively,
while the dot-long dashed line is for the
BBSO+BiSON data combination. Note that while for all the GONG data sets we have
used splitting coefficient from $c_1$ to $c_{11}$,
for the BBSO+BiSON set we have used only the data up to $c_5$.}
\endfigure

\beginfigure{3}
\hbox to 0 pt{\hskip -1.5cm
\vbox to 6.5 true cm{\vskip -1.75 true cm
\epsfysize=10.50 true cm\epsfbox{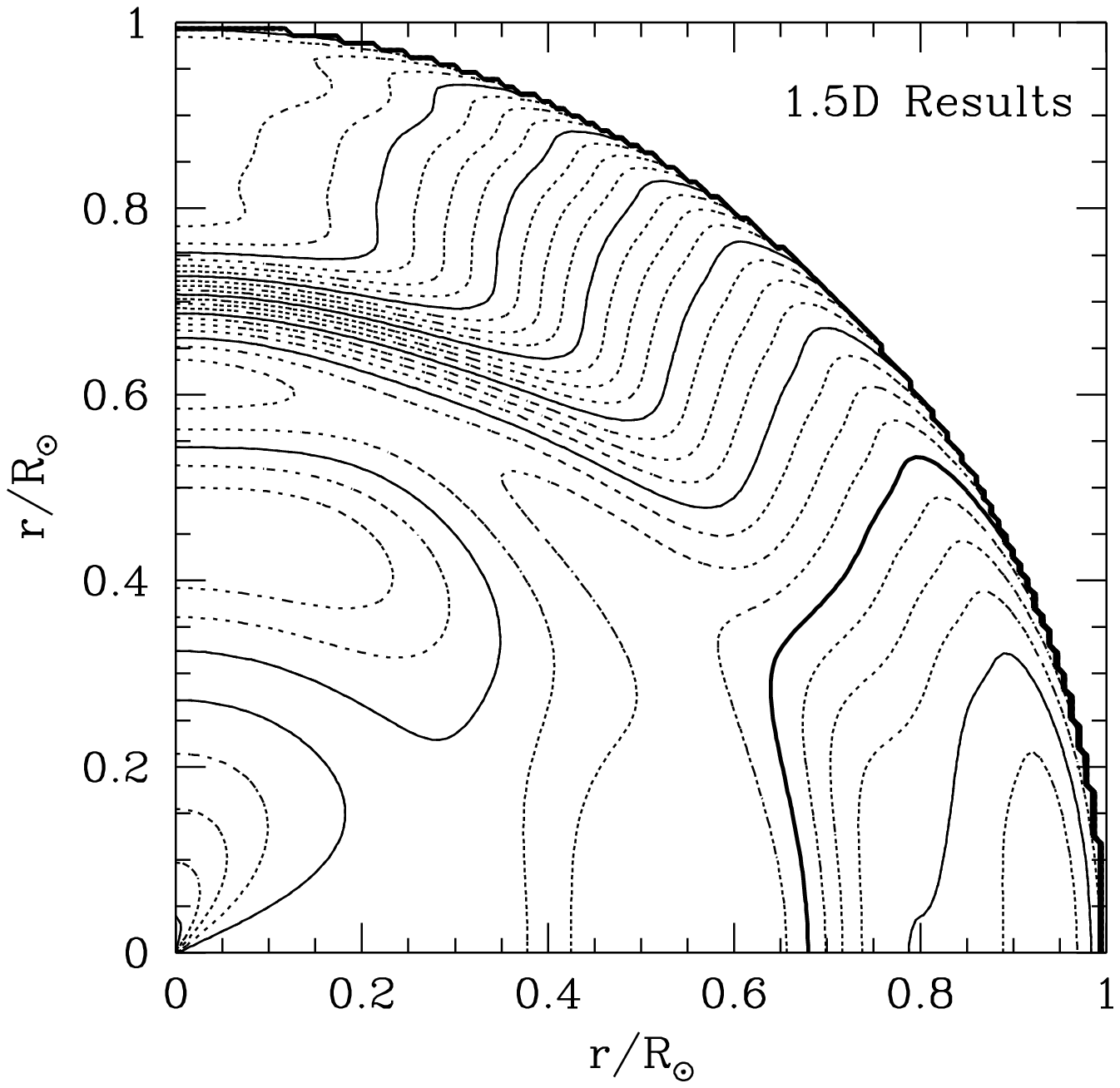}\vskip -2.25 true cm}}
\caption{\bf Figure 3. \rm A contour diagram of the solar rotation rate
as obtained by the 1.5D inversion technique using GONG months 4--14 data.
Due to the symmetry of the inversion results, the rotation rate has been shown
for just one quadrant only. The dotted contours have been drawn at 
intervals of 5 nHz., and the continuous ones at intervals of 
20 nHz. The thick continuous line is the contour at a level of 440 nHz.
The $x$-axis represents the solar equator while the $y$-axis represents
the rotation axis.
}
\endfigure

In order to study the sensitivity of the inversion technique to possible
systematic
errors in the input data sets we have repeated the inversion for various
sets of GONG data and the results are shown in Fig.~2. 
This figure also
includes the results obtained using the averaged  BBSO data 
for years 1986, 1988, 1989 and 1990 (Woodard \& Libbrecht 1993), combined
with the splittings for low degree modes from BiSON (Elsworth et al.~1995).
It is clear
that there is some systematic difference between different data sets, but
the difference is comparable to the error estimates arising from
inversions. This difference is also
comparable to that arising from  different smoothing prescriptions
as displayed in Fig.~1. The spread between various curves in Fig.~2 should
give an estimate of expected errors in inverted profiles, including
those arising from systematic errors in the input data. 
It may be noted that the data from GONG month 10, GONG months 4--7 and
BBSO$+$BiSON have larger errors as compared to that in the GONG months
4--14 data and this is reflected in tachocline region where the
GONG months 4--14 data appear to have higher resolution and hence the
tachocline is sharper.

A contour diagram showing the rotation rate inside the Sun as inferred
using the GONG months 4--14 data is shown in Fig.~3. It can be seen that
 the rotation rate is approximately constant
along
radial lines
in the convection zone. These results are similar to earlier inversions for rotation
rate (Thompson et al.~1996; Kosovichev et al.~1997).
The tachocline is clearly visible in these contour diagram.
Apart from the tachocline there, is another shear layer near the solar surface
where the rotation rate increases with depth. This shear layer appears
to extend to all latitudes and the rotation rate increases by about
17 nHz in this layer at the equator, but the change is lower at
higher latitudes. The maximum value of rotation rate occurs
around $r=0.95R_\odot$. The maximum rotation rate at the equator is
$467.5\pm0.2$ nHz at $r=0.945R_\odot$. The inverted rotation rate at the 
solar surface 
is close to that inferred from Doppler measurements (Snodgrass 1992).
The rotation rate in the radiative interior is more or less constant
and  some of the features seen in the contour diagram
do not appear to be significant. Although there is considerable uncertainty
in the estimate of the rotation rate in the core, but it appears
to be less than the surface equatorial rotation rate.

\subsection{2D inversion}

Although the 1.5D inversion technique described in Section~2.1 is very
efficient in terms of computing resources, it is not  clear
if the expansion of rotation rate given by equation~\rot, imposes any limitation
on the solution. A possible drawback of 1.5D inversion
is the loss of information, since 
the number of splitting coefficients is generally much smaller than the
number of individual splittings, but
it is not clear if the individual splittings
contain any more information than the first few splitting coefficients.
A more serious problem occurs in the inversion of higher order
coefficients, which have useful information only in the outer part of the
convection zone and as a result the solution in most of the interior is
essentially determined by the applied smoothing and boundary conditions.
A 2D representation of rotation rate will hopefully be able to overcome
this problem.
 Another drawback of the 1.5D inversion is that
all components $\Omega_s(r)$ make their maximum contribution at the pole
and further this maximum value is much larger than that in other regions.
This is particularly true for the higher order components. As a result,
the errors in the 1.5D inversions tend to get highly magnified near the pole
and may even give rise to spurious features if very high order terms are
included. In order to overcome these problems we attempt a 2D inversion
technique which does not use the expansion given by equation~\rot, but instead
directly represents the two dimensional function $\Omega(r,\theta)$ in
terms of suitable basis functions.

In order to solve the inversion problem defined by equation~\splitm\ 
 we represent the rotation rate
in terms of B-spline basis functions in $r$ and $\theta$,
$$\Omega(r,\theta)=\sum_{i=1}^{n_r}\sum_{j=1}^{n_\theta}b_{ij}\phi_i(r)
\psi_j(\cos\theta),\eqno\eqname\rottwo$$
where $b_{ij}$ are the coefficients of expansion and
$\phi_i(r)$ are the B-spline basis functions over $r$ and
$\psi_j(\cos\theta)$ are those over $\cos\theta$, $n_r$ and $n_\theta$
are the number of basis functions in $r$ and $\cos\theta$ respectively.
We use a set of
knots which are uniformly spaced in acoustic depth and $\cos\theta$
respectively to define $\phi_i(r)$ and $\psi_j(\cos\theta)$.

\beginfigure{4}
\hbox to 0 pt{\hskip -1.5cm
\vbox to 6.7 true cm{\vskip -1.75 true cm
\epsfysize=10.50 true cm\epsfbox{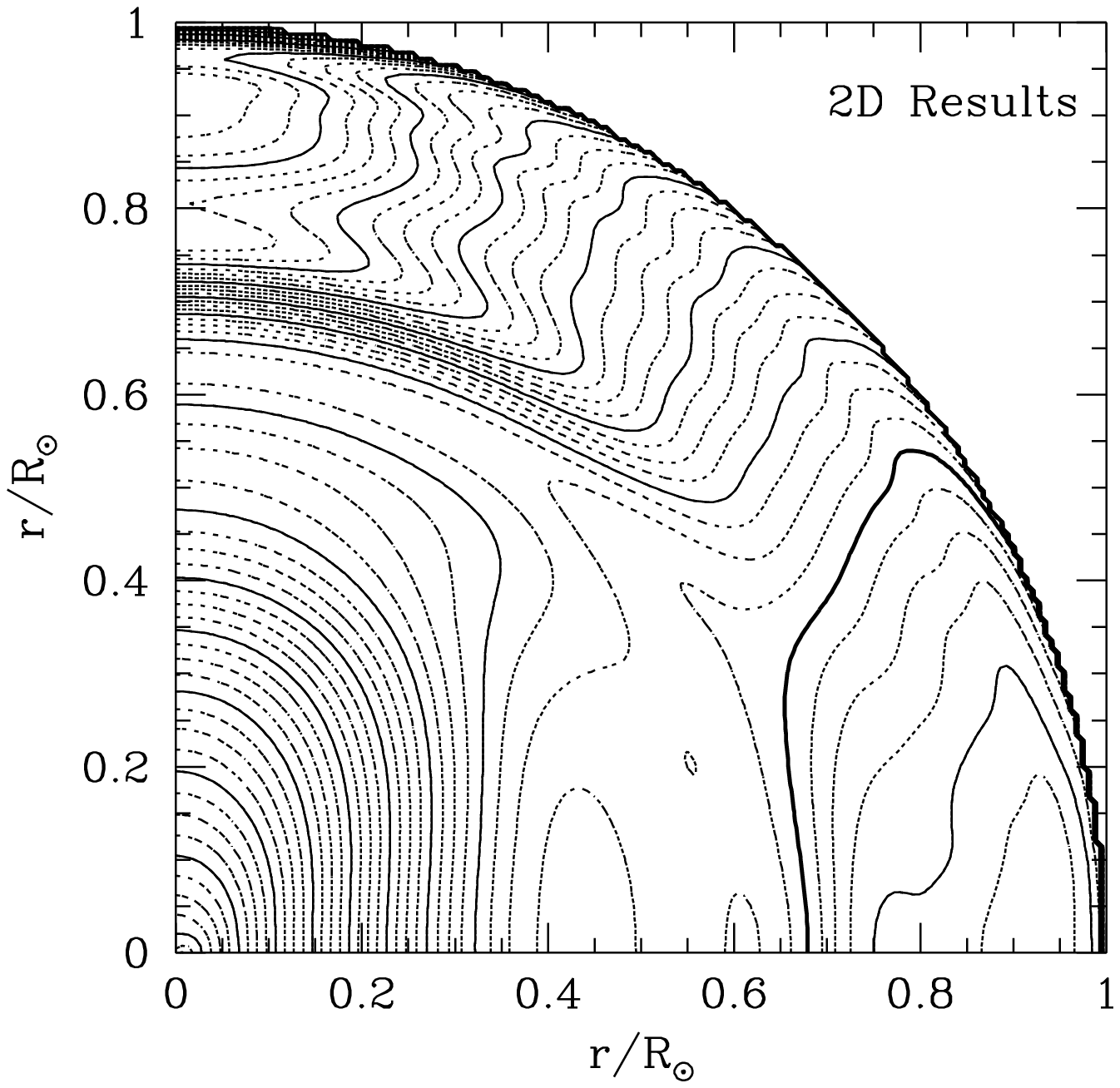}\vskip -2.05 true cm}}
\caption{\bf Figure 4. \rm A contour diagram of the solar rotation rate
as obtained by the 2D inversion of individual splittings using GONG
months 4--14 data.  The format is the same as that for Fig.~3.
}
\endfigure

This inversion problem is also solved using the regularized least squares
technique with iterative refinement. In this case, we have used only
second derivative smoothing, which involves minimizing
$$\eqalignno{
&\sum_{n,\ell,m}\sigma_{n,\ell,m}^{-2}\left[D_{n,\ell,m}-
\int_0^{R_\odot}dr\int_{-1}^1d\cos\theta K_{n,\ell,m}(r,\theta)
\Omega(r,\theta)\right]^2\cr
&\qquad+\lambda_r\int_0^{R_\odot}dr\int_{-1}^1d\cos\theta
r^{-1}\left(\partial^2\Omega\over \partial r^2\right)^2\cr
&\qquad+\lambda_\theta\int_0^{R_\odot}dr\int_{-1}^1d\cos\theta
\sin^2\theta\left(\partial^2\Omega\over \partial\cos\theta^2\right)^2, & 
\eqname\smoothr\cr}
$$
where, $\lambda_r$ and $\lambda_\theta$ are the two regularization parameters
controlling the smoothing. No boundary conditions are applied in this
case to constrain the rotation rate in the core.
We  have used 50 knots in $r$ and 30 knots in
$\cos\theta$ to represent the rotation rate. 

It is also possible to perform 2D inversion for splitting coefficients,
(Schou, \jcd\ \& Thompson 1994; Pijpers 1997) where the rotation rate is
expressed in terms of 2D basis functions (equation~\rottwo) and appropriate
combinations of individual splittings are constructed to relate the
corresponding splitting coefficients to the rotation rate.
In order to see whether the differences between the 1.5D and 2D results
are due to the expansion of rotation rate or the data, we have  done a
two-dimensional inversion for the splitting coefficients also. Thus we have
two sets of results using 2D inversions, one for 2D inversion of individual
splittings $D_{n,\ell,m}$, and the other for 2D inversion of the splitting
coefficients $c_s^{(n,\ell)}$.

\beginfigure{5}
\hbox to 0 pt{\hskip -1.5cm
\vbox to 9.5 true cm{\vskip -0.5 true cm
\epsfysize=10.50 true cm\epsfbox{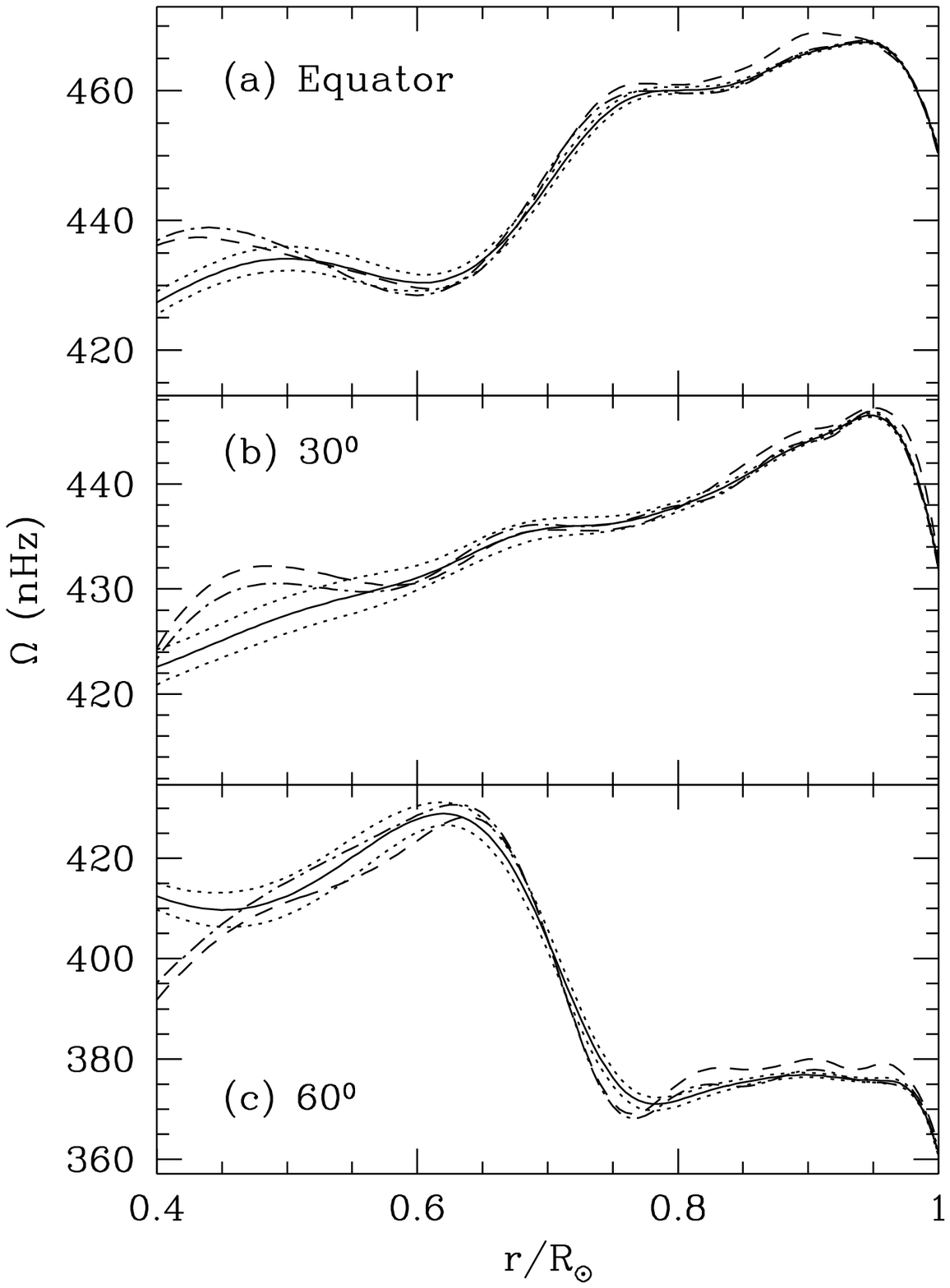}\vskip -0.5 true cm}}
\caption{\bf Figure 5. \rm  Comparison of the rotation inversion results at
fixed latitude  obtained by the 1.5D and  2D inversion methods.
The continuous line shows the results obtained by 
the 1.5D inversion method with the dotted lines showing the 1$\sigma$ error
limits.  The dashed line are those obtained by the 2D inversion of the
individual splittings and the dot-dashed line are those obtained by the
2D inversion of the splitting coefficients.
}
\endfigure

\beginfigure{6}
\hbox to 0 pt{\hskip -1.5cm
\vbox to 9.5 true cm{\vskip -0.5 true cm
\epsfysize=10.50 true cm\epsfbox{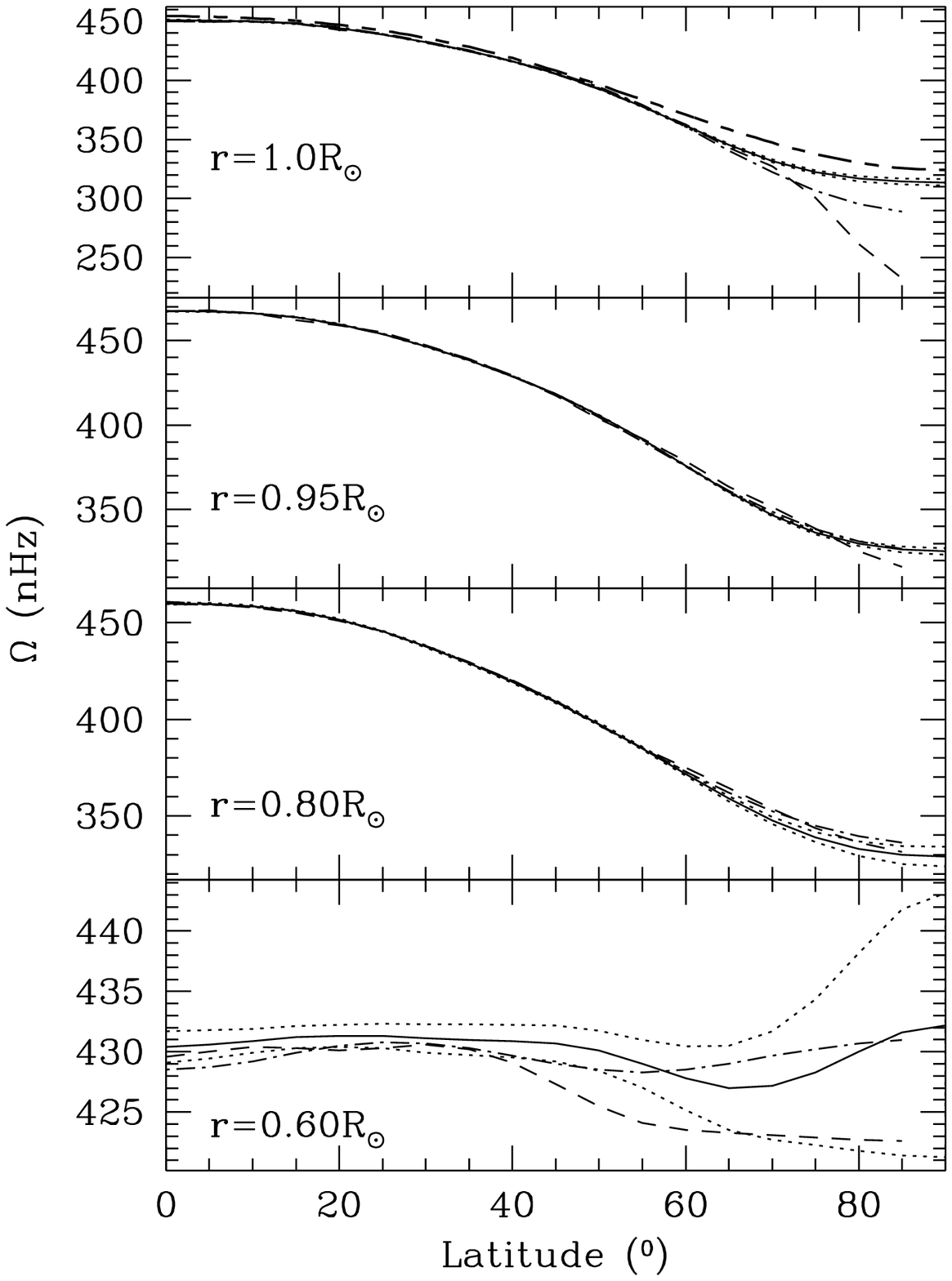}\vskip -0.5 true cm}}
\caption{\bf Figure 6. \rm  Comparison of the rotation inversion
results at fixed radius.  The line styles are the same as in Fig.~5.
In the top panel the heavy short dashed- long dashed line shows the
observed surface rotation rate as estimated by Doppler measurements
(Snodgrass 1992).
}
\endfigure

\subsubsection{Results of 2D inversion}

A contour diagram of the rotation rate inferred by 2D inversion of
GONG months 4--14 data, comprising of about 85000 splittings of individual
modes is shown in Fig.~4. The $\chi^2$ per degree of freedom in this case
is around 1.3.  Note that inside the CZ, the
results are  essentially similar to that obtained by the 1.5D method,
despite having a much larger number of splittings in 2D inversion.
Thus it appears that the
data are well represented by the 6 splitting coefficients.
However, in the radiative interior the solutions obtained using 1.5D
and 2D inversions are significantly different.
A large part of the difference is
due to the boundary conditions imposed in the 1.5D inversion which along
with the smoothing, tend to produce solid body rotation in the interior.
In the absence of any boundary condition, the 2D inversion technique
attempts to fit the splittings for low
degree modes which probably have some systematic errors,
and produces a sharply decreasing rotation rate in the
radiative interior. This decrease may not be real as it could result from
second derivative smoothing coupled with errors in data.

It  appears that the behaviour of the solutions in the polar regions
is also somewhat different. The 2D-solution shows a rapid decrease of the
rotation rate towards the pole similar to that seen in the data from the MDI
instrument on
board the SOHO spacecraft. The 1.5D result
also shows a decrease in the surface rotation rate at the pole, but
the reduction is not as much as in 2D inversion of individual splittings.
The errors in inversion increase
rapidly with latitude near the pole and as a result it is difficult
to discern any features at high latitudes reliably. Thus the
differences at high latitudes possibly reflect our inability to obtain
reliable inversion results in that region. 
However, from the form
of expansion of rotation rate (equation~\rot) in 1.5D inversion, it is clear
that contribution of each component ($\Omega_i$) has a strong maximum
at the pole and any error in these components will be highly magnified
there. This is particularly true of the higher order terms in the expansion.
It thus appears that 2D inversion of individual splittings, which does
not assume any particular expansion of rotation rate, may be able
to give better results near the pole, though the errors will still be large
and the smoothing will play a dominant role in determining the solution
near the poles.

In order to see whether the differences between the 1.5D and 2D results
are due to the representation of rotation rate or the data,
we have also done a two
dimensional inversion of the splitting coefficients. For this we use the
same splitting coefficients as were used in the 1.5D inversion but
expand the rotation rate in terms of 2 dimensional basis functions
using equation~\rottwo.  These results are
also shown in Figs.~5 and 6. Note that in the outer layers of the
Sun the results of the 1.5D and two-dimensional inversion of splitting
coefficients are
very close to each other, but slightly different from the results of the 2D
inversion of the individual splittings. In the deeper layers however, 
the results of 1.5D inversion are quite different from both the 2D inversions.
As already discussed this is due to the boundary
conditions (equation~\bc) that are applied in the 1.5D inversion method.

\beginfigure{7}
\hbox to 0 pt{\hskip -1.5cm
\vbox to 9.5 true cm{\vskip -0.5 true cm
\epsfysize=10.50 true cm\epsfbox{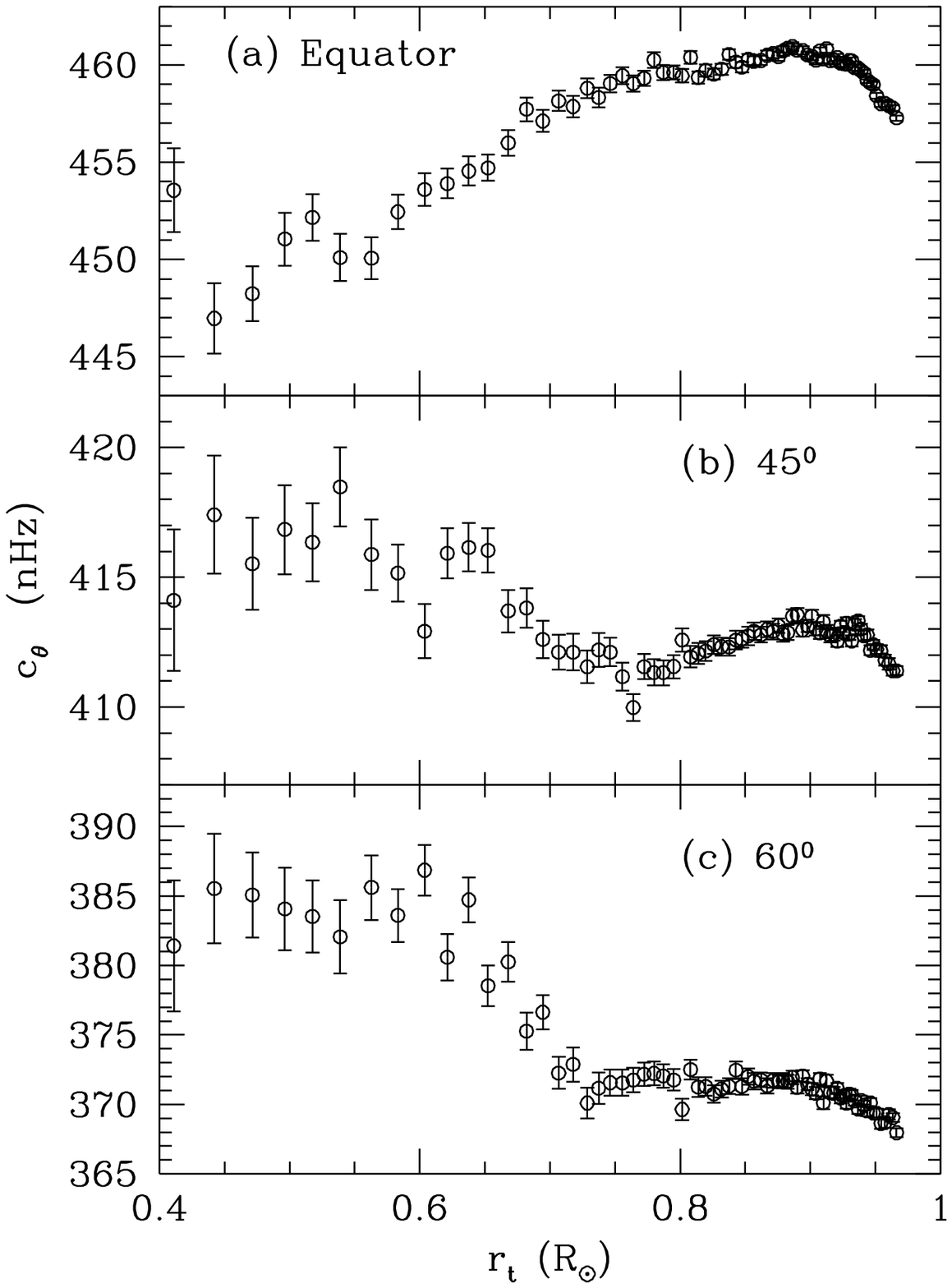}\vskip -0.5 true cm}}
\caption{\bf Figure 7. \rm  The splitting coefficients for GONG months
4--14 combined
to obtain  the data for different latitudes, which are marked in the
figure. The points represent the
combinations binned in groups of 15.
}
\endfigure

\section{The tachocline}

It can be seen from the results of the inversions that there is a
sharp transition close
to the base of the convection zone where the rotation rate changes from
differential rotation in the convection zone to a rotation rate that is
almost independent of latitude. Since the inversions are not able to resolve
the tachocline, other techniques have been employed to study this shear
layer. Even though inversions indicate that the tachocline is slightly
shallower and thicker at high latitudes, it is not clear if this
represents a real variation or is just an artifact of inversion
technique caused by the fact that the extent of jump increases with
latitude and the resolution of inversion techniques deteriorates with
increasing latitude.
For an independent confirmation of this variation
we construct combinations
of rotational splitting coefficients $c_s^{(n,\ell)}$, which give the
rotation velocity at some predefined co-latitude $\theta_0$. Thus
multiplying equation \split\ by $dY_s^0/d\theta|_{\theta=\theta_0}$
and summing over $s$ we get
$$
\sum_{s=0}^{\infty} c_{2s+1}^{(n,\ell)}
\left.{dY_{2s+1}^0\over d\theta}\right|_{\theta=\theta_0} =
-\int_0^{R_\odot}v_{\rm rot}(r,\theta_0) K^{(n,\ell)}(r)r^2\;dr.
\eqno\eqname\comb$$
Here, in principle, the kernel $K^{(\ell,n)}(r)$ also depends on $s$, 
but for simplicity we neglect the $s$ dependence, since the $s$-dependent
term is in general very small, as can be seen from equation~\kernel.
With this approximation equation~\comb\ reduces to an
one dimensional inversion problem at fixed latitude. Since inversion
cannot resolve the tachocline, we use a forward modeling technique
to estimate the parameters defining the tachocline.

We have first made the appropriate combinations of the splittings data 
for different latitudes using the coefficients $c_1$--$c_{11}$.
Fig.~7 shows the data, binned in groups
of 15 modes, plotted as a function of the lower turning point. We show the data
at the equator, $45^\circ$ and $60^\circ$ latitudes.
Note that the equator shows some hint of a jump, while $45^\circ$ and
$60^\circ$ latitudes show a clear jump.
We  concentrate on the latitudes which show
evidence of the transition and try to determine the magnitude of the
jump, position and thickness of the transition at latitudes of
$0^\circ,15^\circ,45^\circ,60^\circ$ and $75^\circ$. 
The data for the $30^\circ$
latitude does not give any clear indication of the 
transition, which is consistent with inversion results.  
The latitudinal variation of the
magnitude of the jump is obvious from the figure, but the change in position 
and thickness are not clear. We use both the calibration method used
by Basu~(1997) and another method based on simulated annealing to study
the tachocline at each latitude separately.

Apart from this we have also tried a 2D fit with some assumed form for
tachocline with latitudinal variation to simultaneously fit all the
splitting coefficients for obtaining the latitudinal variation in the
properties of the tachocline.

\subsection{Calibrating the tachocline}

The method followed here is the same as that used by Basu (1997). 
Data on frequency splittings for modes trapped in the
convection zone and those with turning points around the CZ base
are quite precise. The exact position and thickness of the 
tachocline can be determined by calibrating the difference in splittings 
between the Sun and models with known position and thickness of the tachocline.
While Basu (1997) assumed that the position and thickness of the
tachocline can be determined by the splitting coefficient $c_3$ alone, which
automatically ensures that the position and thickness of tachocline are
independent of latitude. In this work, we explicitly study the
latitudinal variation by applying the same procedure to data for each
of the chosen latitudes.

In order to estimate the position, thickness and the jump in the tachocline
we construct a series of models with known properties.
We parameterize the  calibration models with the rotation profile:
$$
\Omega_{\rm cal}=
{\delta\Omega\over {1+\exp[(r_d-r)/w]}},\eqno\eqname\thi$$
where 
$\delta\Omega$ is the jump in the tachocline,
$w$ is the half-width of the transition layer, and $r_d$ the
mid-point of the transition region.
Thus the rotation rate increases from a factor $1/(1+e)$ of its maximum value
to the factor  $1-1/(1+e)$ of its maximum value in the range $r=r_d-w$
to $r=r_d+w$.
The models described by equation~\thi\ have almost zero rotation rate
in the core, which is, of course, not true for the Sun. To take this
into account we subtract  the contribution
of  a uniform rotation rate $\Omega_c$, the estimated value of
rotation rate in the interior as obtained from the inversion results,
from the observed splittings.

Note that our parameterisation of the tachocline is different
from that of Kosovichev~(1996) and Charbonneau et al.~(1997).
The definition of the position and the jump remains the same,
the thickness of the tachocline as defined by them is roughly 4.9 times
the half-width we have defined, i.e. a half-width of 0.01$R_\odot$ in
our model corresponds to a thickness of $0.049R_\odot$ in their models.
Thus the tachocline thickness of $(0.050\pm 0.012)R_\odot$ as estimated by
Charbonneau et al.~(1997) using their model will be equivalent to a
half-width of
$(0.0102\pm 0.0025))R_\odot$ by our definition, which is consistent
with the value of $(0.0098\pm 0.0026)R_\odot$ determined by Basu (1997).
Of course, the form of variation in rotation rate inside the tachocline
can only be verified by inversions, which do not at present have the required
resolution. However, as long as the thickness of tachocline is small
enough, the exact form of variation within this layer may not be important,
as is seen by the similarity of the results obtained by
Charbonneau et al.~(1997) and Basu~(1997).

If $c_{\theta}$ be the combination of observed splitting coefficients
for a given latitude (after removing the contribution from $\Omega_c$)
and $a_{\theta}$ that of the calibration model, 
then to determine the jump, we make the following fit by treating the
splittings as a function of the lower turning point, $r_t$, of the mode:
$$ c_{\theta}(r_t)= \alpha a_{\theta}(r_t) + \phi(r_t),\eqno\eqname\jump$$
where $\alpha$ is the factor determining the jump and 
$\phi(r_t)$ is a low degree polynomial which takes into account 
any  trend in the real rotation rate not taken into account by the
parameterisation in equation~\thi. The constant term in this polynomial will
also take care of differences arising due to use of incorrect $\Omega_c$
while subtracting out the contribution from core rotation rate.
A polynomial of degree 2 is found to be 
sufficient. The fit is made between $r_t$ of 0.6 and 0.9 $R_\odot$. 
Modes with lower turning point are avoided because of 
large observational errors in these modes. Modes with higher $r_t$ are
not used so that shear layer known to exist just below the solar surface
does not affect the results.
We perform a least squares fit 
with the weights for each mode being the inverse of the
errors in the corresponding splitting.

We follow exactly the same procedure as that of Basu (1997)
for determining the position and thickness of the tachocline.
To recapitulate briefly,
the difference in the splitting coefficients
between a model and similar models which have
discontinuities at different positions have a well defined peak
and the height of the peak is proportional to the difference in
the positions of the discontinuity. Thus the peak height can be
calibrated to find the position of the discontinuity. If the models
are not similar, e.g., have different trends in the convection zone or
in the interior, or have a different width of transition,
the peak lies on a smooth part which can be represented as 
a low degree polynomial. A similar calibration can be used to estimate the 
thickness over which the transition of the rotation rate occurs.
In this case the width of the peak between two models  is proportional
to the thickness of the wider transition, but a simple scaling of the
radius around the peak position can reduce the curves to similar
widths. 

We have constructed models with $r_d$ of $0.68$, $0.69$,
$0.70$, $0.71$ and $0.72$ $R_\odot$ and half-width $w$ of $0$, $0.005$,
$0.01$, $0.015$, $0.02$ and $0.025$ $R_\odot$.
We have used  $\delta\Omega$ of 20 nHz in the calibration models
and determined the actual jump from the splittings as outlined earlier.
The splittings in the calibration models are then scaled to the required
jump.

For the purpose of calibration, we consider the difference in the
splitting coefficients between neighbouring calibration models and fit a
spline through the points
$$
\Phi(r)=\delta a(r)=\sum_i a_i \psi_i(r),
\eqno\eqname\spl
$$
where the $\Phi(r)$'s are the calibration curves and $\psi(r)$
are the cubic B-spline basis-functions. Thus we have 4 calibration curves
from the 5 calibration models.

The difference in splittings between each calibration model and the
observed splittings can be fitted with the form
$$\delta c_{\theta}= c_{\theta}(r)-a_{\theta}(r)=\alpha\Phi(r)+f(r).
\eqno\eqname\pos$$
Here $\Phi(r)$ is the calibration curve defined in equation~\spl\
and $f(r)$ is
a low degree polynomial used to take into account systematic effects
arising from differences in other parameters, like width etc.
 between the observations. As in Basu (1997) we find that a
polynomial of degree two or three is optimum.
The constant $\alpha$ and the
coefficients of the polynomial $f(r)$ are obtained by a
least-squares fit to the data. In practice, we determine
$\alpha$ for all five calibration models and interpolate to find
the points where $\alpha=0$. The four calibration curves
give four results which are then averaged.

Ideally, $\delta\Omega$, $r_d$, and $w$ should be determined simultaneously,
however, for simplicity in this
work we determine these parameters by independent fits. It has been 
shown by Basu (1997) that statistical
errors arising from uncertainties in observed splittings 
dominate over the systematic errors, and therefore, such a procedure may not
introduce significant additional errors. 
 To try and keep the parameters
of the calibration models close to that of the actual tachocline,
the fit in practice is done in two steps. We first determined
the positions and widths of the tachocline at the different
latitudes using the models with parameters found for the
coefficient $c_3$ by Basu~(1997). The process was repeated with 
calibration models constructed with parameters closer to those
determined in the first round of fits.

Note that the models defined by equation~\thi\ have a flat rotation rate in
the CZ, which
is obviously not the case at all latitudes. This is taken care of in our
fitting process described above through the smooth part, $\phi(r)$ in
equation~\jump\ and $f(r)$ in equation~\pos.
However, in order to check how
much difference the flat rotation rate in the CZ makes, we have also
constructed calibration models where the rotation rate in the CZ 
follows the trend revealed by the inversions described in the previous
section.
In addition to the form shown in equation~\thi, these
models have latitude dependent extra terms, with rotation rate defined as:
$$\Omega=\cases{\Omega_{\rm cal}+B(r-0.7)& if $r\le0.95$\cr
                \Omega_{\rm cal}-C(r-0.95)+0.25B& if    $r>0.95$\cr}
\eqno\eqname\trend$$
Here, the coefficients $B$ and $C$  are obtained from the inversion
results and 
the term $\Omega_{\rm cal}$ is defined by equation~\thi.
The models described by equation~\trend\ use an approximate value of the jump;
we still use the fit in equation~\jump\ to find the exact magnitude of the jump.
The procedure to find the  position and thickness of the 
tachocline remains the same. It must be noted that for these models, the 
polynomial $\phi$ of equation~\jump\ and  $f(r)$ in equation~\pos\ are
very small.

For all cases, the error estimates in the tachocline parameters are
obtained using
Monte-Carlo simulations. We have used two sets of GONG data, those from
GONG months 4--10 and GONG months 4--14 for the present work.

\subsection{The method of simulated annealing}

The calibration method described above suffers form the disadvantage
that each parameter defining the tachocline is determined separately
when the rest of the parameters are held fixed.
We therefore tried another forward modeling
method, where the jump, position and width can be found
simultaneously. Since this will require a nonlinear least squares fit, we have
resorted to the method of simulated annealing, which has better chance for
finding the global minimum.  For this purpose the rotation rate at
any given latitude is parameterized by
$$
\Omega_{\rm ann}(r)=\cases{\Omega_c
+B(r-0.7)\cr
\qquad\qquad+{\delta\Omega\over {1+\exp[(r_d-r)/w]}}& if $r\le0.95$\cr
\noalign{\medskip}
\Omega_c+0.25B
-C(r-0.95)\cr
\qquad\qquad+{\delta\Omega\over {1+\exp[(r_d-r)/w]}}& if $r>0.95$\cr}
\eqno\eqname\thian$$
where $\Omega_c$, B, C are the three parameters defining the smooth part of
rotation rate while $\delta\Omega$, $r_d$ and $w$ define the tachocline.
Here $B$ is the average gradient in the lower part of convection zone,
while $C$ is the gradient in the near surface shear layer.
These six parameters are
determined by a non-linear least squares fit to the combinations of
splitting coefficients representing the rotation rate at the required
latitude. Once again we use only those modes which have turning points
in the range 0.6--0.9$R_\odot$ in this fit.
We use the method of simulated annealing (Vanderbilt \& Louie 1984;
Press et al.~1993) to minimize the $\chi^2$ function. Since there
are likely to be many local minima where the minimization tends to get
trapped, 
even with simulated annealing,
 we make 20 attempts using different sequence of random numbers
in the annealing procedure to find the minimum and accept the one that gives
the lowest $\chi^2$.

Instead of fitting the rotation rate at each latitude separately and then
finding the variation in the tachocline properties, we can directly fit
a 2D form of rotation rate with tachocline to obtain this variation.
The form fitted is the same as in the 1D case (equation~\thian) with the
following substitutions
$$
\eqalign{
B=&B_1+B_3P_3(\theta)+B_5P_5(\theta),\cr
\delta\Omega=&\delta\Omega_1+\delta\Omega_3P_3(\theta)+\delta\Omega_5P_5(\theta),\cr
r_d=&r_{d1}+r_{d3}P_3(\theta),\cr
w=&w_1+w_3P_3(\theta),\cr}\eqno\eqname\thitwo
$$
where
$$\eqalign{P_3(\theta)&=5\cos^2\theta-1,\cr
P_5(\theta)&=21\cos^4\theta-14\cos^2\theta+1,\cr}
\eqno\eqname\polth$$
are polynomials used to define the latitude dependence. We have used these
polynomials so as to ensure some degree of separation between contributions
to various splitting coefficients.
This introduces
5 parameters to define the smooth part and another 7 parameters to define
the tachocline. It is not clear if all these parameters are required to
explain the data and we have carried out experiments involving different
combinations to determine which of these parameters are required.
Once again we use the
simulated annealing technique to simultaneously fit the first 3 splitting
coefficients $c_1$--$c_{5}$ for all modes with lower turning point in
the range 0.6--0.9$R_\odot$. The reason for using only the first 3
splitting coefficients is that with the assumed form for tachocline
model given by equation~\thitwo, we do not expect to fit the higher coefficients
properly. If these coefficients are to be fitted then additional
parameters will need to be introduced in $\delta\Omega$ and $B$ in
equation~\thitwo. We have tried this also
but we do not think that the fits with additional parameters and
coefficients are any better than the ones considered here as the higher
order parameters turn out to be rather small. As a result,
in this work we present the results obtained by fitting only the first
three splitting coefficients.

\beginfigure{8}
\hbox to 0 pt{\hskip -1.5cm
\vbox to 9.5 true cm{\vskip -0.5 true cm
\epsfysize=10.50 true cm\epsfbox{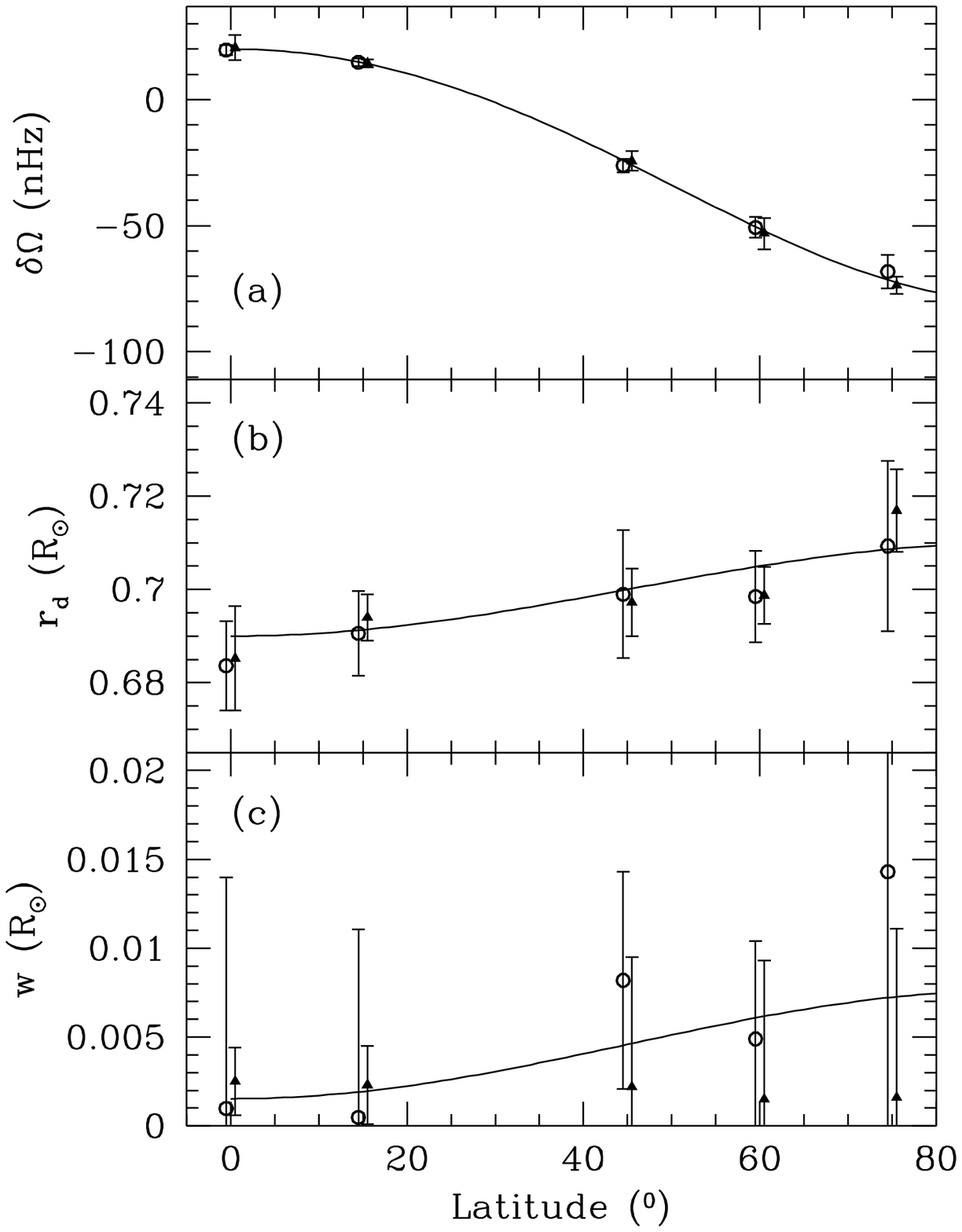}\vskip -0.5 true cm}}
\caption{\bf Figure 8. \rm A summary of the tachocline results for the
test model.
Panels (a),(b) and (c) show the results of the jump, position and half-width 
respectively.
In each panel the continuous line represents the exact value. The circles
show the results of the calibration method and the triangles are the
results obtained by 1D annealing. The symbols are displaced by $\pm0.5^\circ$
about the true latitude for the sake of clarity. In panel (c) the actual
width has been scaled down by a factor of 3.25 to account for the difference
in form of variation within the tachocline.
}
\endfigure

\subsection{Results}

In order to test the procedure outlined above we constructed a test model
with prescribed rotation rate including a tachocline and applied the
techniques for determining the properties. The rotation rate in the test
model was chosen to be
$$\eqalign{
\Omega(r,\theta)&=430+20\sin(\pi r/2)\cr
&\qquad\qquad+10(1-4\cos^2\theta-\cos^4\theta)F_j(r), \cr
\noalign{\medskip}
F_j(r)&=\cases{-1& if $r<r_d-w$\cr
\sin(0.5\pi(r-r_d)/w)& if $r_d-w\le r\le r_d+w$\cr
+1& if $r>r_d+w$\cr}\cr
\noalign{\medskip}
r_d&=(0.69+0.02\cos^2\theta)R_\odot,\cr
w&=(0.005+0.02\cos^2\theta)R_\odot.\cr}
\eqno\eqname\testone
$$
The splitting coefficients computed for this model were perturbed by
adding random errors with the same distribution as that specified by
quoted errors in the GONG months 4--14 data. The perturbed data were
then used to infer the characteristics of the tachocline and the
results using the calibration and 1D annealing methods
are shown in Fig.~8.
Since this model has a different
form of variation within the tachocline as compared to the models we
are using for the fits, we do not expect the thickness of tachocline
as determined by our procedure to agree with the actual thickness.
Comparing the region in which the rotation rate varies from $1/(1+e)$
to $1-1/(1+e)$
of the total jump, it appears that the effective width in the
test model is about $3.25$ times less than what is given by
equation~\testone, when models with form given by equation \thian\ are
used. Thus the estimated values should be compared with
this scaled width as has been done in Fig.~8.
Apart from the form of variation within the tachocline,
the trend in the lower CZ in test model is also far from
linear and hence may not be properly modelled by the tachocline model
used in annealing fits.
Note that all the results are roughly within the error bars of the exact
values, even though
the form of variation within the tachocline as well as the trend in the
CZ in the test model are different from those in the calibration models.
If we use a test model with the same form for tachocline as used in the
calibration models it is possible to obtain much better results.
This gives us confidence that we can indeed determine the
parameters of the solar tachocline.

The variation in the position and thickness of tachocline with latitude
is not totally clear from the estimated parameters, since these
variations were chosen to be comparable to the error estimates as happens
to be the case for observed splittings also. It appears that a variation
of $0.02R_\odot$ in position of tachocline is barely at
the limits of detection with the present data.
The variation in thickness
is not very evident as the estimated thickness appears to be too small
at all latitudes.

\beginfigure{9}
\hbox to 0 pt{\hskip -1.5cm
\vbox to 9.5 true cm{\vskip -0.5 true cm
\epsfysize=10.50 true cm\epsfbox{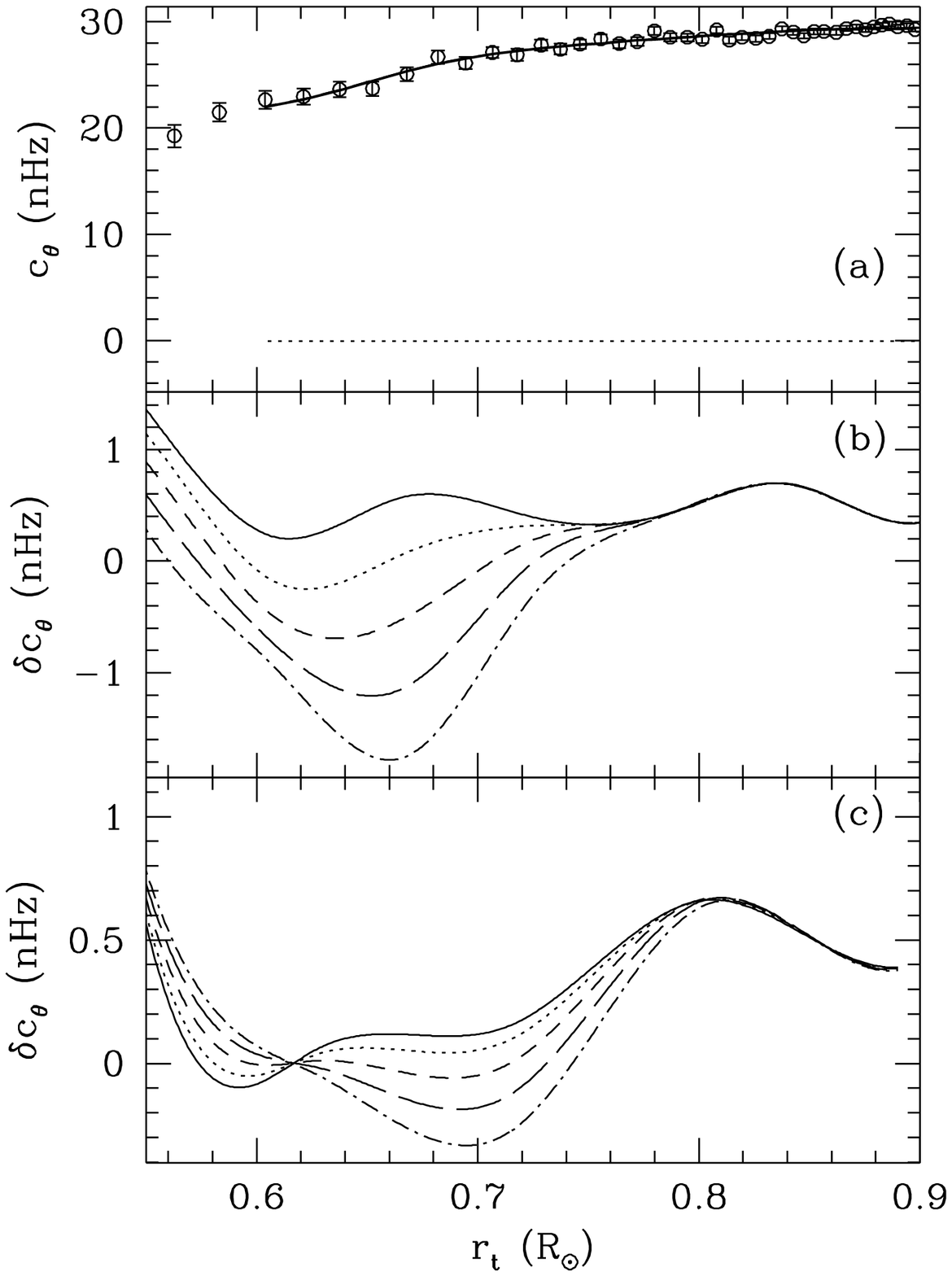}\vskip -0.5 true cm}}
\caption{\bf Figure 9. \rm  (a) Calibration for determining the jump
in the rotation rate in tachocline. The points are the splitting-coefficients
combination for the  equator with the contribution from $\Omega_c$ removed.
The continuous
line is the fit to the data, which can be decomposed into two parts
 --- the dotted line which is the term $\phi(r_t)$ in
equation~\jump\ and  the dashed line which is the calibration 
model scaled to the fitted jump (term $\alpha a_{\theta}(r_t)$).
The data used are the GONG4--14 splittings. The dashed 
and continuous lines  more or less
coincide in this case.
(b) The spline representation  of the 
difference between the data shown above and the 
 the five calibration models used to determine
the tachocline position. The continuous, dotted, small-dashed, long-dashed
and dot-dashed lines are difference with calibration models for $r_d=0.68$,
$0.69$, $0.70$, $0.71$ and $0.72R_\odot$ respectively. All these
calibration models have $w=0.005R_\odot$ (c) The spline 
representation  of the difference between the data and the five 
calibration models used to determine the tachocline width.
The continuous, dotted, small-dashed, long-dashed
and dot-dashed lines are difference with calibration models for
$w=0.005$, $0.01$, $0.015$, $0.02$ and $0.025R_\odot$ respectively. These
calibration models have $r_d=0.69R_\odot$.
}
\endfigure

\begintable*{1}
\caption{\bf Table 1. \rm Solar tachocline parameters as determined by calibration}
{\tablet{15 true cm}{#\hfil&&\hfil $#$ \hfil\cr
\tabmidrule
&\multispan{3}{\hfill Calibration models with flat CZ \hfill }&
\multispan{3}{\hfill Calibration models with trend in CZ \hfill}\cr
\tabmidrule
Lat.&\hbox{Jump}&\hbox{Position}&\hbox{Half-width}&
\hbox{Jump}&\hbox{Position}&\hbox{Half-width}\cr
($^\circ$) & \hbox{(nHz)}& (R_\odot) & (R_\odot)& \hbox{(nHz)}& (R_\odot) & (R_\odot)\cr
\tabmidrule
\multispan{7}{\hfill GONG months 4--10 data\hfill }\cr
\tabmidrule
\phantom{0}0&\phantom{-}21.66\pm 2.57& 0.6921\pm 0.0142& 0.0134\pm 0.0120& \phantom{-}20.92\pm 2.46& 0.6954\pm 0.0113& 0.0029\pm 0.0120\cr
15&\phantom{-}18.52\pm 1.91&0.6941\pm 0.0111& 0.0124\pm 0.0137 & \phantom{-}18.11\pm 0.92& 0.6891\pm 0.0115& 0.0190\pm 0.0131\cr
45&-27.81\pm 2.30& 0.6863\pm 0.0277 & 0.0215\pm 0.0173& -27.17\pm 2.06 &0.6903\pm 0.0263&0.0170\pm 0.0172\cr
60&-60.47\pm 3.52& 0.7025\pm 0.0117 & 0.0070\pm 0.0075& -60.08\pm 1.33 & 0.6990\pm 0.0099& 0.0050\pm 0.0077\cr
75&-86.41\pm 9.99& 0.6968\pm 0.0224 &0.0094\pm 0.0129  & -89.04\pm 2.56& 0.6974\pm 0.0238& 0.0123\pm 0.0135\cr
\tabmidrule
\multispan{7}{\hfill GONG months 4--14 data\hfill }\cr
\tabmidrule
\phantom{0}0 &  \phantom{-}19.39\pm 2.00 & 0.6944\pm 0.0096 & 0.0079\pm 0.0130 & \phantom{-}21.52\pm 0.82& 0.6851\pm 0.0077& 0.0047\pm 0.0083\cr
15&  \phantom{-}18.13\pm 1.55 & 0.6996\pm 0.0091 & 0.0083\pm 0.0106 & \phantom{-}18.29\pm 0.89& 0.6922\pm 0.0097 & 0.0043 \pm 0.0087\cr
45& -28.89\pm 2.66 & 0.7077\pm 0.0137 & 0.0047\pm 0.0061 &-29.87\pm 2.22& 0.7048\pm 0.0148 & 0.0059\pm 0.0067\cr
60& -57.18\pm 4.11 & 0.7058\pm 0.0098 & 0.0031\pm 0.0055 &-57.10\pm 1.51& 0.7082\pm 0.0072 & 0.0051\pm 0.0062\cr
75& -88.21\pm 6.70 & 0.7204\pm 0.0183 & 0.0154\pm 0.0235 &-87.15\pm 1.78& 0.7162\pm 0.0178 & 0.0141 \pm 0.0136\cr
\tabmidrule
}}
\endtable

Having tested our techniques on a test model we now apply the same
procedure to the GONG data for the months 4--10 and 4--14.
Fig.~9 shows the process of determining the tachocline  parameters
for the solar equator.
The results obtained using the calibration methods are 
 listed in Table~1.  A positive jump
implies a rotation rate which is higher in the CZ than in the radiative
interior. 
From Table~1 we note that the results for the two data sets are consistent
with each other within the estimated
errors. However, the GONG months 4--10 data have larger error as compared to
the months 4--14 data and that is reflected in the larger
errors in the tachocline parameters. We thus focus our attention on 
results for data from GONG months 4--14, which have also been used in the
method of simulated annealing.

The change in the tachocline jump as a function of latitude is very clear.
This is not surprising since the change is large enough to be seen by 
normal inversions also. The result obtained seems to depend somewhat on the 
type of calibration model used, although at each latitude they
are consistent within errors. It also appears that the results are
more sensitive
to data errors when models with a flat rotation-rate in the CZ are used.
This is perhaps not surprising as the jump at each latitude is not
assumed to be known beforehand,
while for the models with trend, a first estimate of the
jump is made from the inversion results and only a correction-factor is
obtained by the fit in equation~\jump. The estimated
errors in position and thickness at all latitudes are larger than the
corresponding errors in mean values as estimated by Basu~(1997) using
only the splitting coefficient $c_3$. This is partly due to the fact that
the error estimates in splittings for each latitude is larger than those
in $c_3$ alone. Further, a large part of the variation in tachocline is
determined by $c_3$, which shows a clear jump around the tachocline,
and at the same time has very little variation in the CZ or in radiative
interior thus making it much easier to fit the tachocline parameters.

There appears to be a slight variation in the radial position of the tachocline
with the tachocline moving outwards at higher latitudes. However, the 
change is not very significant -- being only about 1$\sigma$ between 
the equator and $60^\circ$ latitude. But the results seem to
indicate a nearly systematic shift.

The question about variation in the thickness is less clear, however.
In fact for
all latitudes, it appears that the thickness is very small and comparable
to the error estimates. Thus better data with  
reduced  errors are required before the thickness can be determined reliably.
With this method, at the moment we can only put an upper limit on the 
thickness of the tachocline at all latitudes. Basu (1997) had shown that
the thickness measurements are somewhat sensitive to the calibration
models used.  Thus we should try to check these results against those obtained
by the technique of simulated annealing.

\begintable{2}
\caption{\bf Table 2. \rm Tachocline parameters from 1D-annealing}
{\tablet{8.5 true cm}{#\hfil&&\hfil $#$\cr
\tabmidrule
Lat.& \hbox{Jump}\hfil & \hbox{Position}\hfil & \hbox{Half-width}\hfil& \chi^2\hfil \cr
($^\circ$) &\hbox{(nHz)}\hfil & (R_\odot)\hfil& (R_\odot)\hfil&  \cr
\tabmidrule
\phantom{0}0&\phantom{-}17.20\pm 4.96 & 0.6843\pm 0.0112 & 0.0020\pm 0.0019 &   1.0401 \cr
15&\phantom{-}14.84\pm 1.52 & 0.7146\pm0.0050& 0.0028\pm 0.0022 &  1.1271\cr
30&-7.46\pm1.71 & 0.7193\pm0.0196& 0.0242\pm0.0136 &  0.9860\cr
45&-33.83\pm 3.90 & 0.7160\pm 0.0072& 0.0122\pm 0.0073 &0.9276 \cr
60&-59.86\pm6.19& 0.7045\pm 0.0061& 0.0089\pm0.0078 &  1.0712\cr
75&-91.19\pm3.36 & 0.6878\pm0.0089& 0.0216\pm0.0095 &  1.1762&\cr
\tabmidrule
}}
\endtable

In order to obtain an independent measure of variation in properties of
tachocline with latitude, we adopt the technique of simulated annealing
to fit the tachocline parameters to the GONG months 4--14 data for
different latitudes.
Fig.~10 shows the 1D annealing result for the equator. Note that we get a
good fit and the residuals are random and consistent with the error estimates.
The 1D annealing results for various latitudes are listed in Table~2.
As in the case of the calibration technique,
the jump shows a clear change with latitude,
although there appears to be some systematic difference between the value of
the jump obtained by the two techniques.
All the values appear to be reduced in annealing results as compared to
the corresponding values obtained by the calibration method, although
for individual latitudes the results are generally within
error limits from those obtained using calibration method. The reason for
this discrepancy is not altogether obvious,
but there may be some ambiguity in the
definition of jump as a part of the variation across the tachocline may be
accounted for by the smooth trend, defined by the term involving $B$ in
equation~\thian. Note from the
last column which gives the $\chi^2$ per degree of freedom, it is clear
that the fit is reasonably good as all the values are close to unity.

\beginfigure{10}
\hbox to 0 pt{\hskip -1.5cm
\vbox to 6.7 true cm{\vskip -1.75 true cm
\epsfysize=10.50 true cm\epsfbox{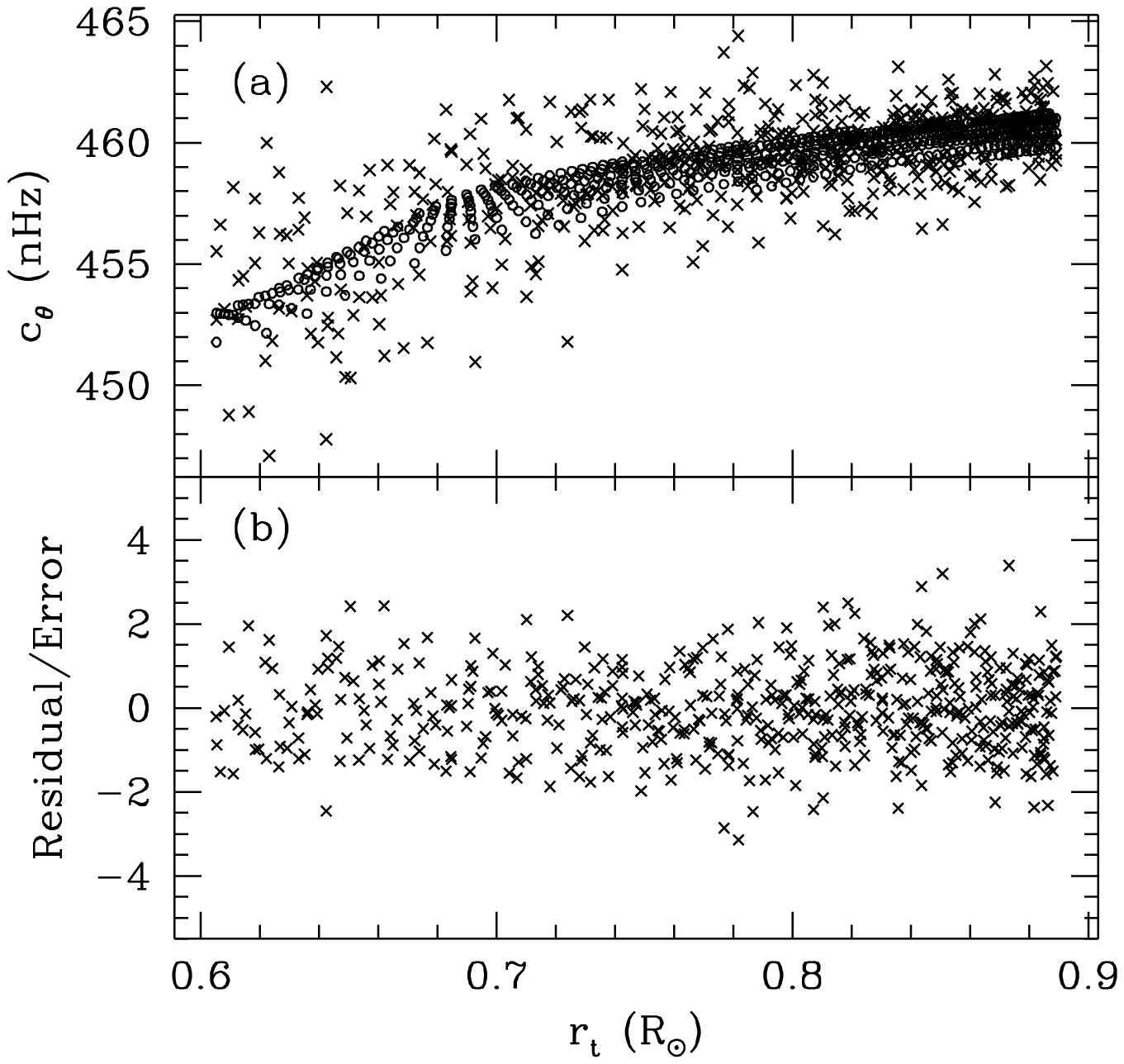}\vskip -2.05 true cm}}
\caption{\bf Figure 10. \rm The 1D simulated annealing results for the solar
equator. In Panel (a) The crosses are the observed splitting combinations
and the circles are those obtained by the fit. Panel (b) shows
the normalised residuals.
}
\endfigure

\beginfigure{11}
\hbox to 0 pt{\hskip -1.5cm
\vbox to 9.5 true cm{\vskip -0.5 true cm
\epsfysize=10.50 true cm\epsfbox{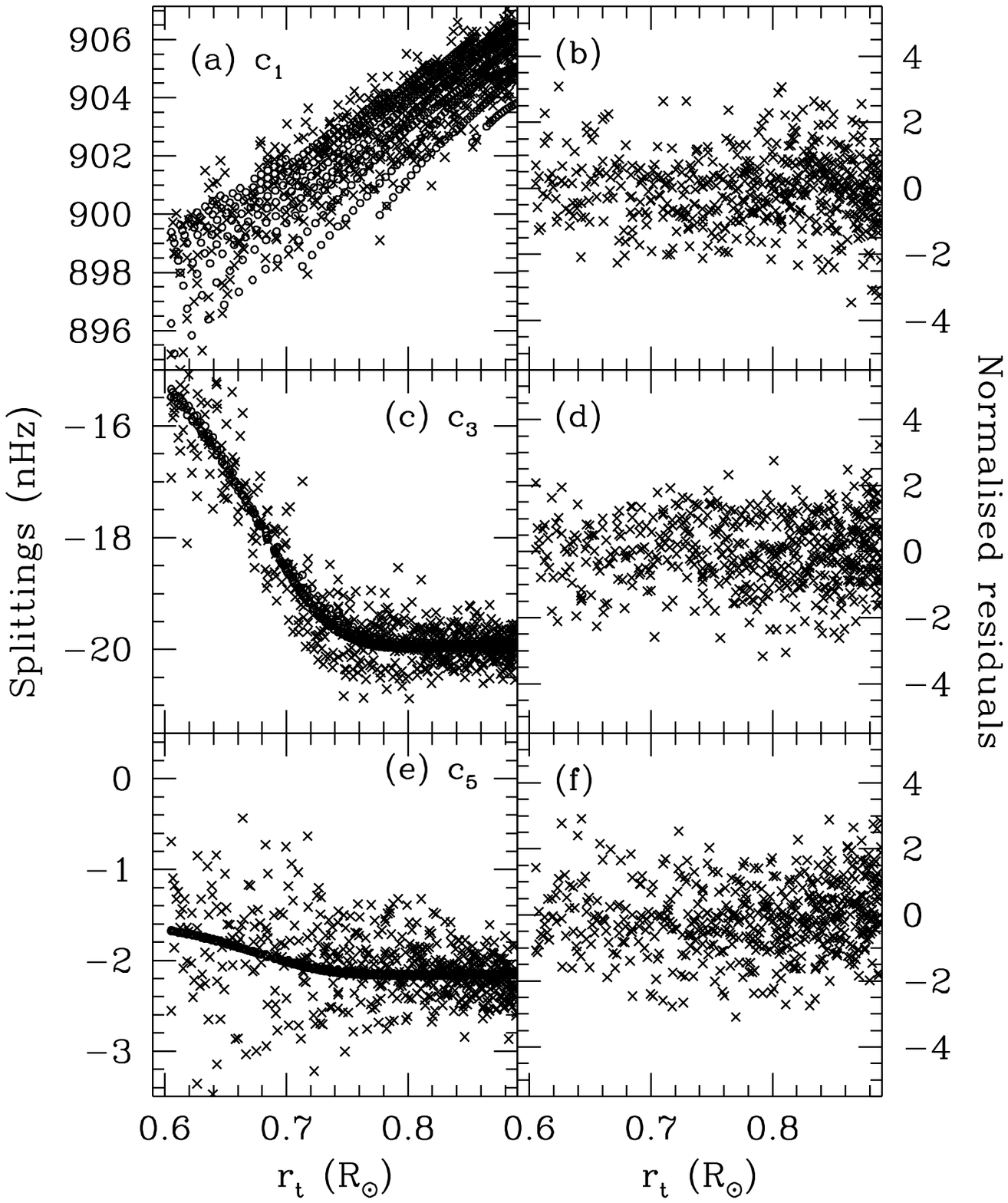}\vskip -0.5 true cm}}
\caption{\bf Figure 11. \rm The fit to the first three splittings
coefficients and the normalized residuals of the fits obtained by
the 2D simulated annealing method
assuming that there is no latitudinal variation in the position and thickness
of the tachocline.
}
\endfigure
\beginfigure{12}
\hbox to 0 pt{\hskip -1.5cm
\vbox to 9.5 true cm{\vskip -0.5 true cm
\epsfysize=10.50 true cm\epsfbox{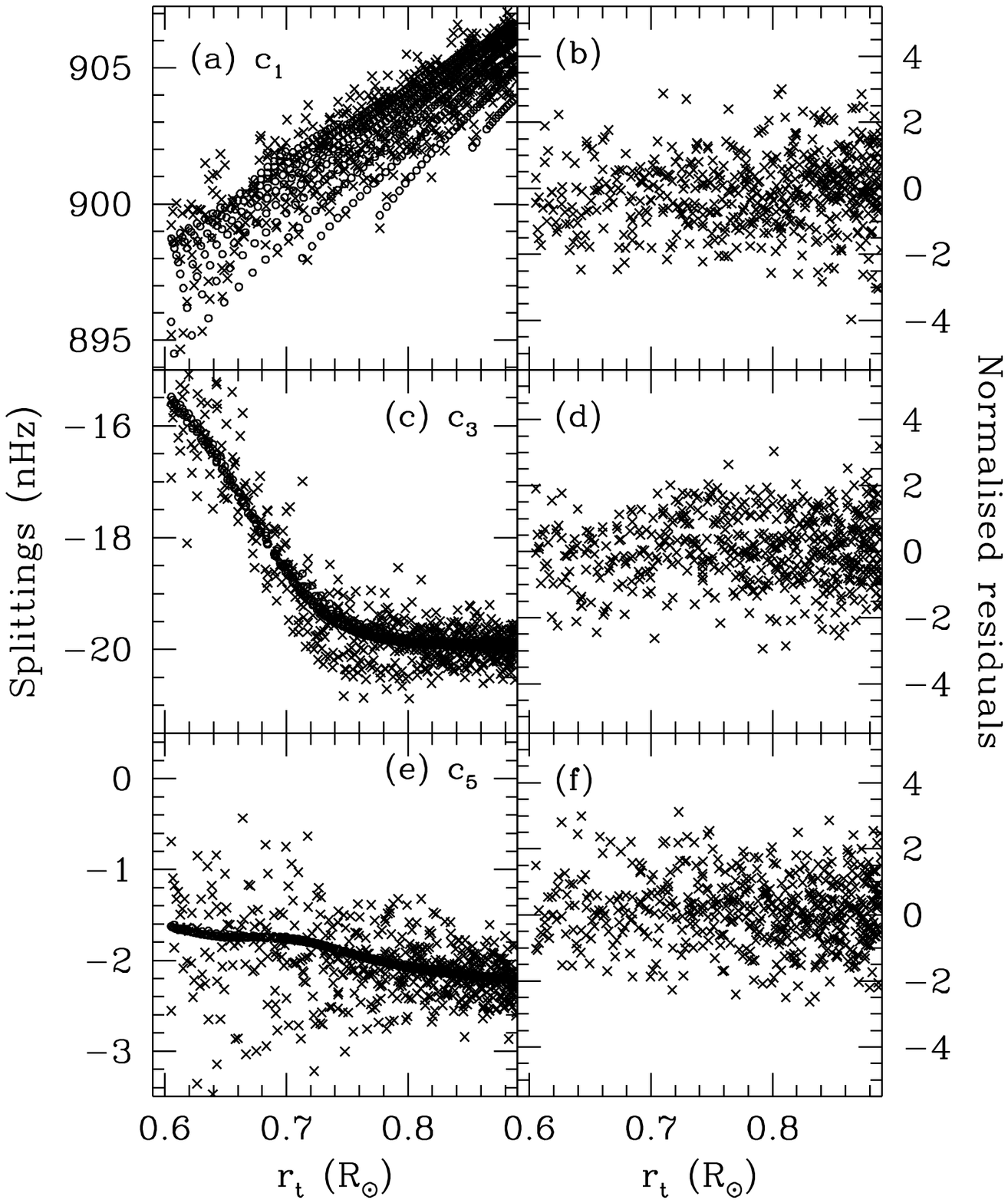}\vskip -0.5 true cm}}
\caption{\bf Figure 12. \rm   Same as Fig.~11, but 
with the possible   latitudinal variation in the position and thickness
of the tachocline taken into account
}
\endfigure
\beginfigure{13}
\hbox to 0 pt{\hskip -1.5cm
\vbox to 9.5 true cm{\vskip -0.5 true cm
\epsfysize=10.50 true cm\epsfbox{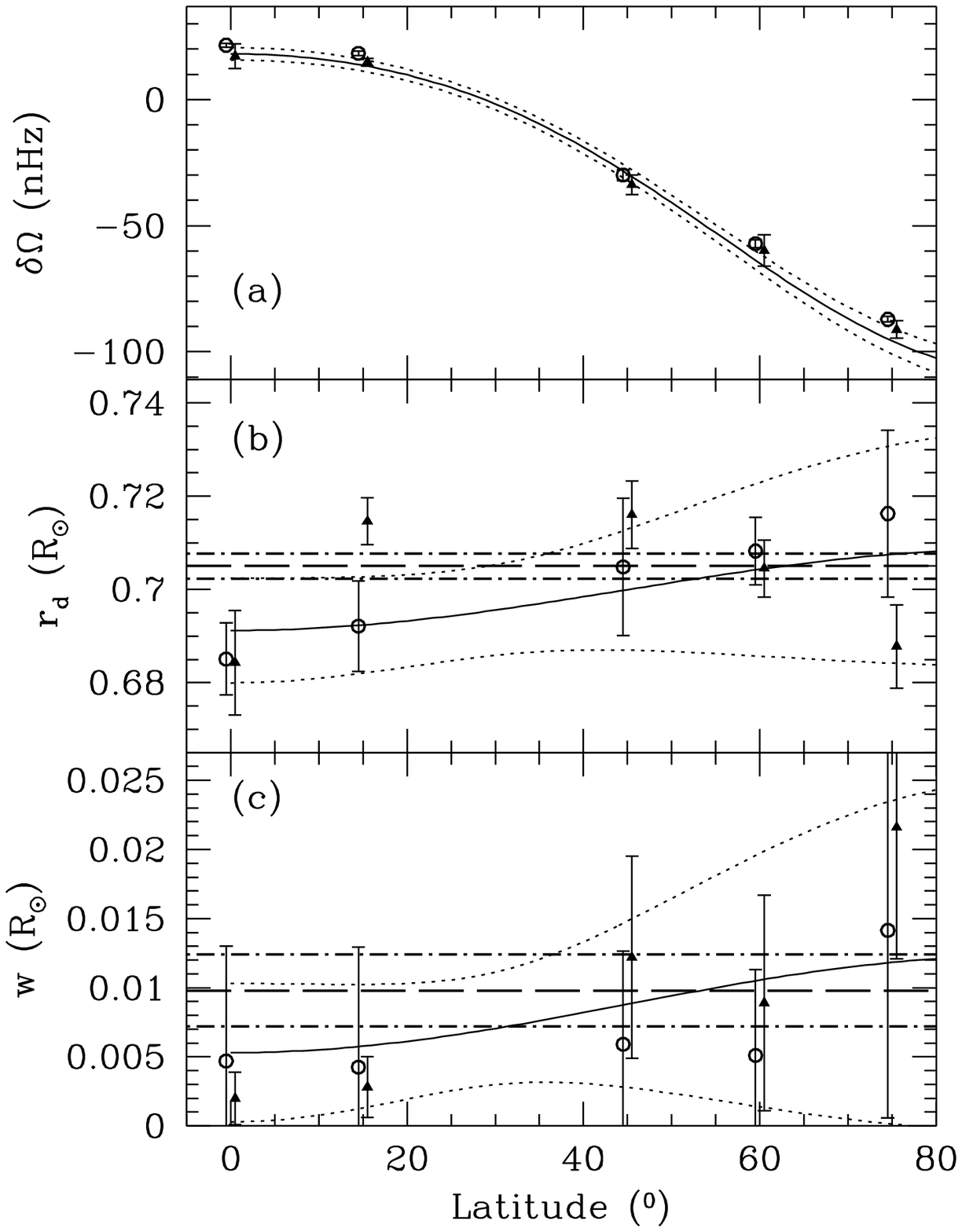}\vskip -0.5 true cm}}
\caption{\bf Figure 13. \rm A summary of the tachocline results obtained
using the GONG months 4--14 data.
Panels (a),(b) and (c) show the results of the jump, position and thickness 
respectively.
In each panel the continuous line is the results of the 2D annealing
(equation~18)
with the $1\sigma$ error bounds shown as the dotted lines. The circles
show the results of the calibration method and the triangles are the
results obtained by 1D annealing. The symbols are displaced by $\pm0.5^\circ$
about the true latitude for the sake of clarity. The dashed line in 
panels (b) and (c) mark the mean values found by Basu~(1997), while the
$1\sigma$ error bounds are shown as dot-dashed lines.
}
\endfigure

The results at $30^\circ$ latitude are not particularly reliable as the
jump is too small to define the tachocline properly and hence the errors
are very large. At high latitudes although the splittings have larger
error, the increase in magnitude of jump makes the tachocline better
defined and in some cases the error estimate also reduces since the fits
are more stable. At low latitudes, however, 
there is some problem in finding a proper fit, as
the estimated values have a larger scatter, which is reflected in
comparatively large errors even though the errors in splitting coefficients
is lowest for these latitudes. This could be due to the fact that the
form of trend assumed in the tachocline model defined by equation~\thian\ is
not sufficient to model the variation in rotation rate. As can be seen
from the contour diagram in Fig.~3, there is some variation at low latitudes
in the convection zone, which perhaps cannot be modelled properly by
a linear trend.

The annealing results also indicate that the thickness of the tachocline is 
very small at low latitudes. The thickness appears to increase at higher
latitudes though the estimated errors also increase and it is not clear
if the increase is significant. Similarly, it is not clear if shift in the
tachocline position with latitude is also significant, though in general
once again the tachocline appears to shift to slightly larger radial
distance at higher latitudes.

We feel that there should be yet another independent test of the
significance of variation in tachocline position and
thickness with latitude. We have therefore tried a 2D annealing fit to
simultaneously determine all the parameters at all latitudes. Since this fit
involves 12 parameters as defined by equations~\thian,\thitwo,
it is not clear if all of them are required.
For verifying this we start with the simplest situation where
the position and width of tachocline are independent of latitude
(i.e., $r_{d3}=0$, $w_3=0$).
This fit yields the mean position of tachocline as $0.7038R_\odot$ and
a half-width of $0.0187R_\odot$ and the fit is shown in Fig.~11.
These values are consistent with the results
obtained by Basu~(1997), though the half-width is slightly higher than
the value found by Basu~(1997). Some of the difference may arise because
while Basu~(1997) used only the splitting coefficient $c_3$ to determine
the tachocline parameters, in this work we are using the first three
splitting coefficients. If this difference is significant it may imply a
latitudinal variation in tachocline position or width.
Further, this fit has a $\chi^2=1.108$ per
degree of freedom and the fit appears to be reasonably good.
When all 12
parameters are included in the fit the minimum $\chi^2$ per degree
of freedom reduces to 1.088.
Thus the reduction in $\chi^2$ is only marginal and it is tempting to
conclude that the GONG months 4--14 data are consistent with no
latitudinal variation
in position or width of the tachocline. Of course, we can not rule out
a small variation in the tachocline properties with latitude.
Monte-Carlo simulation with 12 parameters yields the following
results for the tachocline:
$$\eqalign{r_d&=[(0.6991\pm0.0099)+(0.0030\pm0.0061)P_3(\theta)]R_\odot,\cr
w&=[(0.0084\pm0.0072)+(0.0047\pm0.0042)P_3(\theta)]R_\odot,\cr
\delta\Omega&=(-1.83\pm2.18)-(22.71\pm1.01)P_3(\theta)\cr
&\qquad\qquad-(3.88\pm0.45)P_5(\theta)\;\; \hbox{nHz},\cr}
\eqno\eqname\anntwo$$
It is clear that the variation in position is at a level of $(1/2)\sigma$
only, while the thickness at all latitudes is comparable to the error
estimates and as such it is not clear if the variation in thickness
is significant. Similarly, the first component of $\delta\Omega$
is also not significant. In fact, we find that if this parameter is set
to zero, the fit is more stable in the sense that the $\chi^2$ reduces
to acceptable level in most of the attempts with simulated annealing
and as such we prefer to use that fit. The minimum value of $\chi^2$ per
degree of freedom in this case is 1.105 and the resulting parameters
of the tachocline are
$$
\eqalign{r_d&=(0.6947+0.0035P_3(\theta))R_\odot,\cr
w&=(0.0067+0.0014P_3(\theta))R_\odot,\cr
\delta\Omega&=-21.11P_3(\theta)-2.96P_5(\theta)\;\; \hbox{nHz},\cr}
\eqno\eqname\anntwoc$$
while the fit is shown in Fig.~12. We adopt these values for the
tachocline model for use in the next section and for comparison with
other estimates.  It is clear
that addition of two parameters determining the variation in properties with
latitude does not improve the fit significantly and hence there is
no compelling reason to believe that there is any
variation in position or thickness of tachocline with latitude.
Nevertheless, all the results which attempt to determine the variation
find an increase in thickness with latitude and a shift outwards in
the mean position of tachocline with increasing latitude. Although
this variation does not appear to be significant in terms of the expected
errors, a small variation in properties of
the tachocline with latitude cannot be ruled out.

\beginfigure{14}
\hbox to 0 pt{\hskip -1.5cm
\vbox to 6.7 true cm{\vskip -1.75 true cm
\epsfysize=10.50 true cm\epsfbox{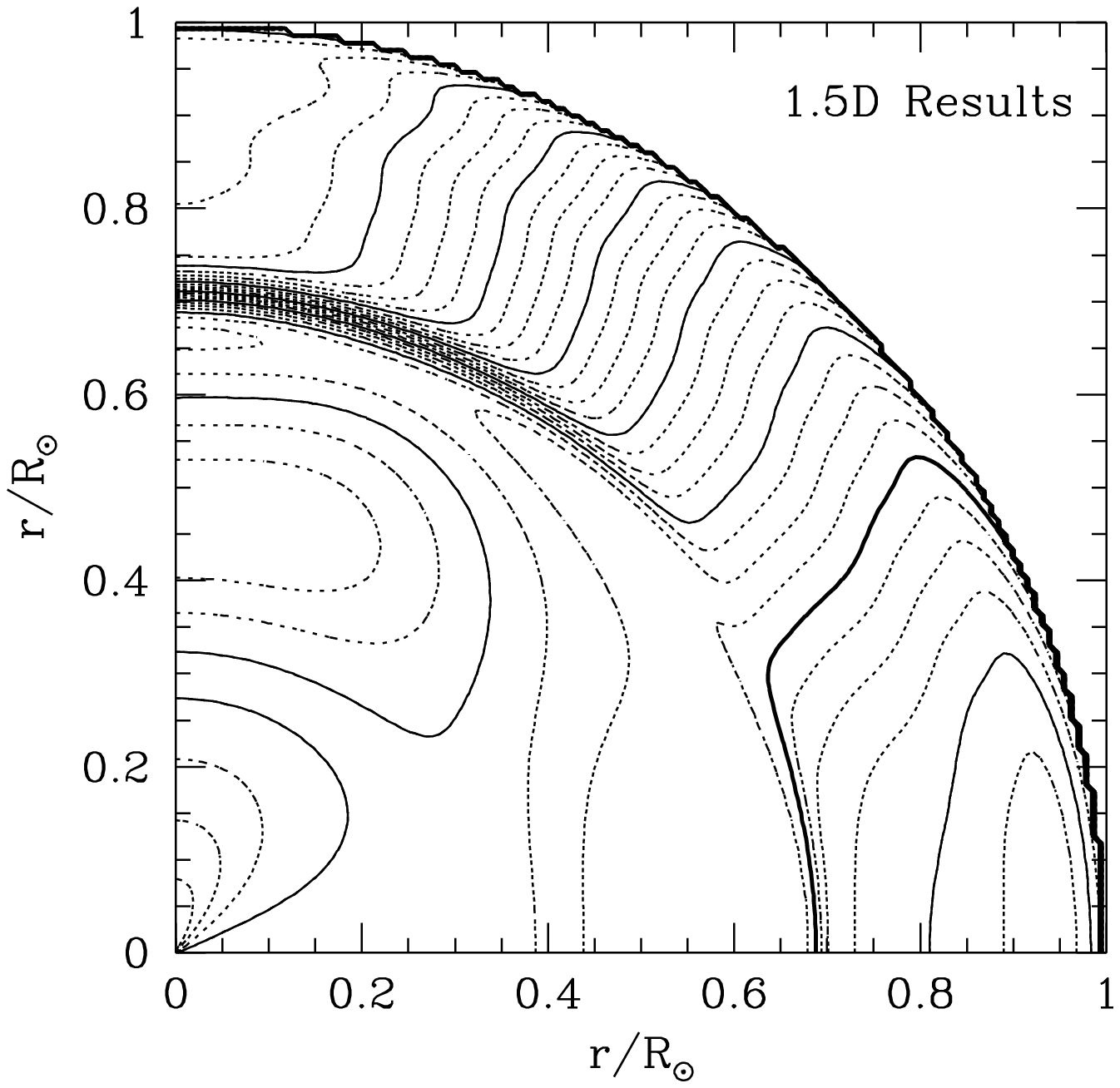}\vskip -2.05 true cm}}
\caption{\bf Figure 14. \rm A contour diagram of the solar rotation rate
as obtained by 1.5D inversion of the GONG months 4--14 data after removal of
the tachocline.  The format is the same as that for Fig.~3.
}
\endfigure

Fig.~13 summarizes the results from all the techniques. This figure
shows the variation in jump, position and thickness of the tachocline
as obtained by different techniques from the GONG months 4--14 data.
We have chosen the results from calibration method using the models with
trend for this purpose. It is clear that there is a general agreement
between the three independent results at most latitudes and all these
results point to an increase in thickness with latitude as well as
an outward radial shift in the position of tachocline with latitude.
This figure also shows the mean position and width of tachocline as
determined by Basu~(1997) and it appears that
all the results are also consistent
with no latitudinal variation in position or width of tachocline. 
Clearly, better data is required to find any possible variation in
tachocline properties with latitude.

\section{Inversion after removing the tachocline signal}

The smoothing used in our inversion procedure tends to smooth out
the steep variation in rotation rate in the tachocline.
Apart from this it may also introduce some ripples away from the position
of tachocline, a feature that is reminiscent of the Gibbs phenomenon in
Fourier transform. In order to overcome this limitation and to use
the tachocline parameters determined in the earlier section for
improving the results of inversion, we attempt an inversion after
removing the signal from tachocline. For this purpose we adopt
tachocline parameters given by equation~\anntwoc, leaving
aside the smooth part and perform a forward calculation to obtain
the splitting coefficients for this model. These splittings are then
subtracted from the observed splittings before inverting the data. The rotation
rate in the tachocline model is then added to the inverted profile to
obtain the actual rotation rate in the Sun.

\beginfigure{15}
\hbox to 0 pt{\hskip -1.5cm
\vbox to 9.5 true cm{\vskip -0.5 true cm
\epsfysize=10.50 true cm\epsfbox{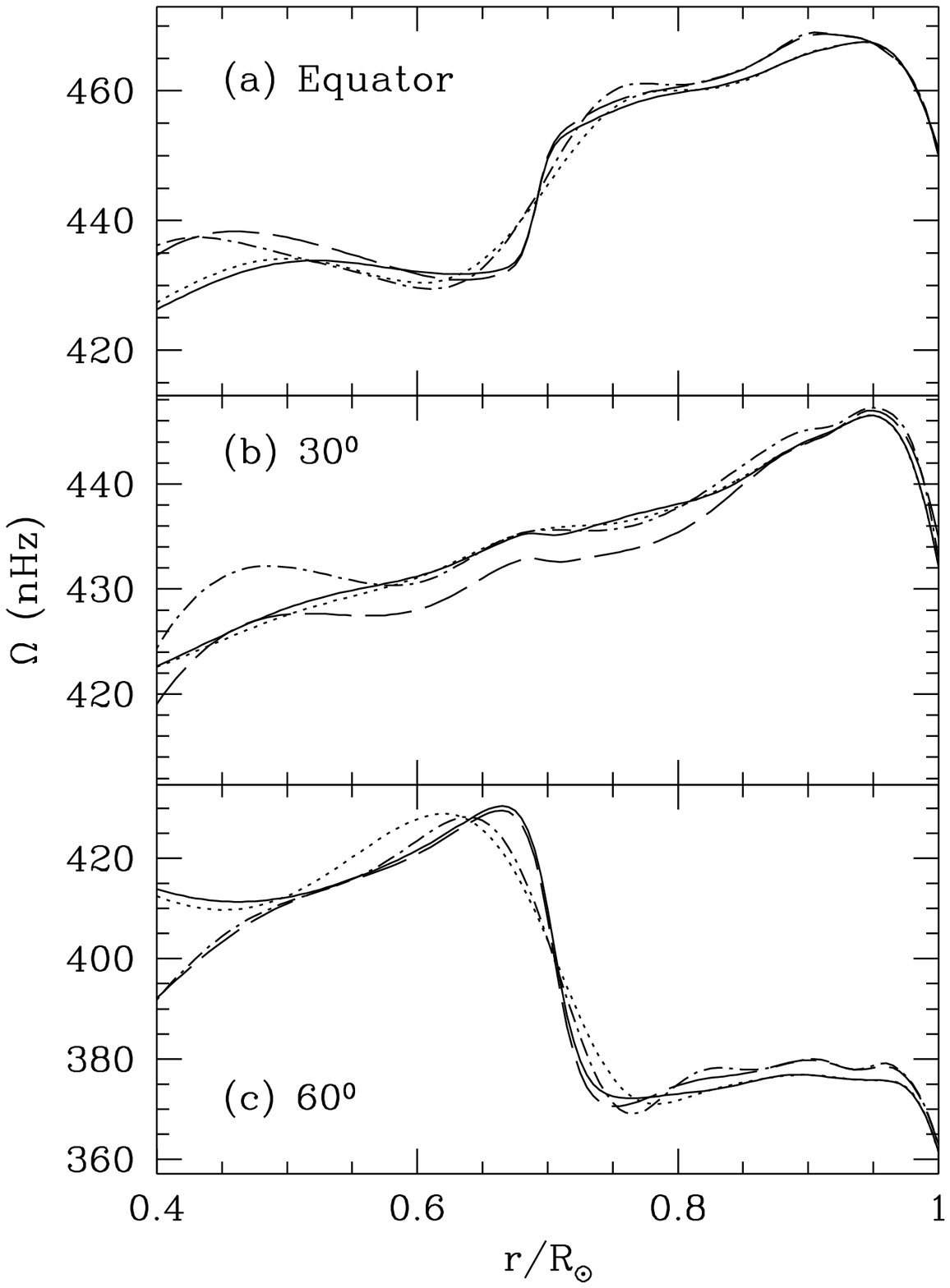}\vskip -0.5 true cm}}
\caption{\bf Figure 15. \rm  A comparison of the inversion results
with and without removal of tachocline. Continuous line shows the
results obtained using 1.5D inversion after removing the tachocline
contribution,
the dotted  line is the 1.5D result of normal inversion.
Similarly, the dashed line show the results
of 2D inversion of individual splittings after removing the tachocline
while the dot-dashed line shows the results from normal 2D inversion.
}
\endfigure

The results in Fig.~14 display the contour diagram
for the rotation rate obtained using 1.5D inversion, while Fig.~15
shows the rotation rate as a function of radial distance at selected
latitudes for the inversions with and without removal of tachocline.
From this figure it appears that at the $30^\circ$ latitude the results obtained
using the 2D inversion after removing the tachocline are different from
other results. The reason for this difference are not altogether clear,
though it appears to manifest at latitudes around the region where
the radial gradient in the tachocline changes sign. 
The results obtained after removing the tachocline appear to be smoother
in general, which is of course, not surprising. Also, the fact that the
simulated annealing was able to obtain a good fit over a large fraction
of radial distance using only 12 parameters suggests that there is probably
not much structure in the rotation rate as a function of latitude or radial
distance (apart from tachocline itself) in the region $0.6<r/R_\odot<0.9$.
However, from the 2D annealing fits there are reasons to believe that these
12 parameters are not completely adequate to represent the trends at all
latitudes adequately. Nevertheless,
it is likely that some features seen in the lower part of CZ
in certain inversion results (e.g., Figs.~3,4) are due to the influence of 
tachocline magnified by errors in data.

\section{Conclusions}

In this work we have attempted to infer the rotation rate inside the
Sun using observed p-mode frequency splittings. We have used a 1.5D inversion
technique to invert the data in the form of splitting coefficients
based on a regularized least squares method with iterative
refinement. We have investigated the influence of using different
prescription for smoothing to find that the differences are comparable
to expected errors in inversion results. Similarly, an adoption of different
sets of observed splitting coefficients shows that the difference in
various results is again roughly consistent with error estimates based
on quoted errors in observed splittings. Apart from the 1.5D inversion
technique, we have also attempted 2D inversion for both the individual
splittings $D_{n,\ell,m}$ and the splitting coefficients $c_s^{(n,\ell)}$.
All these results agree reasonably well inside the convection zone, although
in the deep interior there are some differences between different
inversion results. We believe these reflect our inability to obtain
reliable estimate of rotation rate in the core due to possible systematic
errors in splittings for low degree modes. 

It is clear from our results that the surface differential rotation
 persists through the
solar convection zone, while below the base of convection zone the rotation
rate appears to be relatively independent of latitude. The transition
around the base of the convection zone may not be resolved by the inversion
results.
The core appears to be rotating slower than the surface equatorial
rotation rate as also found by Tomczyk, Schou \& Thompson~(1995) and
Elsworth et al.~(1995). The rotation rate in solar core has been
recently discussed by Gizon et al.~(1997) and Rabello-Soares et al.~(1997)
and it appears that there is significant variation in the inferred rotation
rate in the solar core from different helioseismic data.
The constraints on the rotation rate in the core in the form of the boundary
conditions \bc, in the 1.5D inversion make the core rotate essentially
as a rigid body and reduces the discrepancy in rotation rate inferred using
different data sets.
This may be artificial, but unfortunately we do not have enough data
to resolve any variation in rotation rate in the core and this is
probably the simplest assumption that we can make (Chaplin et al.~1996).
Even a small error in splittings for low degree modes can change the inverted
rotation rate in the core significantly if no constraints are applied.
Further, as can be seen from Fig.~1, the results outside the core
are not too sensitive to the point where
these boundary conditions are applied or the form of smoothing used.
The main problem with reliable
inversion of rotation rate in the core is that the low degree modes
that penetrate into the core have large errors in the splitting
coefficients (possibly including some systematic errors), and this allows
for a wide range of rotation profiles in the core. It is clear that
better quality data are required to make any reliable estimate of
rotation rate in the solar core.

Further, there is a distinct shear layer just underneath the solar
surface where
the rotation rate increases with depth. There is also a hint of this shear
layer becoming less pronounced with latitude, in the sense that the
radial variation of the rotation rate tends to diminish with
increasing latitude.  The
rotation rate has a maximum around $r=0.95R_\odot$ at most latitudes,
except possibly close to the poles.
This feature can also be seen in the raw splitting data (Fig.~7).
The inferred rotation rate at the solar surface
agrees reasonably well with that obtained from Doppler measurements
(Snodgrass 1992). Some differences between the Doppler measurement
and inverted profile (Fig.~6) at higher latitudes could be because the
Doppler measurements determine only terms up to $\cos^4\theta$ in expansion
of rotation rate. The higher order terms included in inversions will
contribute significantly at high latitudes.

The tachocline or the shear layer near the base of convection zone has
been studied in detail using forward techniques to ascertain possible
latitudinal variation in its properties.  Since the tachocline 
cannot be resolved by inversions,
we have used a number of forward modelling techniques
to study the tachocline. With the present data, we do not find any
compelling evidence for any variation in the position or thickness
of the tachocline with latitude. 
The mean position and thickness of tachocline is found to be consistent
with the values found by Basu~(1997).
Our results suggest that
the thickness increases marginally with latitude and the location of
tachocline also appears to shift outwards with increasing latitude, but the
difference between the position and thickness of the tachocline  at the solar 
equator  and at a latitude of $60^\circ$ is comparable 
to the estimated errors. It is therefore, 
not clear if the variation is really significant. Similarly, the 2D annealing
results also show that the variations in the position and thickness
are less than the respective error bars. Further, there is no significant
reduction in the $\chi^2$ for the fit when the variation of position and
width with latitude is included in the 2D fits.
Taking the errors into account we believe
that we get an upper limit of 0.03$R_\odot$ for the variation in the position
of the tachocline and about 0.02$R_\odot$ for the variation of the thickness.
Clearly, more precise 
data are needed to make a better estimate of the parameters reliably.

From our results it appears that the tachocline is centered at a depth
which is below the base of the solar convection zone at all latitudes.
If the latitudinal variation shown by our results is real then at low
latitudes most of the variation in rotation rate within the tachocline
occurs below the CZ base, while at
high latitudes part of the tachocline may extend into the convection zone.

\section*{Acknowledgments}

This work was partly supported  by the Danish National Research
Foundation through its establishment of the Theoretical Astrophysics Center.
This work utilises data obtained by the Global Oscillation
Network Group (GONG) project, managed by the National Solar Observatory, a
Division of the National Optical Astronomy Observatories, which is 
operated by AURA, Inc. under a cooperative agreement with the 
National Science Foundation.

\bye

%% file: mn.tex
%
%
%
%

\catcode `\@=11 

\def\@version{1.6}
\def\@verdate{18th September 1995}

%
%


\newif\ifprod@font

\ifx\@typeface\undefined
  \def\@typeface{Comp. Modern}\prod@fontfalse
\else
  \prod@fonttrue 
\fi

\def\newfam{\alloc@8\fam\chardef\sixt@@n} 

\ifprod@font
\font\fiverm=mtr10 at 5pt
\font\fivebf=mtbx10 at 5pt
\font\fiveit=mtti10 at 5pt
\font\fivesl=mtsl10 at 5pt
\font\fivett=cmtt8 at 5pt     \hyphenchar\fivett=-1
\font\fivecsc=mtcsc10 at 5pt
\font\fivesf=mtss10 at 5pt
\font\fivei=mtmi10 at 5pt      \skewchar\fivei='177
\font\fivesy=mtsy10 at 5pt     \skewchar\fivesy='60

\font\sixrm=mtr10 at 6pt
\font\sixbf=mtbx10 at 6pt
\font\sixit=mtti10 at 6pt
\font\sixsl=mtsl10 at 6pt
\font\sixtt=cmtt8 at 6pt      \hyphenchar\sixtt=-1
\font\sixcsc=mtcsc10 at 6pt
\font\sixsf=mtss10 at 6pt
\font\sixi=mtmi10 at 6pt       \skewchar\sixi='177
\font\sixsy=mtsy10 at 6pt      \skewchar\sixsy='60

\font\sevenrm=mtr10 at 7pt
\font\sevenbf=mtbx10 at 7pt
\font\sevenit=mtti10 at 7pt
\font\sevensl=mtsl10 at 7pt
\font\seventt=cmtt8 at 7pt     \hyphenchar\seventt=-1
\font\sevencsc=mtcsc10 at 7pt
\font\sevensf=mtss10 at 7pt
\font\seveni=mtmi10 at 7pt      \skewchar\seveni='177
\font\sevensy=mtsy10 at 7pt     \skewchar\sevensy='60

\font\eightrm=mtr10 at 8pt
\font\eightbf=mtbx10 at 8pt
\font\eightit=mtti10 at 8pt
\font\eighti=mtmi10 at 8pt      \skewchar\eighti='177
\font\eightsy=mtsy10 at 8pt     \skewchar\eightsy='60
\font\eightsl=mtsl10 at 8pt
\font\eighttt=cmtt8             \hyphenchar\eighttt=-1
\font\eightcsc=mtcsc10 at 8pt
\font\eightsf=mtss10 at 8pt

\font\ninerm=mtr10 at 9pt
\font\ninebf=mtbx10 at 9pt
\font\nineit=mtti10 at 9pt
\font\ninei=mtmi10 at 9pt      \skewchar\ninei='177
\font\ninesy=mtsy10 at 9pt     \skewchar\ninesy='60
\font\ninesl=mtsl10 at 9pt
\font\ninett=cmtt9             \hyphenchar\ninett=-1
\font\ninecsc=mtcsc10 at 9pt
\font\ninesf=mtss10 at 9pt

\font\tenrm=mtr10
\font\tenbf=mtbx10
\font\tenit=mtti10
\font\teni=mtmi10		\skewchar\teni='177
\font\tensy=mtsy10		\skewchar\tensy='60
\font\tenex=cmex10
\font\tensl=mtsl10
\font\tentt=cmtt10		\hyphenchar\tentt=-1
\font\tencsc=mtcsc10
\font\tensf=mtss10

\font\elevenrm=mtr10 at 11pt
\font\elevenbf=mtbx10 at 11pt
\font\elevenit=mtti10 at 11pt
\font\eleveni=mtmi10 at 11pt      \skewchar\eleveni='177
\font\elevensy=mtsy10 at 11pt     \skewchar\elevensy='60
\font\elevensl=mtsl10 at 11pt
\font\eleventt=cmtt10 at 11pt     \hyphenchar\eleventt=-1
\font\elevencsc=mtcsc10 at 11pt
\font\elevensf=mtss10 at 11pt

\font\twelverm=mtr10 at 12pt
\font\twelvebf=mtbx10 at 12pt
\font\twelveit=mtti10 at 12pt
\font\twelvesl=mtsl10 at 12pt
\font\twelvett=cmtt12             \hyphenchar\twelvett=-1
\font\twelvecsc=mtcsc10 at 12pt
\font\twelvesf=mtss10 at 12pt
\font\twelvei=mtmi10 at 12pt      \skewchar\twelvei='177
\font\twelvesy=mtsy10 at 12pt     \skewchar\twelvesy='60

\font\fourteenrm=mtr10 at 14pt
\font\fourteenbf=mtbx10 at 14pt
\font\fourteenit=mtti10 at 14pt
\font\fourteeni=mtmi10 at 14pt      \skewchar\fourteeni='177
\font\fourteensy=mtsy10 at 14pt     \skewchar\fourteensy='60
\font\fourteensl=mtsl10 at 14pt
\font\fourteentt=cmtt12 at 14pt     \hyphenchar\fourteentt=-1
\font\fourteencsc=mtcsc10 at 14pt
\font\fourteensf=mtss10 at 14pt

\font\seventeenrm=mtr10 at 17pt
\font\seventeenbf=mtbx10 at 17pt
\font\seventeenit=mtti10 at 17pt
\font\seventeeni=mtmi10 at 17pt      \skewchar\seventeeni='177
\font\seventeensy=mtsy10 at 17pt     \skewchar\seventeensy='60
\font\seventeensl=mtsl10 at 17pt
\font\seventeentt=cmtt12 at 17pt     \hyphenchar\seventeentt=-1
\font\seventeencsc=mtcsc10 at 17pt
\font\seventeensf=mtss10 at 17pt
\else
\font\fiverm=cmr5
\font\fivei=cmmi5             \skewchar\fivei='177
\font\fivesy=cmsy5            \skewchar\fivesy='60
\font\fivebf=cmbx5

\font\sixrm=cmr6
\font\sixi=cmmi6             \skewchar\sixi='177
\font\sixsy=cmsy6            \skewchar\sixsy='60
\font\sixbf=cmbx6

\font\sevenrm=cmr7
\font\sevenit=cmti7
\font\seveni=cmmi7             \skewchar\seveni='177
\font\sevensy=cmsy7            \skewchar\sevensy='60
\font\sevenbf=cmbx7

\font\eightrm=cmr8
\font\eightbf=cmbx8
\font\eightit=cmti8
\font\eighti=cmmi8			\skewchar\eighti='177
\font\eightsy=cmsy8			\skewchar\eightsy='60
\font\eightsl=cmsl8
\font\eighttt=cmtt8			\hyphenchar\eighttt=-1
\font\eightcsc=cmcsc10 at 8pt
\font\eightsf=cmss8

\font\ninerm=cmr9
\font\ninebf=cmbx9
\font\nineit=cmti9
\font\ninei=cmmi9			\skewchar\ninei='177
\font\ninesy=cmsy9			\skewchar\ninesy='60
\font\ninesl=cmsl9
\font\ninett=cmtt9			\hyphenchar\ninett=-1
\font\ninecsc=cmcsc10 at 9pt
\font\ninesf=cmss9

\font\tenrm=cmr10
\font\tenbf=cmbx10
\font\tenit=cmti10
\font\teni=cmmi10		\skewchar\teni='177
\font\tensy=cmsy10		\skewchar\tensy='60
\font\tenex=cmex10
\font\tensl=cmsl10
\font\tentt=cmtt10		\hyphenchar\tentt=-1
\font\tencsc=cmcsc10
\font\tensf=cmss10

\font\elevenrm=cmr10 scaled \magstephalf
\font\elevenbf=cmbx10 scaled \magstephalf
\font\elevenit=cmti10 scaled \magstephalf
\font\eleveni=cmmi10 scaled \magstephalf	\skewchar\eleveni='177
\font\elevensy=cmsy10 scaled \magstephalf	\skewchar\elevensy='60
\font\elevensl=cmsl10 scaled \magstephalf
\font\eleventt=cmtt10 scaled \magstephalf	\hyphenchar\eleventt=-1
\font\elevencsc=cmcsc10 scaled \magstephalf
\font\elevensf=cmss10 scaled \magstephalf

\font\twelverm=cmr10 scaled \magstep1
\font\twelvebf=cmbx10 scaled \magstep1
\font\twelvei=cmmi10 scaled \magstep1      \skewchar\twelvei='177
\font\twelvesy=cmsy10 scaled \magstep1     \skewchar\twelvesy='60

\font\fourteenrm=cmr10 scaled \magstep2
\font\fourteenbf=cmbx10 scaled \magstep2
\font\fourteenit=cmti10 scaled \magstep2
\font\fourteeni=cmmi10 scaled \magstep2		\skewchar\fourteeni='177
\font\fourteensy=cmsy10 scaled \magstep2	\skewchar\fourteensy='60
\font\fourteensl=cmsl10 scaled \magstep2
\font\fourteentt=cmtt10 scaled \magstep2	\hyphenchar\fourteentt=-1
\font\fourteencsc=cmcsc10 scaled \magstep2
\font\fourteensf=cmss10 scaled \magstep2

\font\seventeenrm=cmr10 scaled \magstep3
\font\seventeenbf=cmbx10 scaled \magstep3
\font\seventeenit=cmti10 scaled \magstep3
\font\seventeeni=cmmi10 scaled \magstep3	\skewchar\seventeeni='177
\font\seventeensy=cmsy10 scaled \magstep3	\skewchar\seventeensy='60
\font\seventeensl=cmsl10 scaled \magstep3
\font\seventeentt=cmtt10 scaled \magstep3	\hyphenchar\seventeentt=-1
\font\seventeencsc=cmcsc10 scaled \magstep3
\font\seventeensf=cmss10 scaled \magstep3
\fi

\def\hexnumber#1{\ifcase#1 0\or1\or2\or3\or4\or5\or6\or7\or8\or9\or
  A\or B\or C\or D\or E\or F\fi}

\def\makestrut{%
  \setbox\strutbox=\hbox{%
    \vrule height.7\baselineskip depth.3\baselineskip width \z@}%
}

\def\baselinestretch{1}
\newskip\tmp@bls

\def\b@ls#1{
  \tmp@bls=#1\relax
  \baselineskip=#1\relax\makestrut
  \normalbaselineskip=\baselinestretch\tmp@bls
  \normalbaselines
}

\def\nostb@ls#1{
  \normalbaselineskip=#1\relax
  \normalbaselines
  \makestrut
}

%

\newfam\scfam  
\newfam\sffam  

\def\mit{\fam\@ne}
\def\cal{\fam\tw@}
\def\em{\ifdim\fontdimen1\font>\z@ \rm\else\it\fi}

\textfont3=\tenex
\scriptfont3=\tenex
\scriptscriptfont3=\tenex

\setbox0=\hbox{\tenex B} \p@renwd=\wd0 

\def\eightpoint{
  \def\rm{\fam0\eightrm}%
  \textfont0=\eightrm \scriptfont0=\sixrm \scriptscriptfont0=\fiverm%
  \textfont1=\eighti  \scriptfont1=\sixi  \scriptscriptfont1=\fivei%
  \textfont2=\eightsy \scriptfont2=\sixsy \scriptscriptfont2=\fivesy%
  \textfont\itfam=\eightit\def\it{\fam\itfam\eightit}%
  \ifprod@font
    \scriptfont\itfam=\sixit
      \scriptscriptfont\itfam=\fiveit
  \else
    \scriptfont\itfam=\eightit
      \scriptscriptfont\itfam=\eightit
  \fi
  \textfont\bffam=\eightbf%
    \scriptfont\bffam=\sixbf%
      \scriptscriptfont\bffam=\fivebf%
  \def\bf{\fam\bffam\eightbf}%
  \textfont\slfam=\eightsl\def\sl{\fam\slfam\eightsl}%
  \ifprod@font
    \scriptfont\slfam=\sixsl
      \scriptscriptfont\slfam=\fivesl
  \else
    \scriptfont\slfam=\eightsl
      \scriptscriptfont\slfam=\eightsl
  \fi
  \textfont\ttfam=\eighttt\def\tt{\fam\ttfam\eighttt}%
  \ifprod@font
    \scriptfont\ttfam=\sixtt
      \scriptscriptfont\ttfam=\fivett
  \else
    \scriptfont\ttfam=\eighttt
      \scriptscriptfont\ttfam=\eighttt
  \fi
  \textfont\scfam=\eightcsc\def\sc{\fam\scfam\eightcsc}%
  \ifprod@font
    \scriptfont\scfam=\sixcsc
      \scriptscriptfont\scfam=\fivecsc
  \else
    \scriptfont\scfam=\eightcsc
      \scriptscriptfont\scfam=\eightcsc
  \fi
  \textfont\sffam=\eightsf\def\sf{\fam\sffam\eightsf}%
  \ifprod@font
    \scriptfont\sffam=\sixsf
      \scriptscriptfont\sffam=\fivesf
  \else
    \scriptfont\sffam=\eightsf
      \scriptscriptfont\sffam=\eightsf
  \fi
  \def\oldstyle{\fam\@ne\eighti}%
  \b@ls{10pt}\rm\@viiipt%
}
\def\@viiipt{}

\def\ninepoint{
  \def\rm{\fam0\ninerm}%
  \textfont0=\ninerm \scriptfont0=\sixrm \scriptscriptfont0=\fiverm%
  \textfont1=\ninei  \scriptfont1=\sixi  \scriptscriptfont1=\fivei%
  \textfont2=\ninesy \scriptfont2=\sixsy \scriptscriptfont2=\fivesy%
  \textfont\itfam=\nineit\def\it{\fam\itfam\nineit}%
  \ifprod@font
    \scriptfont\itfam=\sixit
      \scriptscriptfont\itfam=\fiveit
  \else
    \scriptfont\itfam=\nineit
      \scriptscriptfont\itfam=\nineit
  \fi
  \textfont\bffam=\ninebf%
    \scriptfont\bffam=\sixbf%
      \scriptscriptfont\bffam=\fivebf%
  \def\bf{\fam\bffam\ninebf}%
  \textfont\slfam=\ninesl\def\sl{\fam\slfam\ninesl}%
  \ifprod@font
    \scriptfont\slfam=\sixsl
      \scriptscriptfont\slfam=\fivesl
  \else
    \scriptfont\slfam=\ninesl
      \scriptscriptfont\slfam=\ninesl
  \fi
  \textfont\ttfam=\ninett\def\tt{\fam\ttfam\ninett}%
  \ifprod@font
    \scriptfont\ttfam=\sixtt
      \scriptscriptfont\ttfam=\fivett
  \else
    \scriptfont\ttfam=\ninett
      \scriptscriptfont\ttfam=\ninett
  \fi
  \textfont\scfam=\ninecsc\def\sc{\fam\scfam\ninecsc}%
  \ifprod@font
    \scriptfont\scfam=\sixcsc
      \scriptscriptfont\scfam=\fivecsc
  \else
    \scriptfont\scfam=\ninecsc
      \scriptscriptfont\scfam=\ninecsc
  \fi
  \textfont\sffam=\ninesf\def\sf{\fam\sffam\ninesf}%
  \ifprod@font
    \scriptfont\sffam=\sixsf
      \scriptscriptfont\sffam=\fivesf
  \else
    \scriptfont\sffam=\ninesf
      \scriptscriptfont\sffam=\ninesf
  \fi
  \def\oldstyle{\fam\@ne\ninei}%
  \b@ls{\TextLeading plus \Feathering}\rm\@ixpt%
}
\def\@ixpt{}

\def\tenpoint{
  \def\rm{\fam0\tenrm}%
  \textfont0=\tenrm \scriptfont0=\sevenrm \scriptscriptfont0=\fiverm%
  \textfont1=\teni  \scriptfont1=\seveni  \scriptscriptfont1=\fivei%
  \textfont2=\tensy \scriptfont2=\sevensy \scriptscriptfont2=\fivesy%
  \textfont\itfam=\tenit\def\it{\fam\itfam\tenit}%
  \ifprod@font
    \scriptfont\itfam=\sevenit
      \scriptscriptfont\itfam=\fiveit
  \else
    \scriptfont\itfam=\tenit
      \scriptscriptfont\itfam=\tenit
  \fi
  \textfont\bffam=\tenbf%
    \scriptfont\bffam=\sevenbf%
      \scriptscriptfont\bffam=\fivebf%
  \def\bf{\fam\bffam\tenbf}%
  \textfont\slfam=\tensl\def\sl{\fam\slfam\tensl}%
  \ifprod@font
    \scriptfont\slfam=\sevensl
      \scriptscriptfont\slfam=\fivesl
  \else
    \scriptfont\slfam=\tensl
      \scriptscriptfont\slfam=\tensl
  \fi
  \textfont\ttfam=\tentt\def\tt{\fam\ttfam\tentt}%
  \ifprod@font
    \scriptfont\ttfam=\seventt
      \scriptscriptfont\ttfam=\fivett
  \else
    \scriptfont\ttfam=\tentt
      \scriptscriptfont\ttfam=\tentt
  \fi
  \textfont\scfam=\tencsc\def\sc{\fam\scfam\tencsc}%
  \ifprod@font
    \scriptfont\scfam=\sevencsc
      \scriptscriptfont\scfam=\fivecsc
  \else
    \scriptfont\scfam=\tencsc
      \scriptscriptfont\scfam=\tencsc
  \fi
  \textfont\sffam=\tensf\def\sf{\fam\sffam\tensf}%
  \ifprod@font
    \scriptfont\sffam=\sevensf
      \scriptscriptfont\sffam=\fivesf
  \else
    \scriptfont\sffam=\tensf
      \scriptscriptfont\sffam=\tensf
  \fi
  \def\oldstyle{\fam\@ne\teni}%
  \b@ls{11pt}\rm\@xpt%
}
\def\@xpt{}

\def\elevenpoint{
  \def\rm{\fam0\elevenrm}%
  \textfont0=\elevenrm \scriptfont0=\eightrm \scriptscriptfont0=\sixrm%
  \textfont1=\eleveni  \scriptfont1=\eighti  \scriptscriptfont1=\sixi%
  \textfont2=\elevensy \scriptfont2=\eightsy \scriptscriptfont2=\sixsy%
  \textfont\itfam=\elevenit\def\it{\fam\itfam\elevenit}%
  \ifprod@font
    \scriptfont\itfam=\eightit
      \scriptscriptfont\itfam=\sixit
  \else
    \scriptfont\itfam=\elevenit
      \scriptscriptfont\itfam=\elevenit
  \fi
  \textfont\bffam=\elevenbf%
    \scriptfont\bffam=\eightbf%
      \scriptscriptfont\bffam=\sixbf%
  \def\bf{\fam\bffam\elevenbf}%
  \textfont\slfam=\elevensl\def\sl{\fam\slfam\elevensl}%
  \ifprod@font
    \scriptfont\slfam=\eightsl
      \scriptscriptfont\slfam=\sixsl
  \else
    \scriptfont\slfam=\elevensl
      \scriptscriptfont\slfam=\elevensl
  \fi
  \textfont\ttfam=\eleventt\def\tt{\fam\ttfam\eleventt}%
  \ifprod@font
    \scriptfont\ttfam=\eighttt
      \scriptscriptfont\ttfam=\sixtt
  \else
    \scriptfont\ttfam=\eleventt
      \scriptscriptfont\ttfam=\eleventt
  \fi
  \textfont\scfam=\elevencsc\def\sc{\fam\scfam\elevencsc}%
  \ifprod@font
    \scriptfont\scfam=\eightcsc
      \scriptscriptfont\scfam=\sixcsc
  \else
    \scriptfont\scfam=\elevencsc
      \scriptscriptfont\scfam=\elevencsc
  \fi
  \textfont\sffam=\elevensf\def\sf{\fam\sffam\elevensf}%
  \ifprod@font
    \scriptfont\sffam=\eightsf
      \scriptscriptfont\sffam=\sixsf
  \else
    \scriptfont\sffam=\elevensf
      \scriptscriptfont\sffam=\elevensf
  \fi
  \def\oldstyle{\fam\@ne\eleveni}%
  \b@ls{13pt}\rm\@xipt%
}
\def\@xipt{}

\def\fourteenpoint{
  \def\rm{\fam0\fourteenrm}%
  \textfont0\fourteenrm  \scriptfont0\tenrm  \scriptscriptfont0\sevenrm%
  \textfont1\fourteeni   \scriptfont1\teni   \scriptscriptfont1\seveni%
  \textfont2\fourteensy  \scriptfont2\tensy  \scriptscriptfont2\sevensy%
  \textfont\itfam=\fourteenit\def\it{\fam\itfam\fourteenit}%
  \ifprod@font
    \scriptfont\itfam=\tenit
      \scriptscriptfont\itfam=\sevenit
  \else
    \scriptfont\itfam=\fourteenit
      \scriptscriptfont\itfam=\fourteenit
  \fi
  \textfont\bffam=\fourteenbf%
    \scriptfont\bffam=\tenbf%
      \scriptscriptfont\bffam=\sevenbf%
  \def\bf{\fam\bffam\fourteenbf}%
  \textfont\slfam=\fourteensl\def\sl{\fam\slfam\fourteensl}%
  \ifprod@font
    \scriptfont\slfam=\tensl
      \scriptscriptfont\slfam=\sevensl
  \else
    \scriptfont\slfam=\fourteensl
      \scriptscriptfont\slfam=\fourteensl
  \fi
  \textfont\ttfam=\fourteentt\def\tt{\fam\ttfam\fourteentt}%
  \ifprod@font
    \scriptfont\ttfam=\tentt
      \scriptscriptfont\ttfam=\seventt
  \else
    \scriptfont\ttfam=\fourteentt
      \scriptscriptfont\ttfam=\fourteentt
  \fi
  \textfont\scfam=\fourteencsc\def\sc{\fam\scfam\fourteencsc}%
  \ifprod@font
    \scriptfont\scfam=\tencsc
      \scriptscriptfont\scfam=\sevencsc
  \else
    \scriptfont\scfam=\fourteencsc
      \scriptscriptfont\scfam=\fourteencsc
  \fi
  \textfont\sffam=\fourteensf\def\sf{\fam\sffam\fourteensf}%
  \ifprod@font
    \scriptfont\sffam=\tensf
      \scriptscriptfont\sffam=\sevensf
  \else
    \scriptfont\sffam=\fourteensf
      \scriptscriptfont\sffam=\fourteensf
  \fi
  \def\oldstyle{\fam\@ne\fourteeni}%
  \b@ls{17pt}\rm\@xivpt%
}
\def\@xivpt{}

\def\seventeenpoint{
  \def\rm{\fam0\seventeenrm}%
  \textfont0\seventeenrm  \scriptfont0\twelverm  \scriptscriptfont0\tenrm%
  \textfont1\seventeeni   \scriptfont1\twelvei   \scriptscriptfont1\teni%
  \textfont2\seventeensy  \scriptfont2\twelvesy  \scriptscriptfont2\tensy%
  \textfont\itfam=\seventeenit\def\it{\fam\itfam\seventeenit}%
  \ifprod@font
    \scriptfont\itfam=\twelveit
      \scriptscriptfont\itfam=\tenit
  \else
    \scriptfont\itfam=\seventeenit
      \scriptscriptfont\itfam=\seventeenit
  \fi
  \textfont\bffam=\seventeenbf%
    \scriptfont\bffam=\twelvebf%
      \scriptscriptfont\bffam=\tenbf%
  \def\bf{\fam\bffam\seventeenbf}%
  \textfont\slfam=\seventeensl\def\sl{\fam\slfam\seventeensl}%
  \ifprod@font
    \scriptfont\slfam=\twelvesl
      \scriptscriptfont\slfam=\tensl
  \else
    \scriptfont\slfam=\seventeensl
      \scriptscriptfont\slfam=\seventeensl
  \fi
  \textfont\ttfam=\seventeentt\def\tt{\fam\ttfam\seventeentt}%
  \ifprod@font
    \scriptfont\ttfam=\twelvett
      \scriptscriptfont\ttfam=\tentt
  \else
    \scriptfont\ttfam=\seventeentt
      \scriptscriptfont\ttfam=\seventeentt
  \fi
  \textfont\scfam=\seventeencsc\def\sc{\fam\scfam\seventeencsc}%
  \ifprod@font
    \scriptfont\scfam=\twelvecsc
      \scriptscriptfont\scfam=\tencsc
  \else
    \scriptfont\scfam=\seventeencsc
      \scriptscriptfont\scfam=\seventeencsc
  \fi
  \textfont\sffam=\seventeensf\def\sf{\fam\sffam\seventeensf}%
  \ifprod@font
    \scriptfont\sffam=\twelvesf
      \scriptscriptfont\sffam=\tensf
  \else
    \scriptfont\sffam=\seventeensf
      \scriptscriptfont\sffam=\seventeensf
  \fi
  \def\oldstyle{\fam\@ne\seventeeni}%
  \b@ls{20pt}\rm\@xviipt%
}
\def\@xviipt{}

\lineskip=1pt      \normallineskip=\lineskip
\lineskiplimit=\z@ \normallineskiplimit=\lineskiplimit


\def\,{\relax\ifmmode \mskip\thinmuskip\else \thinspace\fi}
\let\protect=\relax

\long\def\@ifundefined#1#2#3{\expandafter\ifx\csname
  #1\endcsname\relax#2\else#3\fi}




\newtoks\math@groups \math@groups={}
\def\addtom@thgroup#1#2{#1\expandafter{\the#1#2}} 



\def\addtosizeh@ok#1#2#3#4{%
  \expandafter\def\csname @#1pt\endcsname{%
    \def\s@ze{#2}\def\ss@ze{#3}\def\sss@ze{#4}\the\math@groups%
  }%
}



\let\resetsizehook=\addtosizeh@ok


\ifprod@font
  \addtosizeh@ok{viii} {8} {6}  {5}
  \addtosizeh@ok{ix}   {9} {6}  {5}
  \addtosizeh@ok{x}    {10}{7}  {5}
  \addtosizeh@ok{xi}   {11}{8}  {6}
  \addtosizeh@ok{xiv}  {14}{10} {7}
  \addtosizeh@ok{xvii} {17}{12}{10}
\else
  \addtosizeh@ok{viii} {8}     {6}     {5}
  \addtosizeh@ok{ix}   {9}     {6}     {5}
  \addtosizeh@ok{x}    {10}    {7}     {5}
  \addtosizeh@ok{xi}   {10.95} {8}     {6}
  \addtosizeh@ok{xiv}  {14.4}  {10}    {7}
  \addtosizeh@ok{xvii} {17.28} {12}    {10}
\fi

\def\get@font#1#2#3{%
  \edef\fonts@ze{\romannumeral#3}
  \edef\fontn@me{\fonts@ze#1}
  \@ifundefined{\fontn@me}%
    {
     \global\expandafter\font\csname \fontn@me\endcsname=#2 at #3pt}%
    {}%
}

\def\ass@tfont#1#2{%
  \xdef\fam@name{\csname #1\endcsname}%
  \xdef\font@name{\csname #2\endcsname}%
  \let\textfont@name\font@name
  \textfont\fam@name\textfont@name
}

\def\ass@sfont#1#2{%
  \xdef\fam@name{\csname #1\endcsname}%
  \xdef\font@name{\csname #2\endcsname}%
  \let\textfont@name\font@name
  \scriptfont\fam@name\textfont@name
}

\def\ass@ssfont#1#2{%
  \xdef\fam@name{\csname #1\endcsname}%
  \xdef\font@name{\csname #2\endcsname}%
  \let\textfont@name\font@name
  \scriptscriptfont\fam@name\textfont@name
}


\def\NewSymbolFont#1#2{%
  \expandafter\ifx\csname sym#1fam\endcsname\relax 
    \expandafter\newfam\csname sym#1fam\endcsname
    \expandafter\edef\csname sym#1fam\endcsname{\the\allocationnumber}%
    \addtom@thgroup\math@groups{%
      \get@font{#1}{#2}{\s@ze}%
      \ass@tfont{sym#1fam}{\fontn@me}%
      \get@font{#1}{#2}{\ss@ze}%
      \ass@sfont{sym#1fam}{\fontn@me}%
      \get@font{#1}{#2}{\sss@ze}%
      \ass@ssfont{sym#1fam}{\fontn@me}%
    }%
  \else
    \errmessage{Family `#1' already defined}%
  \fi
}


\def\NewMathSymbol#1#2#3#4{%
  \edef\f@mly{\expandafter\hexnumber{\csname sym#3fam\endcsname}}%
  \mathchardef#1="#2\f@mly#4\relax
}


\newif\ifd@f

\def\NewMathDelimiter#1#2#3#4#5#6{%
  \d@ftrue
  \expandafter\ifx\csname sym#3fam\endcsname\relax
    \d@ffalse \errmessage{Family `#3' is not defined}%
  \fi
  \expandafter\ifx\csname sym#5fam\endcsname\relax
    \d@ffalse \errmessage{Family `#5' is not defined}%
  \fi
  \ifd@f
    \edef\f@mly{\expandafter\hexnumber{\csname sym#3fam\endcsname}}%
    \edef\f@mlytw@{\expandafter\hexnumber{\csname sym#5fam\endcsname}}%
    \xdef#1{\delimiter"#2\f@mly #4\f@mlytw@ #6\relax}%
  \fi
}


\def\setboxz@h{\setbox\z@\hbox}
\def\wdz@{\wd\z@}
\def\boxz@{\box\z@}
\def\setbox@ne{\setbox\@ne}
\def\wd@ne{\wd\@ne}

\def\math@atom#1#2{%
   \binrel@{#1}\binrel@@{#2}}
\def\binrel@#1{\setboxz@h{\thinmuskip0mu
  \medmuskip\m@ne mu\thickmuskip\@ne mu$#1\m@th$}%
 \setbox@ne\hbox{\thinmuskip0mu\medmuskip\m@ne mu\thickmuskip
  \@ne mu${}#1{}\m@th$}%
 \setbox\tw@\hbox{\hskip\wd@ne\hskip-\wdz@}}
\def\binrel@@#1{\ifdim\wd2<\z@\mathbin{#1}\else\ifdim\wd\tw@>\z@
 \mathrel{#1}\else{#1}\fi\fi}

\def\m@thit{1}

\def\set@skchar#1{\global\expandafter\skewchar
  \csname\fontn@me\endcsname=#1\relax}

\def\NewMathAlphabet#1#2#3{%
  \def\tst{#3}%
  \ifx\tst\empty\else 
    \expandafter\gdef\csname #1@sc\endcsname{}
  \fi
  \expandafter\def\csname #1\endcsname{
    \protect\csname @#1\endcsname}%
  \expandafter\def\csname @#1\endcsname##1{
    {%
    \begingroup
      \get@font{#1}{#2}{\s@ze}%
      \@ifundefined{#1@sc}{}{\set@skchar{#3}}%
      \ass@tfont{m@thit}{\fontn@me}%
      \get@font{#1}{#2}{\ss@ze}%
      \@ifundefined{#1@sc}{}{\set@skchar{#3}}%
      \ass@sfont{m@thit}{\fontn@me}%
      \get@font{#1}{#2}{\sss@ze}%
      \@ifundefined{#1@sc}{}{\set@skchar{#3}}%
      \ass@ssfont{m@thit}{\fontn@me}%
      \math@atom{##1}{%
      \mathchoice%
        {\hbox{$\m@th\displaystyle##1$}}%
        {\hbox{$\m@th\textstyle##1$}}%
        {\hbox{$\m@th\scriptstyle##1$}}%
        {\hbox{$\m@th\scriptscriptstyle##1$}}}%
    \endgroup
    }%
  }%
}


\newif\iffirstta  \firsttatrue

\def\set@hchar#1{\global\expandafter\hyphenchar
  \csname\fontn@me\endcsname=#1\relax}

\def\NewTextAlphabet#1#2#3{%
  \iffirstta
    \global\firsttafalse
    \newfam\scratchfam
    \edef\scrt@fam{\the\allocationnumber}%
  \fi
  \def\tst{#3}%
  \ifx\tst\empty\else 
    \expandafter\gdef\csname #1@hc\endcsname{}
  \fi
  \expandafter\def\csname #1\endcsname{
    \protect\csname t@#1\endcsname}%
  \long\expandafter\def\csname t@#1\endcsname##1{
    \ifmmode
      \typeout{Warning: do not use \expandafter\string\csname #1\endcsname
        \space in math mode}\fi%
    {%
      \get@font{#1}{#2}{\s@ze}\let\t@xtfnt=\fontn@me\relax
      \@ifundefined{#1@hc}{}{\set@hchar{#3}}%
      \ass@tfont{scrt@fam}{\fontn@me}%
      \get@font{#1}{#2}{\ss@ze}%
      \@ifundefined{#1@hc}{}{\set@hchar{#3}}%
      \ass@sfont{scrt@fam}{\fontn@me}%
      \get@font{#1}{#2}{\sss@ze}%
      \@ifundefined{#1@hc}{}{\set@hchar{#3}}%
      \ass@ssfont{scrt@fam}{\fontn@me}%
      \fam\scratchfam\csname\t@xtfnt\endcsname
    ##1%
    }%
  }%
  \expandafter\def\csname #1shape
    \endcsname{\protect\csname @#1shape\endcsname}%
  \expandafter\def\csname @#1shape\endcsname{
    \ifmmode
      \typeout{Warning: do not use \expandafter\string\csname
        #1shape\endcsname \space in math mode}\fi
      \get@font{#1}{#2}{\s@ze}\let\t@xtfnt=\fontn@me\relax
      \@ifundefined{#1@hc}{}{\set@hchar{#3}}%
      \ass@tfont{scrt@fam}{\fontn@me}%
      \get@font{#1}{#2}{\ss@ze}%
      \@ifundefined{#1@hc}{}{\set@hchar{#3}}%
      \ass@sfont{scrt@fam}{\fontn@me}%
      \get@font{#1}{#2}{\sss@ze}%
      \@ifundefined{#1@hc}{}{\set@hchar{#3}}%
      \ass@ssfont{scrt@fam}{\fontn@me}%
      \fam\scratchfam\csname\t@xtfnt\endcsname
  }%
}


\ifprod@font
  \def\math@itfnt{mtmib10}
  \def\math@syfnt{mtbsy10}
\else
  \def\math@itfnt{cmmib10}
  \def\math@syfnt{cmbsy10}
\fi

\def\m@thsy{2}

\def\bmath{\protect\@bmath}
\def\@bmath#1{%
  {%
  \begingroup
    \get@font{mthit}{\math@itfnt}{\s@ze}\set@skchar{'177}%
    \ass@tfont{m@thit}{\fontn@me}%
    \get@font{mthit}{\math@itfnt}{\ss@ze}\set@skchar{'177}%
    \ass@sfont{m@thit}{\fontn@me}%
    \get@font{mthit}{\math@itfnt}{\sss@ze}\set@skchar{'177}%
    \ass@ssfont{m@thit}{\fontn@me}%
    \get@font{mthsy}{\math@syfnt}{\s@ze}\set@skchar{'60}%
    \ass@tfont{m@thsy}{\fontn@me}%
    \get@font{mthsy}{\math@syfnt}{\ss@ze}\set@skchar{'60}%
    \ass@sfont{m@thsy}{\fontn@me}%
    \get@font{mthsy}{\math@syfnt}{\sss@ze}\set@skchar{'60}%
    \ass@ssfont{m@thsy}{\fontn@me}%
    \math@atom{#1}{%
    \mathchoice%
      {\hbox{$\m@th\displaystyle#1$}}%
      {\hbox{$\m@th\textstyle#1$}}%
      {\hbox{$\m@th\scriptstyle#1$}}%
      {\hbox{$\m@th\scriptscriptstyle#1$}}}%
  \endgroup
  }%
}



\def\diameter{{\ifmmode\mathchoice
{\ooalign{\hfil\hbox{$\displaystyle/$}\hfil\crcr
{\hbox{$\displaystyle\mathchar"20D$}}}}
{\ooalign{\hfil\hbox{$\textstyle/$}\hfil\crcr
{\hbox{$\textstyle\mathchar"20D$}}}}
{\ooalign{\hfil\hbox{$\scriptstyle/$}\hfil\crcr
{\hbox{$\scriptstyle\mathchar"20D$}}}}
{\ooalign{\hfil\hbox{$\scriptscriptstyle/$}\hfil\crcr
{\hbox{$\scriptscriptstyle\mathchar"20D$}}}}
\else{\ooalign{\hfil/\hfil\crcr\mathhexbox20D}}%
\fi}}

\def\sq{\ifmmode\squareforqed\else{\unskip\nobreak\hfil
\penalty50\hskip1em\null\nobreak\hfil\squareforqed
\parfillskip=0pt\finalhyphendemerits=0\endgraf}\fi}
\def\squareforqed{\hbox{\rlap{$\sqcap$}$\sqcup$}}


\def\bbbc{{\mathchoice {\setbox0=\hbox{$\displaystyle\rm C$}\hbox{\hbox
to0pt{\kern0.4\wd0\vrule height0.9\ht0\hss}\box0}}
{\setbox0=\hbox{$\textstyle\rm C$}\hbox{\hbox
to0pt{\kern0.4\wd0\vrule height0.9\ht0\hss}\box0}}
{\setbox0=\hbox{$\scriptstyle\rm C$}\hbox{\hbox
to0pt{\kern0.4\wd0\vrule height0.9\ht0\hss}\box0}}
{\setbox0=\hbox{$\scriptscriptstyle\rm C$}\hbox{\hbox
to0pt{\kern0.4\wd0\vrule height0.9\ht0\hss}\box0}}}}
\def\bbbq{{\mathchoice {\setbox0=\hbox{$\displaystyle\rm
Q$}\hbox{\raise
0.15\ht0\hbox to0pt{\kern0.4\wd0\vrule height0.8\ht0\hss}\box0}}
{\setbox0=\hbox{$\textstyle\rm Q$}\hbox{\raise
0.15\ht0\hbox to0pt{\kern0.4\wd0\vrule height0.8\ht0\hss}\box0}}
{\setbox0=\hbox{$\scriptstyle\rm Q$}\hbox{\raise
0.15\ht0\hbox to0pt{\kern0.4\wd0\vrule height0.7\ht0\hss}\box0}}
{\setbox0=\hbox{$\scriptscriptstyle\rm Q$}\hbox{\raise
0.15\ht0\hbox to0pt{\kern0.4\wd0\vrule height0.7\ht0\hss}\box0}}}}
\def\bbbt{{\mathchoice {\setbox0=\hbox{$\displaystyle\rm
T$}\hbox{\hbox to0pt{\kern0.3\wd0\vrule height0.9\ht0\hss}\box0}}
{\setbox0=\hbox{$\textstyle\rm T$}\hbox{\hbox
to0pt{\kern0.3\wd0\vrule height0.9\ht0\hss}\box0}}
{\setbox0=\hbox{$\scriptstyle\rm T$}\hbox{\hbox
to0pt{\kern0.3\wd0\vrule height0.9\ht0\hss}\box0}}
{\setbox0=\hbox{$\scriptscriptstyle\rm T$}\hbox{\hbox
to0pt{\kern0.3\wd0\vrule height0.9\ht0\hss}\box0}}}}
\def\bbbs{{\mathchoice
{\setbox0=\hbox{$\displaystyle     \rm S$}\hbox{\raise0.5\ht0\hbox
to0pt{\kern0.35\wd0\vrule height0.45\ht0\hss}\hbox
to0pt{\kern0.55\wd0\vrule height0.5\ht0\hss}\box0}}
{\setbox0=\hbox{$\textstyle        \rm S$}\hbox{\raise0.5\ht0\hbox
to0pt{\kern0.35\wd0\vrule height0.45\ht0\hss}\hbox
to0pt{\kern0.55\wd0\vrule height0.5\ht0\hss}\box0}}
{\setbox0=\hbox{$\scriptstyle      \rm S$}\hbox{\raise0.5\ht0\hbox
to0pt{\kern0.35\wd0\vrule height0.45\ht0\hss}\raise0.05\ht0\hbox
to0pt{\kern0.5\wd0\vrule height0.45\ht0\hss}\box0}}
{\setbox0=\hbox{$\scriptscriptstyle\rm S$}\hbox{\raise0.5\ht0\hbox
to0pt{\kern0.4\wd0\vrule height0.45\ht0\hss}\raise0.05\ht0\hbox
to0pt{\kern0.55\wd0\vrule height0.45\ht0\hss}\box0}}}}
\def\bbbz{{\mathchoice {\hbox{$\sf\textstyle Z\kern-0.4em Z$}}
{\hbox{$\sf\textstyle Z\kern-0.4em Z$}}
{\hbox{$\sf\scriptstyle Z\kern-0.3em Z$}}
{\hbox{$\sf\scriptscriptstyle Z\kern-0.2em Z$}}}}


\def\Nulle{0} 
\def\Afe{1}   
\def\Hae{2}   
\def\Hbe{3}   
\def\Hce{4}   
\def\Hde{5}   


\newcount\LastMac       \LastMac=\Nulle

\newskip\half      \half=5.5pt plus 1.5pt minus 2.25pt
\newskip\one       \one=11pt plus 3pt minus 5.5pt
\newskip\onehalf   \onehalf=16.5pt plus 5.5pt minus 8.25pt
\newskip\two       \two=22pt plus 5.5pt minus 11pt

\def\Half{\addvspace{\half}}
\def\One{\addvspace{\one}}
\def\OneHalf{\addvspace{\onehalf}}
\def\Two{\addvspace{\two}}

\def\Raggedright{
  \rightskip=\z@ plus \hsize\relax
}

\def\Fullout{
  \rightskip=\z@\relax
}

\def\Hang#1#2{
  \hangindent=#1%
  \hangafter=#2\relax
}


\newif\ifsp@page
\def\pagestyle#1{\csname ps@#1\endcsname}
\def\thispagestyle#1{\global\sp@pagetrue\gdef\sp@type{#1}}

\def\ps@titlepage{%
  \def\@oddhead{\eightpoint\noindent \the\CatchLine
    \ifprod@font\else\qquad Printed\ \today\qquad
      (MN plain \TeX\ macros\ v\@version)\fi \hfil}%
  \let\@evenhead=\@oddhead
  \def\@oddfoot{\eightpoint\copyright\ \@pubyear\ RAS\hfil}%
  \def\@evenfoot{\hfil\eightpoint\noindent\copyright\ \@pubyear\ RAS}%
}

\def\ps@headings{%
  \def\@oddhead{\elevenpoint\it\noindent
    \hfill\the\RightHeader\hskip1.5em\rm\folio}%
  \def\@evenhead{\elevenpoint\noindent
    \folio\hskip1.5em\it\the\LeftHeader\hfill}%
  \def\@oddfoot{\eightpoint\noindent\copyright\ \@pubyear\ RAS,
    MNRAS {\bf \@volume}, \@pagerange\hfil}%
  \def\@evenfoot{\hfil\eightpoint\copyright\ \@pubyear\ RAS,
    MNRAS {\bf \@volume}, \@pagerange}%
}

\def\ps@plate{%
  \def\@oddhead{\eightpoint\noindent\plt@cap\hfil}%
  \def\@evenhead{\eightpoint\noindent\plt@cap\hfil}%
  \def\@oddfoot{\eightpoint\noindent\copyright\ \@pubyear\ RAS,
    MNRAS {\bf \@volume}, \@pagerange\hfil}%
  \def\@evenfoot{\hfil\eightpoint\copyright\ \@pubyear\ RAS,
    MNRAS {\bf \@volume}, \@pagerange}%
}



\def\title#1{
  \bgroup
    \vbox to 8pt{\vss}%
    \seventeenpoint
    \Raggedright
    \noindent \strut{\bf #1}\par
  \egroup
}

\def\author#1{
  \bgroup
    \ifnum\LastMac=\Afe \OneHalf\else \vskip 21pt\fi
    \fourteenpoint
    \Raggedright
    \noindent \strut #1\par
    \vskip 3pt%
  \egroup
}

\def\affiliation#1{
  \bgroup
    \vskip -4pt%
    \eightpoint
    \Raggedright
    \noindent \strut {\it #1}\par
  \egroup
  \LastMac=\Afe\relax
}

\def\acceptedline#1{
  \bgroup
    \Two
    \eightpoint
    \Raggedright
    \noindent \strut #1\par
  \egroup
}

\long\def\abstract#1{%
  \bgroup
    \vskip 20pt%
    \leftskip 11pc\rightskip\z@
    \noindent{\ninebf ABSTRACT}\par
    \tenpoint
    \Fullout
    \noindent #1\par
  \egroup
}

\long\def\keywords#1{
  \bgroup
    \Half
    \leftskip 11pc\rightskip\z@
    \tenpoint
    \Fullout
    \noindent\hbox{\bf Key words:}\ #1\par
  \egroup
}


\def\maketitle{%
  \EndOpening
  \ifsinglecol \else \MakePage\fi
}



\def\@nameuse#1{\csname #1\endcsname}
\def\arabic#1{\@arabic{\@nameuse{#1}}}
\def\alph#1{\@alph{\@nameuse{#1}}}
\def\Alph#1{\@Alph{\@nameuse{#1}}}
\def\@arabic#1{\number #1}
\def\@Alph#1{\ifcase#1\or A\or B\or C\or D\else\@Ialph{#1}\fi}
\def\@Ialph#1{\ifcase#1\or \or \or \or \or E\or F\or G\or H\or I\or J\or
   K\or L\or M\or N\or O\or P\or Q\or R\or S\or T\or U\or V\or W\or X\or
   Y\or Z\else\errmessage{Counter out of range}\fi}
\def\@alph#1{\ifcase#1\or a\or b\or c\or d\else\@ialph{#1}\fi}
\def\@ialph#1{\ifcase#1\or \or \or \or \or e\or f\or g\or h\or i\or j\or
   k\or l\or m\or n\or o\or p\or q\or r\or s\or t\or u\or v\or w\or x\or y\or
   z\else\errmessage{Counter out of range}\fi}


\newcount\Eqnno
\newcount\SubEqnno

\def\theeq{\arabic{Eqnno}}
\def\thesubeq{\alph{SubEqnno}}

\def\stepeq{\relax
  \global\SubEqnno \z@
  \global\advance\Eqnno \@ne\relax
  {\rm (\theeq)}%
}

\def\startsubeq{\relax
  \global\SubEqnno \z@
  \global\advance\Eqnno \@ne\relax
  \stepsubeq
}

\def\stepsubeq{\relax
  \global\advance\SubEqnno \@ne\relax
  {\rm (\theeq\thesubeq)}%
}


\newcount\Sec        
\newcount\SecSec
\newcount\SecSecSec

\def\thesection{\arabic{Sec}}
\def\thesubsection{\thesection.\arabic{SecSec}}
\def\thesubsubsection{\thesubsection.\arabic{SecSecSec}}

\Sec=\z@

\def\:{\let\@sptoken= } \:  
\def\:{\@xifnch} \expandafter\def\: {\futurelet\@tempc\@ifnch}

\def\@ifnextchar#1#2#3{%
  \let\@tempMACe #1%
  \def\@tempMACa{#2}%
  \def\@tempMACb{#3}%
  \futurelet \@tempMACc\@ifnch%
}

\def\@ifnch{%
\ifx \@tempMACc \@sptoken%
  \let\@tempMACd\@xifnch%
\else%
  \ifx \@tempMACc \@tempMACe%
    \let\@tempMACd\@tempMACa%
  \else%
    \let\@tempMACd\@tempMACb%
  \fi%
\fi%
\@tempMACd%
}

\def\@ifstar#1#2{\@ifnextchar *{\def\@tempMACa*{#1}\@tempMACa}{#2}}

\newskip\@tempskipb

\def\addvspace#1{%
  \ifvmode\else \endgraf\fi%
  \ifdim\lastskip=\z@%
    \vskip #1\relax%
  \else%
    \@tempskipb#1\relax\@xaddvskip%
  \fi%
}

\def\@xaddvskip{%
  \ifdim\lastskip<\@tempskipb%
    \vskip-\lastskip%
    \vskip\@tempskipb\relax%
  \else%
    \ifdim\@tempskipb<\z@%
      \ifdim\lastskip<\z@ \else%
        \advance\@tempskipb\lastskip%
        \vskip-\lastskip\vskip\@tempskipb%
      \fi%
    \fi%
  \fi%
}

\newskip\@tmpSKIP

\def\addpen#1{%
  \ifvmode
    \if@nobreak
    \else
      \ifdim\lastskip=\z@
        \penalty#1\relax
      \else
        \@tmpSKIP=\lastskip
        \vskip -\lastskip
        \penalty#1\vskip\@tmpSKIP
      \fi
    \fi
  \fi
}

\newcount\@clubpen   \@clubpen=\clubpenalty
\newif\if@nobreak    \@nobreakfalse

\def\@noafterindent{%
  \global\@nobreaktrue
  \everypar{\if@nobreak
              \global\@nobreakfalse
              \clubpenalty \@M
              {\setbox\z@\lastbox}%
              \LastMac=\Nulle\relax%
            \else
              \clubpenalty \@clubpen
              \everypar{}%
            \fi}%
}

\newcount\gds@cbrk   \gds@cbrk=-300

\def\@nohdbrk{\interlinepenalty \@M\relax}

\let\@par=\par
\def\@restorepar{\def\par{\@par}}

\newif\if@endpe   \@endpefalse
 
\def\@doendpe{\@endpetrue \@nobreakfalse \LastMac=\Nulle\relax%
     \def\par{\@restorepar\everypar{}\par\@endpefalse}%
              \everypar{\setbox\z@\lastbox\everypar{}\@endpefalse}%
}

\def\section{\@ifstar{\@ssection}{\@section}}

\def\@section#1{
  \if@nobreak
    \everypar{}%
    \ifnum\LastMac=\Hae \addvspace{\half}\fi
  \else
    \addpen{\gds@cbrk}%
    \addvspace{\two}%
  \fi
  \bgroup
    \ninepoint\bf
    \Raggedright
    \global\advance\Sec \@ne
    \ifappendix
      \global\Eqnno=\z@ \global\SubEqnno=\z@\relax
      \def\ch@ck{#1}%
      \ifx\ch@ck\empty \def\c@lon{}\else\def\c@lon{:}\fi
      \noindent\@nohdbrk APPENDIX\ \thesection\c@lon\hskip 0.5em%
        \uppercase{#1}\par
    \else
      \noindent\@nohdbrk\thesection\hskip 1pc \uppercase{#1}\par
    \fi
    \global\SecSec=\z@
  \egroup
  \nobreak
  \vskip\half
  \nobreak
  \@noafterindent
  \LastMac=\Hae\relax
}

\def\@ssection#1{
  \if@nobreak
    \everypar{}%
    \ifnum\LastMac=\Hae \addvspace{\half}\fi
  \else
    \addpen{\gds@cbrk}%
    \addvspace{\two}%
  \fi
  \bgroup
    \ninepoint\bf
    \Raggedright
    \noindent\@nohdbrk\uppercase{#1}\par
  \egroup
  \nobreak
  \vskip\half
  \nobreak
  \@noafterindent
  \LastMac=\Hae\relax
}

\def\subsection{\@ifstar{\@ssubsection}{\@subsection}}

\def\@subsection#1{
  \if@nobreak
    \everypar{}%
    \ifnum\LastMac=\Hae \addvspace{1pt plus 1pt minus .5pt}\fi
  \else
    \addpen{\gds@cbrk}%
    \addvspace{\onehalf}%
  \fi
  \bgroup
    \ninepoint\bf
    \Raggedright
    \global\advance\SecSec \@ne
    \noindent\@nohdbrk\thesubsection \hskip 1pc\relax #1\par
    \global\SecSecSec=\z@
  \egroup
  \nobreak
  \vskip\half
  \nobreak
  \@noafterindent
  \LastMac=\Hbe\relax
}

\def\@ssubsection#1{
  \if@nobreak
    \everypar{}%
    \ifnum\LastMac=\Hae \addvspace{1pt plus 1pt minus .5pt}\fi
  \else
    \addpen{\gds@cbrk}%
    \addvspace{\onehalf}%
  \fi
  \bgroup
    \ninepoint\bf
    \Raggedright
    \noindent\@nohdbrk #1\par
  \egroup
  \nobreak
  \vskip\half
  \nobreak
  \@noafterindent
  \LastMac=\Hbe\relax
}

\def\subsubsection{\@ifstar{\@ssubsubsection}{\@subsubsection}}

\def\@subsubsection#1{
  \if@nobreak
    \everypar{}%
    \ifnum\LastMac=\Hbe \addvspace{1pt plus 1pt minus .5pt}\fi
  \else
    \addpen{\gds@cbrk}%
    \addvspace{\onehalf}%
  \fi
  \bgroup
    \ninepoint\it
    \Raggedright
    \global\advance\SecSecSec \@ne
    \noindent\@nohdbrk\thesubsubsection \hskip 1pc\relax #1\par
  \egroup
  \nobreak
  \vskip\half
  \nobreak
  \@noafterindent
  \LastMac=\Hce\relax
}

\def\@ssubsubsection#1{
  \if@nobreak
    \everypar{}%
    \ifnum\LastMac=\Hbe \addvspace{1pt plus 1pt minus .5pt}\fi
  \else
    \addpen{\gds@cbrk}%
    \addvspace{\onehalf}%
  \fi
  \bgroup
    \ninepoint\it
    \Raggedright
    \noindent\@nohdbrk #1\par
  \egroup
  \nobreak
  \vskip\half
  \nobreak
  \@noafterindent
  \LastMac=\Hce\relax
}

\def\paragraph#1{
  \if@nobreak
    \everypar{}%
  \else
    \addpen{\gds@cbrk}%
    \addvspace{\one}%
  \fi%
  \bgroup%
    \ninepoint\it
    \noindent #1\ \nobreak%
  \egroup
  \LastMac=\Hde\relax
  \ignorespaces
}


\newif\ifappendix

\def\appendix{%
  \global\appendixtrue
  \def\thesection{\Alph{Sec}}%
  \def\thesubsection{\thesection\arabic{SecSec}}%
  \def\theeq{\thesection\arabic{Eqnno}}%
  \Sec=\z@ \SecSec=\z@ \SecSecSec=\z@ \Eqnno=\z@ \SubEqnno=\z@\relax
}




\def\beginlist{%
  \par\if@nobreak \else\addvspace{\half}\fi%
  \bgroup%
    \ninepoint
    \let\item=\list@item%
}

\def\list@item{%
  \par\noindent\hskip 1em\relax%
  \ignorespaces%
}

\def\endlist{\par\egroup\addvspace{\half}\@doendpe}


\def\beginrefs{%
  \par
  \bgroup
    \eightpoint
    \Fullout
    \let\bibitem=\bib@item
}

\def\bib@item{%
  \par\parindent=1.5em\Hang{1.5em}{1}%
  \everypar={\Hang{1.5em}{1}\ignorespaces}%
  \noindent\ignorespaces
}

\def\endrefs{\par\egroup\@doendpe}


\newtoks\CatchLine

\def\@journal{Mon.\ Not.\ R.\ Astron.\ Soc.\ }  
\def\@pubyear{1994}        
\def\@pagerange{000--000}  
\def\@volume{000}          
\def\@microfiche{}         %

\def\pubyear#1{\gdef\@pubyear{#1}\@makecatchline}
\def\pagerange#1{\gdef\@pagerange{#1}\@makecatchline}
\def\volume#1{\gdef\@volume{#1}\@makecatchline}
\def\microfiche#1{\gdef\@microfiche{and Microfiche\ #1}\@makecatchline}

\def\@makecatchline{%
  \global\CatchLine{%
    {\rm \@journal {\bf \@volume},\ \@pagerange\ (\@pubyear)\ \@microfiche}}%
}

\@makecatchline 

\newtoks\LeftHeader

\newtoks\RightHeader

\def\PageHead{
  \begingroup
    \ifsp@page
      \csname ps@\sp@type\endcsname
    \fi
    \ifodd\pageno
      \let\the@head=\@oddhead
    \else
      \let\the@head=\@evenhead
    \fi
    \vbox to \z@{\vskip-22.5\p@%
      \hbox to \PageWidth{\vbox to8.5\p@{}%
        \the@head
      }%
    \vss}%
  \endgroup
  \nointerlineskip
}

\gdef\PageFoot{%
  \nointerlineskip%
  \begingroup
  \ifsp@page
    \csname ps@\sp@type\endcsname
    \global\sp@pagefalse
  \fi
  \vbox to 22pt{\vfil%
    \hbox to \PageWidth{%
      \eightpoint\strut\noindent
      \ifodd\pageno
        \@oddfoot
      \else
        \@evenfoot
      \fi
    }%
  }%
  \endgroup
}

\def\today{%
  \number\day\space
  \ifcase\month\or January\or February\or March\or April\or May\or June\or
    July\or August\or September\or October\or November\or December\fi
  \space\number\year%
}

\def\authorcomment#1{%
  \gdef\PageFoot{%
    \nointerlineskip%
    \vbox to 20pt{\vfil%
      \hbox to \PageWidth{\elevenpoint\noindent \hfil #1 \hfil}}%
  }%
}


\newif\ifplate@page
\newbox\plt@box

\def\beginplatepage{%
  \let\plate=\plate@head
  \let\caption=\fig@caption
  \global\setbox\plt@box=\vbox\bgroup
  \TEMPDIMEN=\PageWidth 
  \hsize=\PageWidth\relax
}

\def\endplatepage{\par\egroup\global\plate@pagetrue}
\def\plate@head#1{\gdef\plt@cap{#1}}


\def\letters{%
  \gdef\folio{\ifnum\pageno<\z@ L\romannumeral-\pageno
    \else L\number\pageno \fi}%
}


\newdimen\mathindent

\global\mathindent=\z@
\global\everydisplay{\global\@dspwd=\displaywidth\displaysetup}


\def\@displaylines#1{
  {}$\displ@y\hbox{\vbox{\halign{$\@lign\hfil\displaystyle##\hfil$\crcr
  #1\crcr}}}${}%
}

\def\@eqalign#1{\null\vcenter{\openup\jot\m@th
  \ialign{\strut\hfil$\displaystyle{##}$&$\displaystyle{{}##}$\hfil
      \crcr#1\crcr}}%
}

\def\@eqalignno#1{
  \global\advance\@dspwd by -\mathindent%
  {}$\displ@y\hbox{\vbox{\halign to\@dspwd%
  {\hfil$\@lign\displaystyle{##}$\tabskip\z@skip
  &$\@lign\displaystyle{{}##}$\hfil\tabskip\centering
  &\llap{$\@lign##$}\tabskip\z@skip\crcr
  #1\crcr}}}${}%
}


\global\let\displaylines=\@displaylines
\global\let\eqalign=\@eqalign
\global\let\eqalignno=\@eqalignno
\global\let\leqalignno=\@eqalignno

\newdimen\@dspwd   \@dspwd=\z@
\newif\if@eqno
\newif\if@leqno
\newtoks\@eqn
\newtoks\@eq

\def\displaysetup#1$${\displaytest#1\eqno\eqno\displaytest}

\def\displaytest#1\eqno#2\eqno#3\displaytest{%
 \if!#3!\ldisplaytest#1\leqno\leqno\ldisplaytest
 \else\@eqnotrue\@leqnofalse\@eqn={#2}\@eq={#1}\fi
 \generaldisplay$$}

\def\ldisplaytest#1\leqno#2\leqno#3\ldisplaytest{%
\@eq={#1}%
 \if!#3!\@eqnofalse\else\@eqnotrue\@leqnotrue
  \@eqn={#2}\fi}

\def\generaldisplay{%
  \if@eqno
    \if@leqno
      \hbox to \displaywidth{\noindent
        \rlap{$\displaystyle\the\@eqn$}%
        \hskip\mathindent$\displaystyle\the\@eq$\hfil}%
    \else
      \hbox to \displaywidth{\noindent
        \hskip\mathindent
        $\displaystyle\the\@eq$\hfil$\displaystyle\the\@eqn$}%
    \fi
  \else
    \hbox to \displaywidth{\noindent
      \hskip\mathindent$\displaystyle\the\@eq$\hfil}%
  \fi
}


\def\@notice{%
  \par\Two%
  \noindent{\b@ls{11pt}\ninerm This paper has been produced using the
    Royal Astronomical Society/Blackwell Science \TeX\ macros.\par}%
}

\outer\def\bye{\@notice\par\vfill\supereject\end}


\def\start@mess{%
  Monthly notices of the RAS journal style (\@typeface)\space
    v\@version,\space \@verdate.%
}

\everyjob{\Warn{\start@mess}}



\newif\if@debug \@debugfalse  

\def\Print#1{\if@debug\immediate\write16{#1}\else \fi}
\def\Warn#1{\immediate\write16{#1}}
\def\wlog#1{}

\newcount\Iteration 

\def\Single{0} \def\Double{1}                 
\def\Figure{0} \def\Table{1}                  

\def\InStack{0}  
\def\InZoneA{1}
\def\InZoneB{2}
\def\InZoneC{3}

\newcount\TEMPCOUNT 
\newdimen\TEMPDIMEN 
\newbox\TEMPBOX     
\newbox\VOIDBOX     

\newcount\LengthOfStack 
\newcount\MaxItems      
\newcount\StackPointer
\newcount\Point         
\newcount\NextFigure    
\newcount\NextTable     
\newcount\NextItem      

\newcount\StatusStack   
\newcount\NumStack      
\newcount\TypeStack     
\newcount\SpanStack     
\newcount\BoxStack      

\newcount\ItemSTATUS    
\newcount\ItemNUMBER    
\newcount\ItemTYPE      
\newcount\ItemSPAN      
\newbox\ItemBOX         
\newdimen\ItemSIZE      

\newdimen\PageHeight    
\newdimen\TextLeading   
\newdimen\Feathering    
\newcount\LinesPerPage  
\newdimen\ColumnWidth   
\newdimen\ColumnGap     
\newdimen\PageWidth     
\newdimen\BodgeHeight   
\newcount\Leading       

\newdimen\ZoneBSize  
\newdimen\TextSize   
\newbox\ZoneABOX     
\newbox\ZoneBBOX     
\newbox\ZoneCBOX     

\newif\ifFirstSingleItem
\newif\ifFirstZoneA
\newif\ifMakePageInComplete
\newif\ifMoreFigures \MoreFiguresfalse 
\newif\ifMoreTables  \MoreTablesfalse  

\newif\ifFigInZoneB 
\newif\ifFigInZoneC 
\newif\ifTabInZoneB 
\newif\ifTabInZoneC

\newif\ifZoneAFullPage

\newbox\MidBOX    
\newbox\LeftBOX
\newbox\RightBOX
\newbox\PageBOX   

\newif\ifLeftCOL  
\LeftCOLtrue

\newdimen\ZoneBAdjust

\newcount\ItemFits
\def\Yes{1}
\def\No{2}


\MaxItems=15
\NextFigure=\z@        
\NextTable=\@ne

\BodgeHeight=6pt
\TextLeading=11pt    
\Leading=11
\Feathering=\z@      
\LinesPerPage=61     
\topskip=\TextLeading
\ColumnWidth=20pc    
\ColumnGap=2pc       

\newskip\ItemSepamount  
\ItemSepamount=\TextLeading plus \TextLeading minus 4pt

\parskip=\z@ plus .1pt
\parindent=18pt
\widowpenalty=\z@
\clubpenalty=10000
\tolerance=1500
\hbadness=1500
\abovedisplayskip=6pt plus 2pt minus 1pt
\belowdisplayskip=6pt plus 2pt minus 1pt
\abovedisplayshortskip=6pt plus 2pt minus 1pt
\belowdisplayshortskip=6pt plus 2pt minus 1pt

\frenchspacing

\ninepoint 

\PageHeight=682pt
\PageWidth=2\ColumnWidth
\advance\PageWidth by \ColumnGap

\pagestyle{headings}




\newcount\DUMMY \StatusStack=\allocationnumber
\newcount\DUMMY \newcount\DUMMY \newcount\DUMMY 
\newcount\DUMMY \newcount\DUMMY \newcount\DUMMY 
\newcount\DUMMY \newcount\DUMMY \newcount\DUMMY
\newcount\DUMMY \newcount\DUMMY \newcount\DUMMY 
\newcount\DUMMY \newcount\DUMMY \newcount\DUMMY

\newcount\DUMMY \NumStack=\allocationnumber
\newcount\DUMMY \newcount\DUMMY \newcount\DUMMY 
\newcount\DUMMY \newcount\DUMMY \newcount\DUMMY 
\newcount\DUMMY \newcount\DUMMY \newcount\DUMMY 
\newcount\DUMMY \newcount\DUMMY \newcount\DUMMY 
\newcount\DUMMY \newcount\DUMMY \newcount\DUMMY

\newcount\DUMMY \TypeStack=\allocationnumber
\newcount\DUMMY \newcount\DUMMY \newcount\DUMMY 
\newcount\DUMMY \newcount\DUMMY \newcount\DUMMY 
\newcount\DUMMY \newcount\DUMMY \newcount\DUMMY 
\newcount\DUMMY \newcount\DUMMY \newcount\DUMMY 
\newcount\DUMMY \newcount\DUMMY \newcount\DUMMY

\newcount\DUMMY \SpanStack=\allocationnumber
\newcount\DUMMY \newcount\DUMMY \newcount\DUMMY 
\newcount\DUMMY \newcount\DUMMY \newcount\DUMMY 
\newcount\DUMMY \newcount\DUMMY \newcount\DUMMY 
\newcount\DUMMY \newcount\DUMMY \newcount\DUMMY 
\newcount\DUMMY \newcount\DUMMY \newcount\DUMMY

\newbox\DUMMY   \BoxStack=\allocationnumber
\newbox\DUMMY   \newbox\DUMMY \newbox\DUMMY 
\newbox\DUMMY   \newbox\DUMMY \newbox\DUMMY 
\newbox\DUMMY   \newbox\DUMMY \newbox\DUMMY 
\newbox\DUMMY   \newbox\DUMMY \newbox\DUMMY 
\newbox\DUMMY   \newbox\DUMMY \newbox\DUMMY

\def\wlog{\immediate\write\m@ne}


\def\GetItemAll#1{%
 \GetItemSTATUS{#1}
 \GetItemNUMBER{#1}
 \GetItemTYPE{#1}
 \GetItemSPAN{#1}
 \GetItemBOX{#1}
}

\def\GetItemSTATUS#1{%
 \Point=\StatusStack
 \advance\Point by #1
 \global\ItemSTATUS=\count\Point
}

\def\GetItemNUMBER#1{%
 \Point=\NumStack
 \advance\Point by #1
 \global\ItemNUMBER=\count\Point
}

\def\GetItemTYPE#1{%
 \Point=\TypeStack
 \advance\Point by #1
 \global\ItemTYPE=\count\Point
}

\def\GetItemSPAN#1{%
 \Point\SpanStack
 \advance\Point by #1
 \global\ItemSPAN=\count\Point
}

\def\GetItemBOX#1{%
 \Point=\BoxStack
 \advance\Point by #1
 \global\setbox\ItemBOX=\vbox{\copy\Point}
 \global\ItemSIZE=\ht\ItemBOX
 \global\advance\ItemSIZE by \dp\ItemBOX
 \TEMPCOUNT=\ItemSIZE
 \divide\TEMPCOUNT by \Leading
 \divide\TEMPCOUNT by 65536
 \advance\TEMPCOUNT \@ne
 \ItemSIZE=\TEMPCOUNT pt
 \global\multiply\ItemSIZE by \Leading
}


\def\JoinStack{%
 \ifnum\LengthOfStack=\MaxItems 
  \Warn{WARNING: Stack is full...some items will be lost!}
 \else
  \Point=\StatusStack
  \advance\Point by \LengthOfStack
  \global\count\Point=\ItemSTATUS
  \Point=\NumStack
  \advance\Point by \LengthOfStack
  \global\count\Point=\ItemNUMBER
  \Point=\TypeStack
  \advance\Point by \LengthOfStack
  \global\count\Point=\ItemTYPE
  \Point\SpanStack
  \advance\Point by \LengthOfStack
  \global\count\Point=\ItemSPAN
  \Point=\BoxStack
  \advance\Point by \LengthOfStack
  \global\setbox\Point=\vbox{\copy\ItemBOX}
  \global\advance\LengthOfStack \@ne
  \ifnum\ItemTYPE=\Figure 
   \global\MoreFigurestrue
  \else
   \global\MoreTablestrue
  \fi
 \fi
}


\def\LeaveStack#1{%
 {\Iteration=#1
 \loop
 \ifnum\Iteration<\LengthOfStack
  \advance\Iteration \@ne
  \GetItemSTATUS{\Iteration}
   \advance\Point by \m@ne
   \global\count\Point=\ItemSTATUS
  \GetItemNUMBER{\Iteration}
   \advance\Point by \m@ne
   \global\count\Point=\ItemNUMBER
  \GetItemTYPE{\Iteration}
   \advance\Point by \m@ne
   \global\count\Point=\ItemTYPE
  \GetItemSPAN{\Iteration}
   \advance\Point by \m@ne
   \global\count\Point=\ItemSPAN
  \GetItemBOX{\Iteration}
   \advance\Point by \m@ne
   \global\setbox\Point=\vbox{\copy\ItemBOX}
 \repeat}
 \global\advance\LengthOfStack by \m@ne
}


\newif\ifStackNotClean

\def\CleanStack{%
 \StackNotCleantrue
 {\Iteration=\z@
  \loop
   \ifStackNotClean
    \GetItemSTATUS{\Iteration}
    \ifnum\ItemSTATUS=\InStack
     \advance\Iteration \@ne
     \else
      \LeaveStack{\Iteration}
    \fi
   \ifnum\LengthOfStack<\Iteration
    \StackNotCleanfalse
   \fi
 \repeat}
}


\def\FindItem#1#2{%
 \global\StackPointer=\m@ne 
 {\Iteration=\z@
  \loop
  \ifnum\Iteration<\LengthOfStack
   \GetItemSTATUS{\Iteration}
   \ifnum\ItemSTATUS=\InStack
    \GetItemTYPE{\Iteration}
    \ifnum\ItemTYPE=#1
     \GetItemNUMBER{\Iteration}
     \ifnum\ItemNUMBER=#2
      \global\StackPointer=\Iteration
      \Iteration=\LengthOfStack 
     \fi
    \fi
   \fi
  \advance\Iteration \@ne
 \repeat}
}


\def\FindNext{%
 \global\StackPointer=\m@ne 
 {\Iteration=\z@
  \loop
  \ifnum\Iteration<\LengthOfStack
   \GetItemSTATUS{\Iteration}
   \ifnum\ItemSTATUS=\InStack
    \GetItemTYPE{\Iteration}
   \ifnum\ItemTYPE=\Figure
    \ifMoreFigures
      \global\NextItem=\Figure
      \global\StackPointer=\Iteration
      \Iteration=\LengthOfStack 
    \fi
   \fi
   \ifnum\ItemTYPE=\Table
    \ifMoreTables
      \global\NextItem=\Table
      \global\StackPointer=\Iteration
      \Iteration=\LengthOfStack 
    \fi
   \fi
  \fi
  \advance\Iteration \@ne
 \repeat}
}


\def\ChangeStatus#1#2{%
 \Point=\StatusStack
 \advance\Point by #1
 \global\count\Point=#2
}



\def\Zone{\InZoneA}

\ZoneBAdjust=\z@

\def\MakePage{
 \global\ZoneBSize=\PageHeight
 \global\TextSize=\ZoneBSize
 \global\ZoneAFullPagefalse
 \global\topskip=\TextLeading
 \MakePageInCompletetrue
 \MoreFigurestrue
 \MoreTablestrue
 \FigInZoneBfalse
 \FigInZoneCfalse
 \TabInZoneBfalse
 \TabInZoneCfalse
 \global\FirstSingleItemtrue
 \global\FirstZoneAtrue
 \global\setbox\ZoneABOX=\box\VOIDBOX
 \global\setbox\ZoneBBOX=\box\VOIDBOX
 \global\setbox\ZoneCBOX=\box\VOIDBOX
 \loop
  \ifMakePageInComplete
 \FindNext
 \ifnum\StackPointer=\m@ne
  \NextItem=\m@ne
  \MoreFiguresfalse
  \MoreTablesfalse
 \fi
 \ifnum\NextItem=\Figure
   \FindItem{\Figure}{\NextFigure}
   \ifnum\StackPointer=\m@ne \global\MoreFiguresfalse
   \else
    \GetItemSPAN{\StackPointer}
    \ifnum\ItemSPAN=\Single \def\Zone{\InZoneB}\relax
     \ifFigInZoneC \global\MoreFiguresfalse\fi
    \else
     \def\Zone{\InZoneA}
     \ifFigInZoneB \def\Zone{\InZoneC}\fi
    \fi
   \fi
   \ifMoreFigures\Print{}\FigureItems\fi
 \fi
\ifnum\NextItem=\Table
   \FindItem{\Table}{\NextTable}
   \ifnum\StackPointer=\m@ne \global\MoreTablesfalse
   \else
    \GetItemSPAN{\StackPointer}
    \ifnum\ItemSPAN=\Single\relax
     \ifTabInZoneC \global\MoreTablesfalse\fi
    \else
     \def\Zone{\InZoneA}
     \ifTabInZoneB \def\Zone{\InZoneC}\fi
    \fi
   \fi
   \ifMoreTables\Print{}\TableItems\fi
 \fi
   \MakePageInCompletefalse 
   \ifMoreFigures\MakePageInCompletetrue\fi
   \ifMoreTables\MakePageInCompletetrue\fi
 \repeat
 \ifZoneAFullPage
  \global\TextSize=\z@
  \global\ZoneBSize=\z@
  \global\vsize=\z@\relax
  \global\topskip=\z@\relax
  \vbox to \z@{\vss}
  \eject
 \else
 \global\advance\ZoneBSize by -\ZoneBAdjust
 \global\vsize=\ZoneBSize
 \global\hsize=\ColumnWidth
 \global\ZoneBAdjust=\z@
 \ifdim\TextSize<23pt
 \Warn{}
 \Warn{* Making column fall short: TextSize=\the\TextSize *}
 \vskip-\lastskip\eject\fi
 \fi
}

\def\MakeRightCol{
 \global\TextSize=\ZoneBSize
 \MakePageInCompletetrue
 \MoreFigurestrue
 \MoreTablestrue
 \global\FirstSingleItemtrue
 \global\setbox\ZoneBBOX=\box\VOIDBOX
 \def\Zone{\InZoneB}
 \loop
  \ifMakePageInComplete
 \FindNext
 \ifnum\StackPointer=\m@ne
  \NextItem=\m@ne
  \MoreFiguresfalse
  \MoreTablesfalse
 \fi
 \ifnum\NextItem=\Figure
   \FindItem{\Figure}{\NextFigure}
   \ifnum\StackPointer=\m@ne \MoreFiguresfalse
   \else
    \GetItemSPAN{\StackPointer}
    \ifnum\ItemSPAN=\Double\relax
     \MoreFiguresfalse\fi
   \fi
   \ifMoreFigures\Print{}\FigureItems\fi
 \fi
 \ifnum\NextItem=\Table
   \FindItem{\Table}{\NextTable}
   \ifnum\StackPointer=\m@ne \MoreTablesfalse
   \else
    \GetItemSPAN{\StackPointer}
    \ifnum\ItemSPAN=\Double\relax
     \MoreTablesfalse\fi
   \fi
   \ifMoreTables\Print{}\TableItems\fi
 \fi
   \MakePageInCompletefalse 
   \ifMoreFigures\MakePageInCompletetrue\fi
   \ifMoreTables\MakePageInCompletetrue\fi
 \repeat
 \ifZoneAFullPage
  \global\TextSize=\z@
  \global\ZoneBSize=\z@
  \global\vsize=\z@\relax
  \global\topskip=\z@\relax
  \vbox to \z@{\vss}
  \eject
 \else
 \global\vsize=\ZoneBSize
 \global\hsize=\ColumnWidth
 \ifdim\TextSize<23pt
 \Warn{}
 \Warn{* Making column fall short: TextSize=\the\TextSize *}
 \vskip-\lastskip\eject\fi
\fi
}

\def\FigureItems{
 \Print{Considering...}
 \ShowItem{\StackPointer}
 \GetItemBOX{\StackPointer} 
 \GetItemSPAN{\StackPointer}
  \CheckFitInZone 
  \ifnum\ItemFits=\Yes
   \ifnum\ItemSPAN=\Single
     \ChangeStatus{\StackPointer}{\InZoneB} 
     \global\FigInZoneBtrue
     \ifFirstSingleItem
      \hbox{}\vskip-\BodgeHeight
     \global\advance\ItemSIZE by \TextLeading
     \fi
     \unvbox\ItemBOX\ItemSep
     \global\FirstSingleItemfalse
     \global\advance\TextSize by -\ItemSIZE
     \global\advance\TextSize by -\TextLeading
   \else
    \ifFirstZoneA
     \global\advance\ItemSIZE by \TextLeading
     \global\FirstZoneAfalse\fi
    \global\advance\TextSize by -\ItemSIZE
    \global\advance\TextSize by -\TextLeading
    \global\advance\ZoneBSize by -\ItemSIZE
    \global\advance\ZoneBSize by -\TextLeading
    \ifFigInZoneB\relax
     \else
     \ifdim\TextSize<3\TextLeading
     \global\ZoneAFullPagetrue
     \fi
    \fi
    \ChangeStatus{\StackPointer}{\Zone}
    \ifnum\Zone=\InZoneC \global\FigInZoneCtrue\fi
  \fi
   \Print{TextSize=\the\TextSize}
   \Print{ZoneBSize=\the\ZoneBSize}
  \global\advance\NextFigure \@ne
   \Print{This figure has been placed.}
  \else
   \Print{No space available for this figure...holding over.}
   \Print{}
   \global\MoreFiguresfalse
  \fi
}

\def\TableItems{
 \Print{Considering...}
 \ShowItem{\StackPointer}
 \GetItemBOX{\StackPointer} 
 \GetItemSPAN{\StackPointer}
  \CheckFitInZone 
  \ifnum\ItemFits=\Yes
   \ifnum\ItemSPAN=\Single
    \ChangeStatus{\StackPointer}{\InZoneB}
     \global\TabInZoneBtrue
     \ifFirstSingleItem
      \hbox{}\vskip-\BodgeHeight
     \global\advance\ItemSIZE by \TextLeading
     \fi
     \unvbox\ItemBOX\ItemSep
     \global\FirstSingleItemfalse
     \global\advance\TextSize by -\ItemSIZE
     \global\advance\TextSize by -\TextLeading
   \else
    \ifFirstZoneA
    \global\advance\ItemSIZE by \TextLeading
    \global\FirstZoneAfalse\fi
    \global\advance\TextSize by -\ItemSIZE
    \global\advance\TextSize by -\TextLeading
    \global\advance\ZoneBSize by -\ItemSIZE
    \global\advance\ZoneBSize by -\TextLeading
    \ifFigInZoneB\relax
     \else
     \ifdim\TextSize<3\TextLeading
     \global\ZoneAFullPagetrue
     \fi
    \fi
    \ChangeStatus{\StackPointer}{\Zone}
    \ifnum\Zone=\InZoneC \global\TabInZoneCtrue\fi
   \fi
  \global\advance\NextTable \@ne
   \Print{This table has been placed.}
  \else
  \Print{No space available for this table...holding over.}
   \Print{}
   \global\MoreTablesfalse
  \fi
}


\def\CheckFitInZone{%
{\advance\TextSize by -\ItemSIZE
 \advance\TextSize by -\TextLeading
 \ifFirstSingleItem
  \advance\TextSize by \TextLeading
 \fi
 \ifnum\Zone=\InZoneA\relax
  \else \advance\TextSize by -\ZoneBAdjust
 \fi
 \ifdim\TextSize<3\TextLeading \global\ItemFits=\No
 \else \global\ItemFits=\Yes\fi}
}

\def\BeginOpening{%
  \ninepoint
  \thispagestyle{titlepage}%
  \global\setbox\ItemBOX=\vbox\bgroup%
    \hsize=\PageWidth%
    \hrule height \z@
    \ifsinglecol\vskip 6pt\fi 
}

\let\begintopmatter=\BeginOpening  

\def\EndOpening{%
  \One
  \egroup
  \ifsinglecol
    \box\ItemBOX%
    \vskip\TextLeading plus 2\TextLeading
    \@noafterindent
  \else
    \ItemNUMBER=\z@%
    \ItemTYPE=\Figure
    \ItemSPAN=\Double
    \ItemSTATUS=\InStack
    \JoinStack
  \fi
}


\newif\if@here  \@herefalse

\def\no@float{\global\@heretrue}
\let\nofloat=\relax 

\def\beginfigure{%
  \@ifstar{\global\@dfloattrue \@bfigure}{\global\@dfloatfalse \@bfigure}%
}

\def\@bfigure#1{%
  \par
  \if@dfloat
    \ItemSPAN=\Double
    \TEMPDIMEN=\PageWidth
  \else
    \ItemSPAN=\Single
    \TEMPDIMEN=\ColumnWidth
  \fi
  \ifsinglecol
    \TEMPDIMEN=\PageWidth
  \else
    \ItemSTATUS=\InStack
    \ItemNUMBER=#1%
    \ItemTYPE=\Figure
  \fi
  \bgroup
    \hsize=\TEMPDIMEN
    \global\setbox\ItemBOX=\vbox\bgroup
      \eightpoint\nostb@ls{10pt}%
      \let\caption=\fig@caption
      \ifsinglecol \let\nofloat=\no@float\fi
}

\def\fig@caption#1{%
  \vskip 5.5pt plus 6pt%
  \bgroup 
    \eightpoint\nostb@ls{10pt}%
    \setbox\TEMPBOX=\hbox{#1}%
    \ifdim\wd\TEMPBOX>\TEMPDIMEN
      \noindent \unhbox\TEMPBOX\par
    \else
      \hbox to \hsize{\hfil\unhbox\TEMPBOX\hfil}%
    \fi
  \egroup
}

\def\endfigure{%
  \par\egroup 
  \egroup
  \ifsinglecol
    \if@here \midinsert\global\@herefalse\else \topinsert\fi
      \unvbox\ItemBOX
    \endinsert
  \else
    \JoinStack
    \Print{Processing source for figure \the\ItemNUMBER}%
  \fi
}


\newbox\tab@cap@box
\def\tab@caption#1{\global\setbox\tab@cap@box=\hbox{#1\par}}

\newtoks\tab@txt@toks
\long\def\tab@txt#1{\global\tab@txt@toks={#1}\global\table@txttrue}

\newif\iftable@txt  \table@txtfalse
\newif\if@dfloat    \@dfloatfalse

\def\begintable{%
  \@ifstar{\global\@dfloattrue \@btable}{\global\@dfloatfalse \@btable}%
}

\def\@btable#1{%
  \par
  \if@dfloat
    \ItemSPAN=\Double
    \TEMPDIMEN=\PageWidth
  \else
    \ItemSPAN=\Single
    \TEMPDIMEN=\ColumnWidth
  \fi
  \ifsinglecol
    \TEMPDIMEN=\PageWidth
  \else
    \ItemSTATUS=\InStack
    \ItemNUMBER=#1%
    \ItemTYPE=\Table
  \fi
  \bgroup
    \eightpoint\nostb@ls{10pt}%
    \global\setbox\ItemBOX=\vbox\bgroup
      \let\caption=\tab@caption
      \let\tabletext=\tab@txt
      \ifsinglecol \let\nofloat=\no@float\fi
}

\def\endtable{%
  \par\egroup 
  \egroup
  \setbox\TEMPBOX=\hbox to \TEMPDIMEN{%
    \eightpoint\nostb@ls{10pt}%
    \hss
    \vbox{%
      \hsize=\wd\ItemBOX
      \ifvoid\tab@cap@box
      \else
        \noindent\unhbox\tab@cap@box
        \vskip 5.5pt plus 6pt%
      \fi
      \box\ItemBOX
      \iftable@txt
        \vskip 10pt%
        \noindent\the\tab@txt@toks
        \global\table@txtfalse
      \fi
    }%
    \hss
  }%
  \ifsinglecol
    \if@here \midinsert\global\@herefalse\else \topinsert\fi
      \box\TEMPBOX
    \endinsert
  \else
    \global\setbox\ItemBOX=\box\TEMPBOX
    \JoinStack
    \Print{Processing source for table \the\ItemNUMBER}%
  \fi
}

\def\UnloadZoneA{%
\FirstZoneAtrue
 \Iteration=\z@
  \loop
   \ifnum\Iteration<\LengthOfStack
    \GetItemSTATUS{\Iteration}
    \ifnum\ItemSTATUS=\InZoneA
     \GetItemBOX{\Iteration}
     \ifFirstZoneA \vbox to \BodgeHeight{\vfil}%
     \FirstZoneAfalse\fi
     \unvbox\ItemBOX\ItemSep
     \LeaveStack{\Iteration}
     \else
     \advance\Iteration \@ne
   \fi
 \repeat
}

\def\UnloadZoneC{%
\Iteration=\z@
  \loop
   \ifnum\Iteration<\LengthOfStack
    \GetItemSTATUS{\Iteration}
    \ifnum\ItemSTATUS=\InZoneC
     \GetItemBOX{\Iteration}
     \ItemSep\unvbox\ItemBOX
     \LeaveStack{\Iteration}
     \else
     \advance\Iteration \@ne
   \fi
 \repeat
}


\def\ShowItem#1{
  {\GetItemAll{#1}
  \Print{\the#1:
  {TYPE=\ifnum\ItemTYPE=\Figure Figure\else Table\fi}
  {NUMBER=\the\ItemNUMBER}
  {SPAN=\ifnum\ItemSPAN=\Single Single\else Double\fi}
  {SIZE=\the\ItemSIZE}}}
}

\def\ShowStack{%
 \Print{}
 \Print{LengthOfStack = \the\LengthOfStack}
 \ifnum\LengthOfStack=\z@ \Print{Stack is empty}\fi
 \Iteration=\z@
 \loop
 \ifnum\Iteration<\LengthOfStack
  \ShowItem{\Iteration}
  \advance\Iteration \@ne
 \repeat
}

\def\B#1#2{%
\hbox{\vrule\kern-0.4pt\vbox to #2{%
\hrule width #1\vfill\hrule}\kern-0.4pt\vrule}
}


\newif\ifsinglecol   \singlecolfalse

\def\onecolumn{%
  \global\output={\singlecoloutput}%
  \global\hsize=\PageWidth
  \global\vsize=\PageHeight
  \global\ColumnWidth=\hsize
  \global\TextLeading=12pt
  \global\Leading=12
  \global\singlecoltrue
  \global\let\onecolumn=\relax
  \global\let\footnote=\sing@footnote
  \global\let\vfootnote=\sing@vfootnote
  \ninepoint 
  \message{(Single column)}%
}

\def\singlecoloutput{%
  \shipout\vbox{\PageHead\vbox to \PageHeight{\pagebody\vss}\PageFoot}%
  \advancepageno
  \ifplate@page
    \shipout\vbox{%
      \sp@pagetrue
      \def\sp@type{plate}%
      \global\plate@pagefalse
      \PageHead\vbox to \PageHeight{\unvbox\plt@box\vfil}\PageFoot%
    }%
    \message{[plate]}%
    \advancepageno
  \fi
  \ifnum\outputpenalty>-\@MM \else\dosupereject\fi%
}

\def\ItemSep{\vskip\ItemSepamount\relax}

\def\ItemSepbreak{\par\ifdim\lastskip<\ItemSepamount
  \removelastskip\penalty-200\ItemSep\fi%
}


\let\@@endinsert=\endinsert 

\def\endinsert{\egroup 
  \if@mid \dimen@\ht\z@ \advance\dimen@\dp\z@ \advance\dimen@12\p@
    \advance\dimen@\pagetotal \advance\dimen@-\pageshrink
    \ifdim\dimen@>\pagegoal\@midfalse\p@gefalse\fi\fi
  \if@mid \ItemSep\box\z@\ItemSepbreak
  \else\insert\topins{\penalty100 
    \splittopskip\z@skip
    \splitmaxdepth\maxdimen \floatingpenalty\z@
    \ifp@ge \dimen@\dp\z@
    \vbox to\vsize{\unvbox\z@\kern-\dimen@}
    \else \box\z@\nobreak\ItemSep\fi}\fi\endgroup%
}


\def\gobbleone#1{}
\def\gobbletwo#1#2{}
\let\footnote=\gobbletwo 
\let\vfootnote=\gobbleone

\def\sing@footnote#1{\let\@sf\empty 
  \ifhmode\edef\@sf{\spacefactor\the\spacefactor}\/\fi
  \hbox{$^{\hbox{\eightpoint #1}}$}\@sf\sing@vfootnote{#1}%
}

\def\sing@vfootnote#1{\insert\footins\bgroup\eightpoint\b@ls{9pt}%
  \interlinepenalty\interfootnotelinepenalty
  \splittopskip\ht\strutbox 
  \splitmaxdepth\dp\strutbox \floatingpenalty\@MM
  \leftskip\z@skip \rightskip\z@skip \spaceskip\z@skip \xspaceskip\z@skip
  \noindent $^{\scriptstyle\hbox{#1}}$\hskip 4pt%
    \footstrut\futurelet\next\fo@t%
}

\def\footnoterule{\kern-3\p@ \hrule height \z@ \kern 3\p@}

\skip\footins=19.5pt plus 12pt minus 1pt
\count\footins=1000
\dimen\footins=\maxdimen

\def\note#1#2{%
  \let\@sf=\empty \ifhmode\edef\@sf{\spacefactor\the\spacefactor}\/\fi
  #1\insert\footins\bgroup
    \eightpoint\b@ls{10pt}\rm
    \interlinepenalty\interfootnotelinepenalty
    \splitmaxdepth\dp\strutbox \floatingpenalty\@MM
    \leftskip\z@skip \rightskip\z@skip \spaceskip\z@skip \xspaceskip\z@skip
    \noindent\footstrut #1$\,$#2\strut\par
  \egroup
  \@sf\relax}


\def\landscape{%
  \global\TEMPDIMEN=\PageWidth
  \global\PageWidth=\PageHeight
  \global\PageHeight=\TEMPDIMEN
  \global\let\landscape=\relax
  \onecolumn
  \message{(landscape)}%
  \raggedbottom
}


\output{%
  \ifLeftCOL
    \global\setbox\LeftBOX=\vbox to \ZoneBSize{\box255\unvbox\ZoneBBOX
      \ifvoid\footins\else
        \vskip\skip\footins\unvbox\footins\fi
    }%
    \global\LeftCOLfalse
    \MakeRightCol
  \else
    \setbox\RightBOX=\vbox to \ZoneBSize{\box255\unvbox\ZoneBBOX
      \ifvoid\footins\else
        \vskip\skip\footins\unvbox\footins\fi
    }%
    \setbox\MidBOX=\hbox{\box\LeftBOX\hskip\ColumnGap\box\RightBOX}%
    \setbox\PageBOX=\vbox to \PageHeight{%
      \UnloadZoneA\box\MidBOX\UnloadZoneC}%
    \shipout\vbox{\PageHead\vbox to \PageHeight{\box\PageBOX\vss}\PageFoot}%
    \advancepageno
    \ifplate@page
      \shipout\vbox{%
        \sp@pagetrue
        \def\sp@type{plate}%
        \global\plate@pagefalse
        \PageHead\vbox to \PageHeight{\unvbox\plt@box\vfil}\PageFoot%
      }%
      \message{[plate]}%
      \advancepageno
    \fi
    \global\LeftCOLtrue
    \CleanStack
    \MakePage
  \fi
}


\Warn{\start@mess}

\newif\ifCUPmtplainloaded 
\ifprod@font
  \global\CUPmtplainloadedtrue
\fi


\catcode `\@=12 

